\PassOptionsToPackage{pdfpagelabels=false}{hyperref}
\documentclass[useAMS,usenatbib]{mnras}
\pdfoutput=1
\usepackage[pdftex]{graphicx}
\usepackage{amsmath}
\usepackage{natbib}
\usepackage{color}
\usepackage{times}

\def \MSUN{{\rm M}_{\odot}}

\def \RTC{R_{\rm 200c}}
\def \MTC{M_{\rm 200c}}
\def \MSTC{M^{\rm stars}_{200c}}
\def \RFC{R_{\rm 500c}}
\def \MFC{M_{\rm 500c}}

\newcommand{\ap}[1]{\textcolor{magenta}{#1}}

\title[IllustrisTNG: the stellar mass content of galaxy groups and clusters]{First results from the IllustrisTNG simulations: \\the stellar mass content of groups and clusters of galaxies}

\author[Pillepich et al.] {Annalisa Pillepich$^1$$^,$$^2$\thanks{E-mail: pillepich@mpia-hd.mpg.de},
Dylan Nelson$^3$, 
Lars Hernquist$^2$,
Volker Springel$^4$$^,$$^5$,
\newauthor
R{\"u}diger Pakmor$^4$,
Paul Torrey$^6$\thanks{Hubble Fellow},
Rainer Weinberger$^4$, 
Shy Genel$^7$$^,$$^8$,
Jill P. Naiman$^2$,
\newauthor
Federico Marinacci$^6$,
and Mark Vogelsberger$^6$\thanks{Alfred P. Sloan Fellow}
\vspace{2mm}
\\
$^1${Max-Planck-Institut f{\"u}r Astronomie, K{\"o}nigstuhl 17, 69117 Heidelberg, Germany}\\
$^2${Harvard--Smithsonian Center for Astrophysics, 60 Garden Street, Cambridge, MA 02138}\\
$^3${Max-Planck-Institut f{\"u}r Astrophysik, Karl-Schwarzschild-Str. 1, 85741 Garching, Germany}\\
$^4${Heidelberg Institute for Theoretical Studies, Schloss-Wolfsbrunnenweg 35, D-69118 Heidelberg, Germany}\\
$^5${Zentrum f{\"u}r Astronomie der Universit{\"a}t Heidelberg, ARI, M{\"o}nchhofstr. 12-14, D-69120 Heidelberg, Germany}\\
$^6${Kavli Institute for Astrophysics and Space Research, Massachusetts Institute of Technology, Cambridge, MA 02139, USA}\\
$^7${Center for Computational Astrophysics, Flatiron Institute, 162 Fifth Avenue, New York, NY 10010, USA}\\
$^8${Columbia Astrophysics Laboratory, Columbia University, 550 West 120th Street, New York, NY 10027, USA}\\
}

\begin{document}
\maketitle
\begin{abstract}
The IllustrisTNG project is a new suite of cosmological magneto-hydrodynamical simulations of galaxy formation performed with the \textsc{Arepo} code and updated models for feedback physics. Here we introduce the first two simulations of the series, TNG100 and TNG300, and quantify the stellar mass content of about 4000 massive galaxy groups and clusters ($10^{13} \leq \MTC/\MSUN \leq 10^{15}$) at recent times ($z \leq 1$). 
The richest clusters have half of their total stellar mass bound to satellite galaxies, with the other half being associated with the central galaxy and the diffuse intra-cluster light.
Haloes more massive than about $5\times 10^{14}\MSUN$ have more diffuse stellar mass outside 100 kpc than within 100 kpc, with power-law slopes of the radial mass density distribution as shallow as the dark matter's ( $-3.5 \la \alpha_{\rm 3D} \la -3$).
Total halo mass is a very good predictor of stellar mass, and vice versa: at $z=0$, the 3D stellar mass measured within 30 kpc scales as $\propto (\MFC)^{0.49}$ with a $\sim 0.12$ dex scatter. This is possibly too steep in comparison to the available observational constraints, even though the abundance of TNG less massive galaxies ($\la 10^{11}\MSUN$ in stars) is in good agreement with the measured galaxy stellar mass functions at recent epochs. 
The 3D sizes of massive galaxies fall too on a tight ($\sim$0.16 dex scatter) power-law relation with halo mass, with $r^{\rm stars}_{\rm 0.5} \propto (\MTC)^{0.53}$.
Even more fundamentally, halo mass alone is a good predictor for the whole stellar mass profiles beyond the inner few kpc, and we show how on average these can be precisely recovered given a single mass measurement of the galaxy or its halo.
\end{abstract}


\section{Introduction}
\label{sec:intro}

The $\Lambda$-Cold Dark Matter ($\Lambda$CDM) cosmological paradigm \citep{PlanckXIII:2015} predicts a bottom-up hierarchical growth of structures, where massive systems such as groups and clusters of galaxies assembled recently through a series of mergers and the accretion of smaller units that previously formed due to gravitational instability within the dark matter (DM) dominated density field. As demonstrated by large-volume, gravity-only simulations \citep[e.g.][]{Evrard:2002, Angulo:2012, Potter:2016}, the abundance of the most massive haloes is a steeply falling function of halo mass. At the current epoch, objects of mass $10^{14} \,(10^{15})\, \MSUN$ are about 50 (a few thousand) times rarer than Milky-Way like haloes \citep[$10^{12}\,\MSUN$; e.g.][]{Skillman:2014}, and they were about ten (a hundred) times more infrequent at $z\sim1$ than today \citep[e.g.][]{Bocquet:2016}.

Although only a few per cent of all galaxies at the present day reside in these very massive haloes, observations have shown that such clusters contain a rich mixture of galaxy types, ranging from hundreds or even thousands of dwarf galaxies (as seen, e.g., in the Virgo, Fornax and Coma surveys, the Next Generation Virgo cluster Survey, the 2dF Fornax cluster survey, and SAURON) to the most massive galaxies in the Universe situated at the centers of their deep gravitational potential wells (studied in, e.g., ATLAS3D, The Massive Survey, among others). This galaxy population is thought to be shaped by intense mutual interactions in groups and clusters, such as galaxy harassment \citep{Moore:1996}, stripping \citep{Gunn:1972} or strangulation \citep{Larson:1980}, making these environments particularly interesting for studies of galaxy evolution. Also, the tidal stripping of stars from galaxies and the ingestion of smaller systems by the central cluster galaxy produces a diffuse intra-cluster light (ICL) component in the background halo. The overall prominence and radial extent of this low surface brightness envelope around the central galaxy forms an important integral constraint for cosmic structure formation.

Observationally constraining the ICL component is challenging, as it requires both wide and deep observations capable of capturing spatially extended low-surface brightness regions. It has been attempted for the SDSS with stacking methods \citep{Zibetti:2005}, with targeted observations of individual objects in the local Universe \citep[e.g.][]{Mihos:2005}, as well as with HST at intermediate redshifts in the Frontier Field and CLASH\footnote{Cluster Lensing and Supernova Survey with Hubble} clusters \citep{Montes:2014, DeMaio:2015, Burke:2015, Morishita:2016, Montes:2017}. Recently, \citealt{Huang:2017} have taken a leap forward by measuring the light profiles out to 100 kpc of a few thousand massive galaxies at intermediate redshifts using the Hyper Suprime Camera. The results thus far show significant quantitative discrepancies, but are of substantial importance, for example, for the theoretical interpretation of abundance matching constraints at the cluster scale \citep{Giodini:2009, Andreon:2012, Budzynski:2012, Gonzales:2013,Kravtsov:2014,Andreon:2015, Chiu:2016}. As the brightness profile of central galaxies often smoothly blends into the ICL, a particular challenge lies in arriving at an unambiguous definition of the stellar mass of the central galaxy, which has been a recurrent theme in previous modeling attempts of the ICL based on hydrodynamical simulations \citep{Puchwein:2010, Contini:2014,Cui:2014}, semi-analytical and controlled N-body experiments \citep{DeLucia:2007,Conroy:2007,Laporte:2013, Cooper:2015} or semi-empirical models \citep{Behroozi:2013, Moster:2017}. Properly assigning mass to the central galaxies in large haloes is also critical for meaningfully comparing to observational results for the massive end of the galaxy stellar mass function \citep{Baldry:2012, Bernardi:2013, Bernardi:2017, DSouza:2015}.

To model the evolution of massive haloes and their embedded galaxies in a physically meaningful way is challenging as it depends on a complex combination of many processes, coupling gravity, star formation, radiative gas cooling, heating, and stellar feedback in a highly non-linear fashion. Specific mechanisms that need to be considered include cosmological gas accretion, feedback and outflows driven by Active Galactic Nuclei (AGN), turbulence, thermal conduction, thermal instabilities, magnetic field amplification and transport of charged particles as well as galaxy-galaxy interactions, galaxy mergers, tidal disruption, interactions with the intra-cluster medium (ICM) itself, e.g., via ram-pressure stripping, and repeated high-speed galaxy encounters within the cluster gravitational potential \citep[see][for a review]{Kravtsov:2012}.  This diversity of the involved physical processes can be followed in full generality only through cosmological hydrodynamical simulations. However, the enormous range of scales (from sub-pc to many Mpc scales) makes this a formidable computational challenge \citep[see][for a review]{Borgani:2011rev}. Simulating this in a truly self-consistent fashion is still a distant goal, even though the ever increasing power of modern supercomputers and improvements in the physical fidelity of numerical codes have already allowed remarkable progress in this direction.

In recent years, efforts in galaxy formation simulations have mostly focused on individual haloes up to Milky-Way sized mass \citep[e.g.][]{Guedes:2011, Stinson:2013, Marinacci:2014, Hopkins:2014, Wang:2015, Christensen:2016, Grand:2017, Agertz:2016, Hopkins:2017} or on large cosmological volumes about $100^3$ Mpc$^3$ across \citep{Vogelsberger:2014a, Schaye:2015, Dubois:2014, Khandai:2015}. The latter have by construction spatial and mass resolutions limited to about the kpc length and $10^6\MSUN$ mass scales, respectively. Still, projects like Illustris \citep{Vogelsberger:2014a, Vogelsberger:2014b, Genel:2014, Sijacki:2015} and EAGLE \citep{Schaye:2015, Crain:2015} have shown that hydrodynamical simulations with sub-grid models tailored to such resolutions can reproduce many structural properties and scaling relations of observed galaxies across a wide range of masses. While some empirical input is used to set the free parameters of the models (such as feedback efficiencies), the large array of diverse physical measurements admitted by these calculations gives them considerable predictive power and makes them an invaluable tool to test galaxy formation theory. However, the limited volume covered by these simulations hinders their usage for an exploration of astrophysical processes acting on the largest scales of the Universe. In particular, this precludes the study of very rare objects, like rich clusters of galaxies, and the properties of their stellar mass content.

At present just a handful of hydrodynamical simulations exist which are capable of accessing the realm of {\it statistical} samples of haloes more massive than $10^{14}\MSUN$ and including some form of feedback from AGN \citep[][]{LeBrun:2014, Dolag:2016, McCarthy:2016}. Yet, none of them has the spatial and mass resolution needed to unveil the structural details of the cluster galaxies. In fact, some of those models have been specifically constructed to reproduce the global scaling relations of massive haloes (e.g.~the fraction of hot halo gas and the stellar mass - BH mass relations) by means of adjusting the adopted subgrid prescriptions accordingly, without paying much attention to the cluster galaxies themselves. Other numerical works have focused on simulating smaller cluster samples to gain insight on specific aspects of the numerical implementations \citep[e.g. the nIFTy comparison project,][]{Sembolini:2016} or of the cluster physics, like the impact of AGN feedback on the thermodynamical properties of the ICM \citep[e.g.][]{Puchwein:2008, Planelles:2014}, the stellar properties and profiles of the brightest central galaxies \citep{Puchwein:2010, Martizzi:2012b, Ragone:2013}, or the cool core - non cool core duality \citep[e.g.][]{Rasia:2015,Hahn:2017}.

\begin{figure*}
\centering                                                                      
\includegraphics[width=15.5cm]{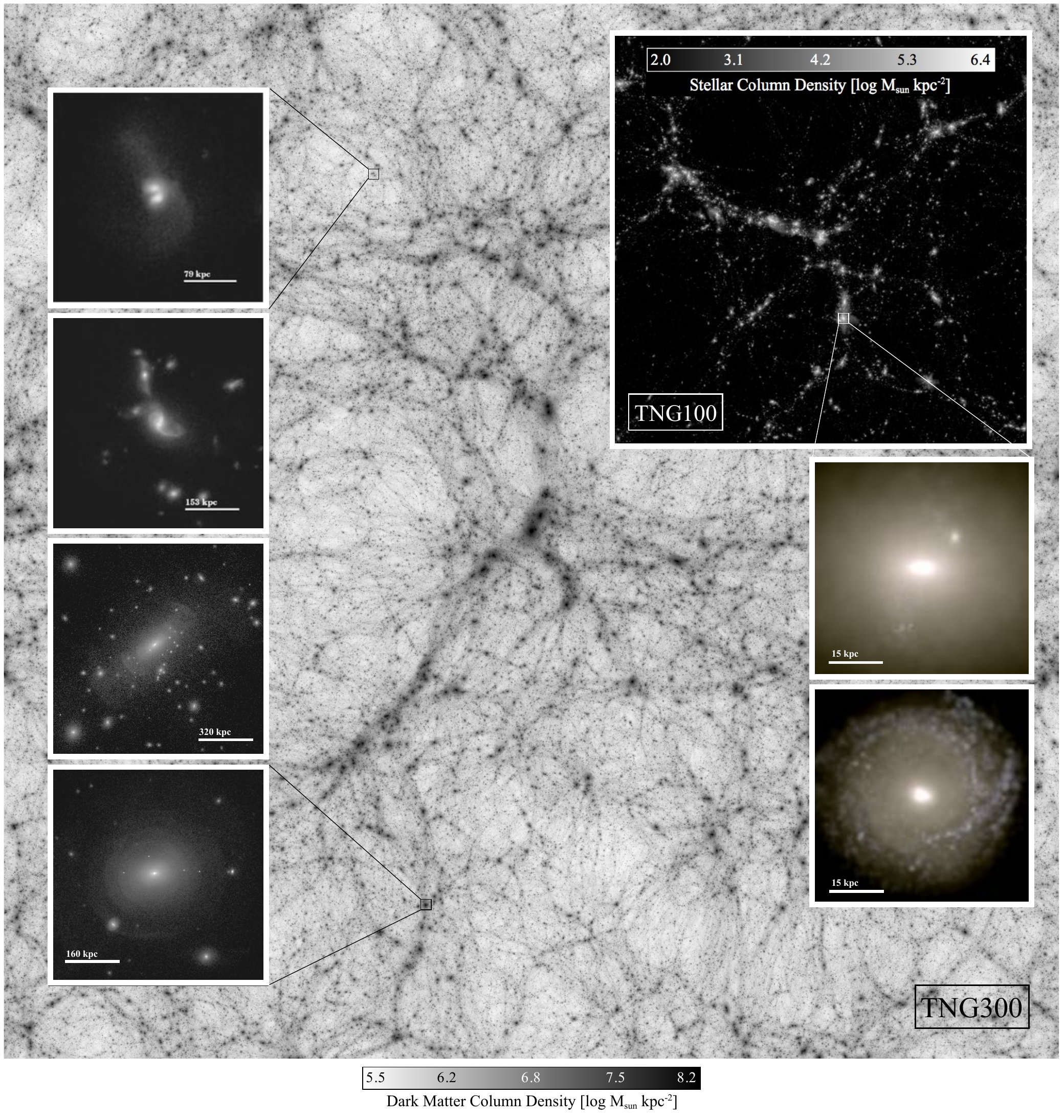}
\caption{\label{fig:tng}The IllustrisTNG Simulations: $z=0$ visual representation of the scope and spatial volumes encompassed by the TNG100 and TNG300 runs presented in this paper. The background represents the DM density field across the $\sim$ 300 Mpc volume of TNG300, while the upper right inset shows the distribution of stellar mass across the entire $\sim$ 100 Mpc volume of TNG100, each projected through a slice a third of the box in depth. Panels on the left show two examples of galaxy-galaxy interactions, and two examples of fine-grained structure of the extended stellar haloes -- shells, tidal tails, and luminous satellites -- around two massive ellipticals at $z=0$, in projected stellar mass density. The bottom right insets show the stellar light on scales of 60 kpc per side (face-on) of two randomly-selected $z=0$ galaxies with a stellar mass larger than $10^{11}\MSUN$, from the high-resolution TNG100 box. }
\end{figure*}

Recently, \cite{Bahe:2017} and \cite{Barnes:2017} have started to fill the gap between these approaches by producing a suite of 24 (Hydrangea) plus 6 additional (C-EAGLE) zoom-in simulations of massive galaxy clusters between $10^{14}$ and $10^{15.4}\MSUN$ with the resolution and galaxy physics model of the EAGLE simulation. In our new IllustrisTNG project, we go beyond this approach by pushing it further with a much larger, uniformly sampled cosmological
volume. We use an updated modeling technique that excels in the use of magneto-hydrodynamics (MHD) and improved treatments of AGN feedback, galactic winds and metal enrichment, and we aim for an unprecedented mix of statistical power, dynamic range in mass, numerical resolution, and included physics.

IllustrisTNG (Figure \ref{fig:tng}) is the follow up project of the Illustris simulation. Upon completion, it will consist of three large cosmological volumes, TNG50, TNG100, and TNG300 of about 50, 100, and 300 Mpc on a side, respectively. TNG100 features the same resolution and initial condition phases as Illustris, while TNG50 and TNG300 extend the series to a better resolution and a larger volume, respectively.
In this paper, we use TNG100 and TNG300 to provide a census of the stellar mass content and its spatial distribution within massive galaxy groups and clusters ($10^{13} \leq \MTC/\MSUN \la 10^{15}$). We focus on their central galaxies, diffuse ICL, satellite populations, and total stellar mass out to the virial radius.
In our simulations, galaxies and haloes are extended, spatially resolved objects, unlike in semi-analytical and semi-empirical models. The tidal truncation and eventual disruption of galaxies is followed consistently in the simulations, hence the results for the spatial distribution of the various stellar components in massive haloes provide a highly non-trivial prediction of the calculations. 

{\renewcommand{\arraystretch}{1.2}
\begin{table*}
  \caption{The TNG100 and TNG300 simulations of IllustrisTNG: table of physical and numerical parameters for the three resolution levels of the two simulations presented in this paper. A third simulation with side length 35$h^{-1} \sim 50$ Mpc (TNG50) completes the series: it is currently ongoing and will be presented in the future. The parameters are: the box side-length, the initial number of gas cells and dark matter particles, the target baryon mass, roughly equal to the average initial stellar particle mass, the dark matter particle mass, the $z$\,=\,0 Plummer equivalent gravitational softening of the collisionless component, and the minimum comoving value of the adaptive gas gravitational softenings. Lastly, the total run time including substructure identification in CPU core hours, and the number of compute cores used. For details on the adaptive mass and spatial resolution of the gas component see \protect\cite{Pillepich:2017}, Appendix A.}
  \label{tab:sims}
  \begin{center}
    \begin{tabular}{llllllllll}
     \hline\hline
     
 Run Name & $L_{\rm box}$ & $N_{\rm GAS}$ & $N_{\rm DM}$ & $m_{\rm baryon}$ & $m_{\rm DM}$ & $\epsilon_{\rm DM,stars}^{z=0}$ & $\epsilon_{\rm gas,min}$ & CPU Time & $N_{\rm cores}$ \\
  & [\,cMpc\,] & - & - & [\,10$^6$\,M$_\odot$\,] & [\,10$^6$\,M$_\odot$\,] & [\,kpc\,] & [\,ckpc\,] & [\,Mh\,] & - \\ \hline

 \textbf{TNG100}(-1) & $110.7$ & $1820^3$ & $1820^3$ & 1.4  & 7.5   & 0.74 & 0.19 & 18.0    & 10752 \\
 TNG100-2            & $110.7$ & $910^3$  & $910^3$  & 11.2 & 59.7  & 1.48 & 0.37 & 0.6     & 2688  \\
 TNG100-3            & $110.7$ & $455^3$  & $455^3$  & 89.2 & 477.7 & 2.95 & 0.74 & $\ll$ 1 & 336   \\
 \textbf{TNG300}(-1) & $302.6$ & $2500^3$ & $2500^3$ & 11   & 59    & 1.48 & 0.37 & 34.9    & 24000 \\
 TNG300-2            & $302.6$ & $1250^3$ & $1250^3$ & 88   & 470   & 2.95 & 0.74 & 1.3     & 6000  \\
 TNG300-3            & $302.6$ & $625^3$  & $625^3$  & 703  & 3764  & 5.90 & 1.47 & $\ll$ 1 & 768   \\
 \hline
 
    \end{tabular}
  \end{center}
\end{table*}}

The structure of the paper is hence as follows. In Section~\ref{sec:tng}, we introduce all the technical details of the IllustrisTNG simulations presented in this paper. Subtleties of the adopted methods are explained in Section~\ref{sec:methods}. The basic demographics of the high-mass end sample enabled by the TNG100 and TNG300 simulations is provided in Section~\ref{sec:sample}. We show the results of our models in terms of 3D enclosed stellar mass profiles as well as 3D stellar mass density profiles in Section~\ref{sec:profiles}, and give analytic fits to them that require as sole input either the total halo mass or the galaxy stellar mass within a given aperture. The mass budget in TNG groups and clusters is quantified in Section~\ref{sec:budget} in terms of scaling relations between stellar mass and total halo mass for different fixed spatial apertures. In the same section we assess the fractional contributions of the various cluster components to the total stellar mass budget and to the total halo mass, including a discussion on the scatter. We highlight selected theoretical insights enabled by the simulations in Section~\ref{sec:theory}, and summarize and discuss our general conclusions in Section~\ref{sec:summary}.


\section{The TNG Simulations}
\label{sec:tng}

In this paper, we analyze two recently finished simulations of the {\it The Next Generation} Illustris project: IllustrisTNG\footnote{\url{http://www.tng-project.org}}. IllustrisTNG (Fig.~\ref{fig:tng}) is an ongoing suite of magneto-hydrodynamical cosmological simulations that model the formation and evolution of galaxies within the $\Lambda$CDM paradigm. It builds upon the scientific achievements of the Illustris simulation \citep{Vogelsberger:2014a, Vogelsberger:2014b, Genel:2014} and improves upon Illustris by 1) extending the mass range of the simulated galaxies and haloes, 2) adopting an improved numerical and astrophysical modelling, and 3) addressing the identified shortcomings of the previous generation simulations \citep[summarized in][]{Nelson:2015b}. 

The galaxy formation model underlying the TNG simulations includes a new kinetic black hole feedback, magneto-hydrodynamics (MHD), and a revised scheme for galactic winds, among other changes. These are all described in full detail in the project's two method papers by \cite{Pillepich:2017} and \cite{Weinberger:2017}, where the former covers the aspects responsible for shaping low- to intermediate-mass galaxies and gives a detailed summary of the parameters used in the model, while the latter focuses on the high-mass end of the stellar mass function and introduces a new BH-driven wind feedback model, a fundamental change in the feedback modeling compared with Illustris. 

In the following, we will refer to the galaxy formation physics specified in these two studies as the TNG model. For the sake of brevity, we only mention its most salient features here. The TNG model builds upon the infrastructure developed for Illustris \citep{Vogelsberger:2013b,Torrey:2014a}: it follows radiative gas cooling modulated by a time-variable UV background, star formation regulated by a subgrid model for the interstellar medium \citep{Springel:2003}, galactic wind feedback powered by supernova explosions, a detailed metal enrichment model that tracks 9 elements and uses metallicity-dependent yields from SN-II, SN-Ia, and AGB stars, growth of supermassive black holes through Bondi gas accretion and black hole mergers, as well as thermal quasar feedback in high accretion rate states, and kinetic black hole wind feedback in low accretion rate states of the black holes. The magnetic fields are followed with ideal magneto-hydrodynamics and are dynamically coupled to the gas through the magnetic pressure. We point out that all TNG model parameters of the IllustrisTNG simulations are exactly as described in the default model of \cite{Pillepich:2017}, and their values are kept the same for all of our simulations, independent of numerical resolution.

IllustrisTNG uses the \textsc{Arepo} code \citep{Springel:2010} which employs a tree-particle-mesh algorithm to solve Poisson's equation for gravity and a second-order accurate finite-volume Godunov scheme on a moving, unstructured Voronoi-mesh for the equations of ideal magnetohydrodynamics. The divergence constraint of the magnetic field  is taken care of by an 8-wave Powell cleaning scheme described in \citet{Pakmor:2013}. A uniform magnetic field of strength $10^{-14}$~G (comoving) is set up in the initial conditions and functions as the seed for the subsequent self-consistent amplification of the magnetic fields through small-scale dynamo processes \citep{Rieder:2016,Pakmor:2017}. Previous work has shown that our results are insensitive to the precise value of this initial seed field \citep{Marinacci:2015}.

Compared to the code version used for the Illustris simulation, a number of improvements have been implemented and utilized in the TNG runs. These include an improved gradient estimate \citep{Pakmor:2016a} that leads to better convergence properties of the hydrodynamics scheme as well as to an improved angular momentum conservation, a refined advection scheme for passive scalars used to track metals \citep{Pillepich:2017}, and a modified, more efficient hierarchical time integration method for gravitational interactions (\textcolor{blue}{Springel et al. in prep}). Through local refinement and de-refinement operations acting upon the gas mesh, we ensure that the baryonic mass resolution of gas and stars always stays within a factor of two of the initial mass resolution (see Table~\ref{tab:sims}).

The IllustrisTNG runs studied in this work are two uniform mass resolution cosmological volume simulations with side lengths $75\,h^{-1} \approx 100$~Mpc and $205\,h^{-1}\approx 300$~Mpc, referred to as TNG100 and TNG300 in the following. A third set with side length $35\,h^{-1}\approx50$~Mpc (TNG50) is currently still in progress and will not be discussed in this paper. TNG100 has been performed at resolution similar to the original Illustris simulation, while TNG300 has a factor 8 (2) worse mass (spatial) resolution. TNG100 and TNG300 are each augmented by a series of lower resolution realizations (TNG100-2, TNG100-3, and TNG300-2, TNG300-3, respectively) of the same volumes, each respectively with 8 and 64 times more massive DM particles than their respective flagship counterpart. Table~\ref{tab:sims} provides an overview of the primary characteristics of the simulations used in this study. 


The initial conditions of all simulations have been set at $z=127$ using the Zeldovich approximation. The adopted cosmological parameters are
given by a matter density $\Omega_{\rm m} =\Omega_{\rm dm} + \Omega_{\rm b} = 0.3089$, baryonic density $\Omega_{\rm b} = 0.0486$, cosmological constant $\Omega_\Lambda=0.6911$, Hubble constant $H_0 = 100\,h\, {\rm km\, s^{-1}Mpc^{-1}}$ with $h=0.6774$, normalisation $\sigma_8 = 0.8159$ and spectral index $n_s=0.9667$ \citep[taken from Planck,][]{PlanckXIII:2015}.

In this paper, we focus on the stellar mass content of massive gravitationally collapsed haloes. It is one in a series of five papers that introduce the IllustrisTNG project, all analyzing different aspects of the new simulations, highlighting their wide scientific scope. In \textcolor{blue}{Nelson et al. (2017)}, we show that the color bimodality of TNG galaxies is in good agreement with data from SDSS, demonstrating that the included feedback mechanisms quench galaxies at the appropriate stellar mass scale. We also find that the spatial clustering of our simulated galaxies matches observational constraints both in the local universe and at higher redshift (\textcolor{blue}{Springel et al. 2017}). In \textcolor{blue}{Marinacci et al. (2017)}, we study the predicted magnetic field strengths in haloes and derive maps of radio synchrotron emission from massive galaxy clusters. Finally, in \textcolor{blue}{Naiman et al. (2017)} we trace the enrichment from r-processes and explore the chemical evolution of europium and magnesium directly in cosmological hydrodynamical simulations of Milky Way-like galaxies.


\section{Methods}
\label{sec:methods}

\subsection{Identification of simulated haloes and galaxies}
\label{sec:haloes}

Haloes, subhaloes, and their basic properties are obtained with the {\sc FOF} and {\sc subfind} algorithms \citep{Davis:1985, Springel:2001, Dolag:2009}, at each of 100 snapshots saved from $z\sim 20$ to $z=0$. Throughout, we call galaxy any luminous (sub)halo, i.e. any gravitationally bound object with nonzero stellar component. To identify (sub)haloes, first a standard friends-of-friends group finder is run to identify {\sc FOF} haloes (linking length 0.2) within which gravitationally bound substructures are then located and characterized hierarchically. The {\sc subfind} object catalog includes both central and satellite subhaloes: the position of centrals coincides with the {\sc FOF} centers (defined as the minimum of the gravitational potential), and centrals may contain one or more child subhaloes; satellite subhaloes may be either dark or luminous, and are members of their parent {\sc FOF} group regardless of their distance from the centers. For the halo masses examined in this work, every {\sc FOF} halo is univocally  associated to a central {\sc subfind} haloes. 

Throughout this paper, we refer to galaxies that do not reside within $\RTC$ or $\RFC$\footnote{$R_{\Delta}$ is defined as the radius within which the mean enclosed mass density is $\Delta$ times the critical value $\rho_c$ i.e. $\bar \rho_{\rm halo} = \Delta \rho_c$.} of a larger halo as \textit{centrals}. For any given central, \textit{satellites} are all {\sc subfind} luminous objects which reside within a certain (3D or 2D) spherical aperture. Member galaxies are either centrals or satellites.

Host haloes and galaxies are characterized by a total halo mass; here we adopt the spherical-overdensity mass $\MTC$ ($\MFC$), obtained by summing the mass of {\it all} particles and cells enclosed within $\RTC$ ($\RFC$) -- including dark matter, gas, stars and black holes.
At $z=0$, we measure the total mass of a satellite galaxy by summing the mass of all its gravitationally bound particles and cells, according to {\sc subfind}, regardless of their distance. The same applies to the stellar mass content alone: the {\sc Subfind} algorithm determines which stellar particles are gravitationally bound to which objects -- a stellar particle can be associated to only one galaxy (central or satellite). Therefore, the stars of a satellite galaxy belong to the satellite rather than to the host galaxy around which the satellite orbits. As a result, satellite stellar masses are not included in the stellar mass of their hosts. Where needed, we adopt a \cite{Chabrier:2003} initial mass function (IMF), consistent with the choice used for the simulations themselves.

Given the collection of gravitationally bound stellar particles assigned to a galaxy, the stellar half mass radius ($r_{\rm stars,1/2}$) is measured as the three dimensional radius containing half of the stellar mass of all constituent stars. 

\begin{table*}
 \begin{tabular}{|p{3.5cm}|p{7.2cm}|p{5.7cm} }
\hline
Name & Definition & Note\\
\hline
Groups and clusters of galaxies & Objects with a total spherical-overdensity halo mass of $\MTC \geq 10^{13}\MSUN$, described in this work also in terms of   $\MFC$. & The total mass includes contribution from DM, gas, stars, both smooth and within the central galaxy as well as including subhaloes/satellites and possibly gravitationally-unbound resolution elements.\\
& &\\
Central galaxy & Galaxy at the potential minimum of a group or cluster that does not reside within the virial radius (e.g. $\RTC$ or $\RFC$) of a larger halo. Its stellar mass is defined as the sum of all stellar particles mass {\it within} fixed apertures, e.g. 3D 10, 30, 100 physical or comoving kpc, excising satellites. & Here we use the terms central to refer to what others might also call host, parent galaxy or BCG (brightest cluster galaxy): in this work the mass of the central is {\it always} defined within a fixed (3D) aperture.\\ 
& &\\
Satellite galaxy& Any galaxy within the virial radius ($\MTC$ or $\MFC$)  of a central. Its stellar mass is defined as the sum of the mass of all stellar particles which are gravitationally bound according to the {\sc Subfind} algorithm. & This separation between centrals and satellites and hence the definition of a satellite stellar mass are not directly applicable to observations.\\
& &\\
ICL (intra-cluster light) & All stellar mass {\it beyond} a fixed aperture from the center of the group or cluster to a maximum boundary set by the virial radius (e.g. $\RTC$ or $\RFC$) of the hosting halo, regardless of origin and excising satellites and gravitationally unbound stars. & This is also referred to as diffuse envelopes, stellar haloes, or intra-halo light. \\
& &\\
diffuse mass (central + ICL) & All stellar mass in a group and cluster within a fixed aperture (here usually the virial radius) which is {\it not} locked in satellites, namely the sum of the stellar mass in the central galaxy and intra-cluster light. & This is what we have labeled in Section \ref{sec:profiles} $\MSTC$. It is very close to the mass of what is sometimes labeled as `BCG' in those observational analyses where the BCG total stellar mass/light budget is estimated by integrating the stellar profile to infinity. \\
\hline
\end{tabular}
\caption{\label{tab:defs} Operational definitions of the group and cluster components adopted in this work. As it will be justified in the next Sections, we advocate that a separation between central galaxies and intra-cluster light (ICL) cannot be generalized across mass scales, and that fixed-aperture boundaries are the best practical way to compare mass (or light) estimates across theoretical and observational compilations. The total stellar mass of a group or cluster is the sum of all components (central+ICL+satellites), out to a specified spherical or circular aperture. In this study, we adopt throughout as a boundary the virial radius of the underlying total halo, $\RTC$ or $\RFC$. The proposed separation can be extended to galaxies/haloes of any mass.}
\end{table*}
\subsection{On a galaxy's stellar mass and cluster components}
\label{sec:components}

Before we proceed any further and quantify the amount of stellar mass in TNG groups and clusters, it is helpful to outline a set of operational definitions for the different components. In Table~\ref{tab:defs}, we clarify the nomenclature applied throughout this work for referring to the various stellar components of massive haloes: the central galaxy (or BCG), satellite galaxies, the stellar halo or ICL, as well as combinations thereof. These terms are commonly used in the literature, although with varying definitions.

We proceed in our analysis by labeling all central galaxies of our sample with a series of stellar mass measurements performed within {\it fixed} spatial 3D apertures. Alternatively, one could have used 2D circularized apertures. The mass in the ICL is, conversely, defined as the sum of all mass residing beyond the same fixed spatial aperture. This operational definition eliminates any ambiguity in theoretical mass estimates from simulations, which is clearly desirable. Similarly, despite the complexity inherent in observational techniques to measure stellar mass or light, we hope that fixed-aperture estimates could become a well-defined standard practice. As it will be shown via the analysis of the stellar mass profiles in Section~\ref{sec:profiles}, no physical transition between central galaxies and ICL can be identified and generalized across mass scales, and hence fixed-aperture boundaries are the best practical and most unambiguous way to compare mass (or light) estimates across theoretical and observational compilations.

Two complications still remain: the definition of satellite masses, and the outer boundary of a halo. In both cases we still implement definitions which are theoretically motivated. Namely, the mass of a satellite galaxy is taken as the sum of all its gravitationally bound particles, while observationally its light would be mixed with the background ICL, requiring a 2D based separation. Similarly, the outer boundary of our haloes is adopted as the minimum between the virial radius and the distance of the farthest gravitationally bound stellar particle, while observationally this will be set by a surface brightness sensitivity limit. In advocating for fixed aperture stellar mass definitions, these need to be properly reconciled with a consideration of the relevant boundaries of objects.

\subsection{Resolution effects and rTNG300}
\label{sec:res}

As can be seen from Table \ref{tab:sims}, the TNG300 simulation is performed at a factor of 8 (2) lower in mass (spatial) resolution when compared to the flagship run TNG100. This is our typical step between two resolution levels. Therefore, TNG100-2 is realized at the same resolution as the flagship TNG300 run, while TNG100-3 corresponds to TNG300-2. As a reminder, TNG100 has the same resolution as the original Illustris simulation (aside from a small shift in the adopted cosmological parameters), with a mean baryonic mass of $1.4\times 10^6\MSUN$.

Other than the gravitational softening length parameters, all TNG runs use the default model parameter values given in \cite{Pillepich:2017}, with no adjustments with resolution. This approach is equivalent to aiming for ``strong resolution convergence'' \citep[in the language of][]{Schaye:2015}, in contrast to the ``weak resolution convergence'' approach where parameters are intentionally re-scaled in order to obtain better converged simulation output. As a result, and as explored in \citet[][Appendix A]{Pillepich:2017} and \citet[][Appendix B]{Weinberger:2017}, properties of the simulated galaxies may not be necessarily fully converged at the mass and spatial resolutions adopted in TNG100 and TNG300. In particular, in our galaxy physics model, stellar masses and star formation rates typically increase with better resolution. This is particularly problematic for TNG300, because we would like to exploit the powerful statistics of the large volume, but with galaxy properties as would be obtained at TNG100 resolution.

Thanks to the whole set of ancillary realizations, we have clear constraints on how the model results differ at different resolutions. We demonstrate this explicitly in Appendix \ref{sec:app_res} for all observables studied in this paper. In fact, despite the changes in box size and initial conditions phases, average results from TNG100-2 are in excellent agreement with those of TNG300, and the same applies at the next lower levels, TNG100-3 vs TNG300-2. 
We therefore account for possible resolution shifts by judicial application of a rescaling procedure. In practice, we assume that the model outcome at TNG100 resolution (the best in this work and the target one for the model development) is our best estimate of reality. For any observable studied in this paper, curves denoted `rTNG300' represent the TNG300 results {\it multiplied} by a resolution correction function informed by the comparison between TNG100 and TNG100-2 on the same statistics or observable (see Appendix~\ref{sec:app_res} and Eq.~(\ref{eq:appendix})). 

\begin{figure}
\centering
\includegraphics[width=8.6cm]{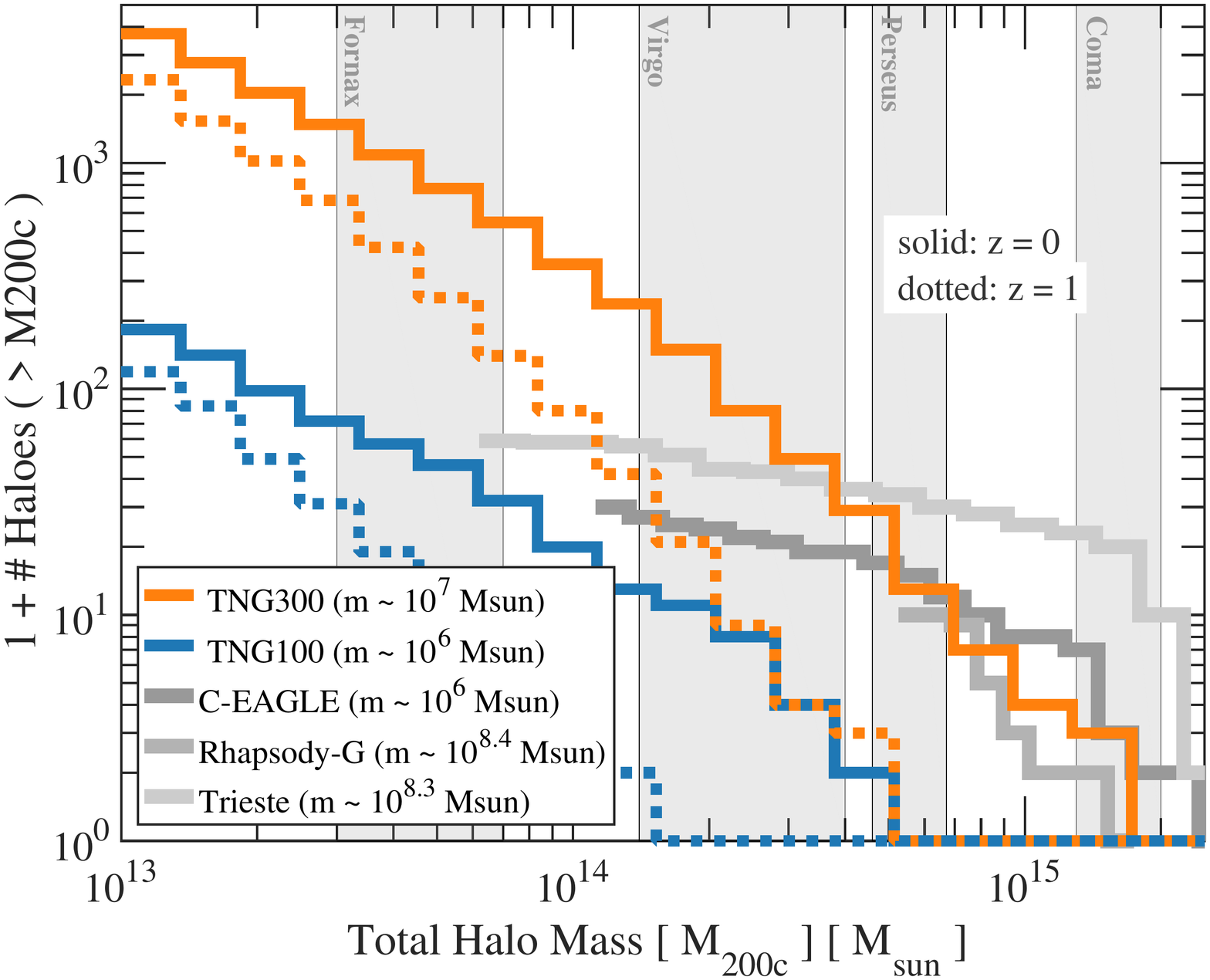}
\includegraphics[width=8.5cm]{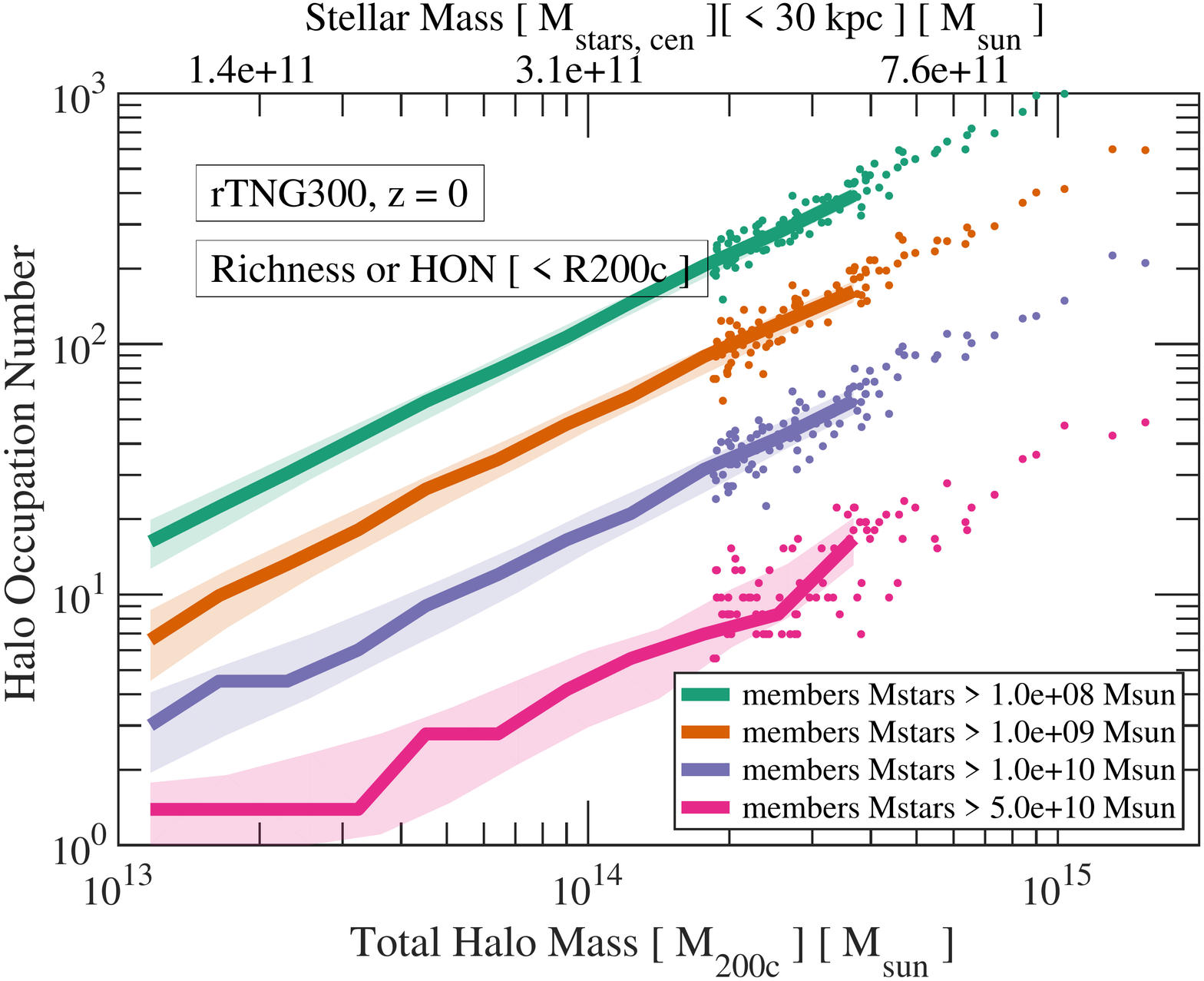}
\caption{\label{fig:stats} Demographics of the haloes and galaxies in the TNG100 and TNG300 simulations. Top: cumulative halo mass functions at different redshifts, in absolute numbers from the $100^3$ (TNG100) and $300^3$ Mpc$^3$ (TNG300) simulated volumes. The size of the TNG massive end is compared to the C-Eagle/Hydrangea \citep{Barnes:2017,Bahe:2017}, the Rhapsody-G \citep{Hahn:2017} and the Trieste \citep{Ragone:2013} zoom-in projects: currently the only simulations at comparable resolution (expressed in terms of the baryonic resolution element mass, $m$). Vertical grey bands denote the mass estimates of the local Fornax, Virgo, Perseus, and Coma clusters \citep{Weinmann:2011}. Bottom: Richness of haloes in the rescaled rTNG300, i.e. average number of member galaxies (including the central) within $\RTC$ as a function of halo mass (bottom axis) or central stellar mass (top axis).}
\end{figure}

\begin{figure*}
\centering                                      
\includegraphics[width=17.8cm]{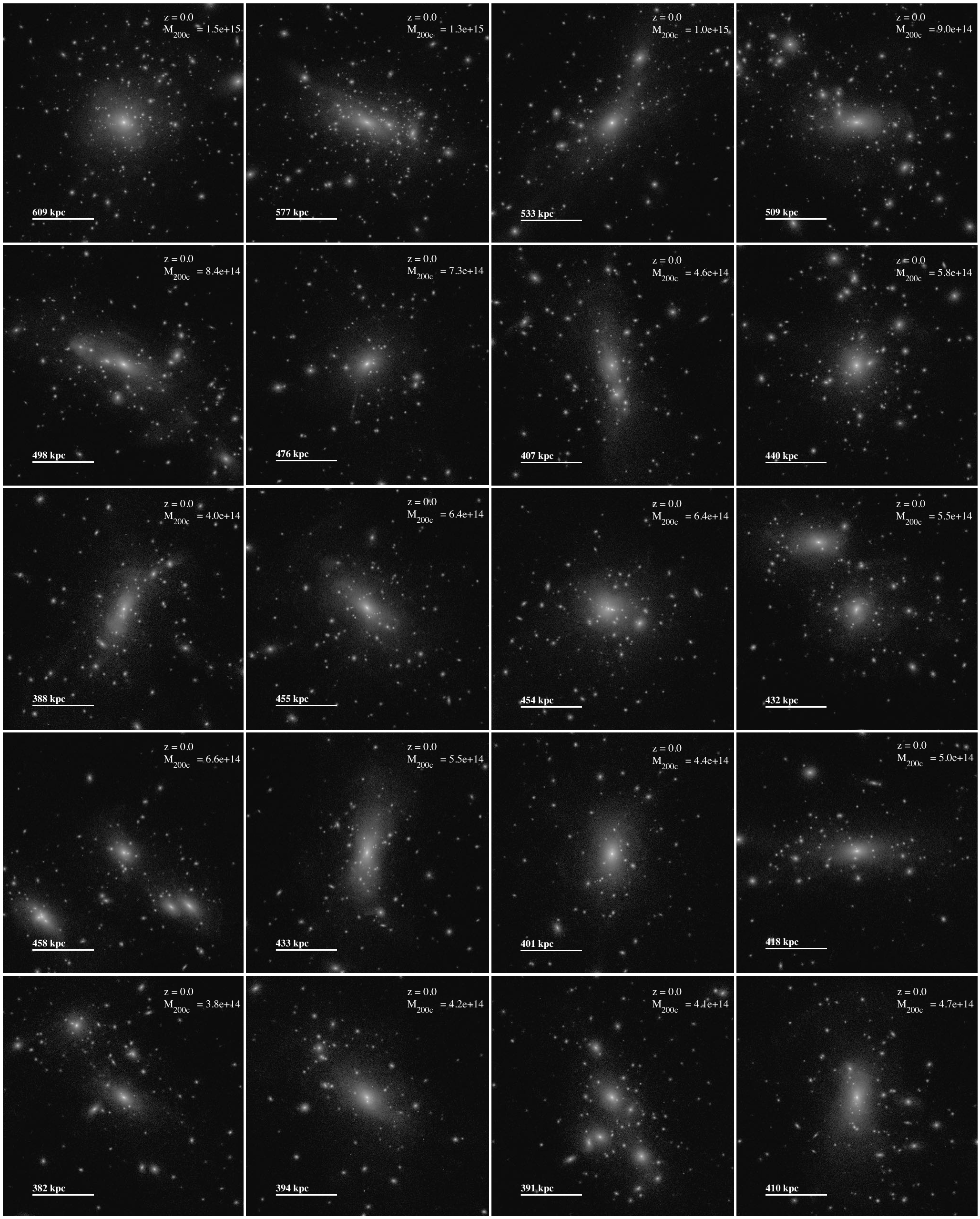}
\caption{\label{fig:stamps1} Stellar mass density projections of the 20 most massive objects in the TNG300 simulation, spanning total halo masses from $1.5\times 10^{15}\MSUN$ to about $4\times 10^{14}\MSUN$. The surface mass densities range from 0.1 to $10^{10} \MSUN$ kpc$^{-2}$ and the stamps measure $\RTC$ on a side. The most massive galaxies in the Universe occupy the centers of these massive systems and are surrounded by hundreds (even thousands) of less luminous satellite galaxies as well as by a cloud of diffuse stellar material extending to very large distances: the intra-cluster light.} 
\end{figure*}

\begin{figure*}
\centering
\includegraphics[width=17.8cm]{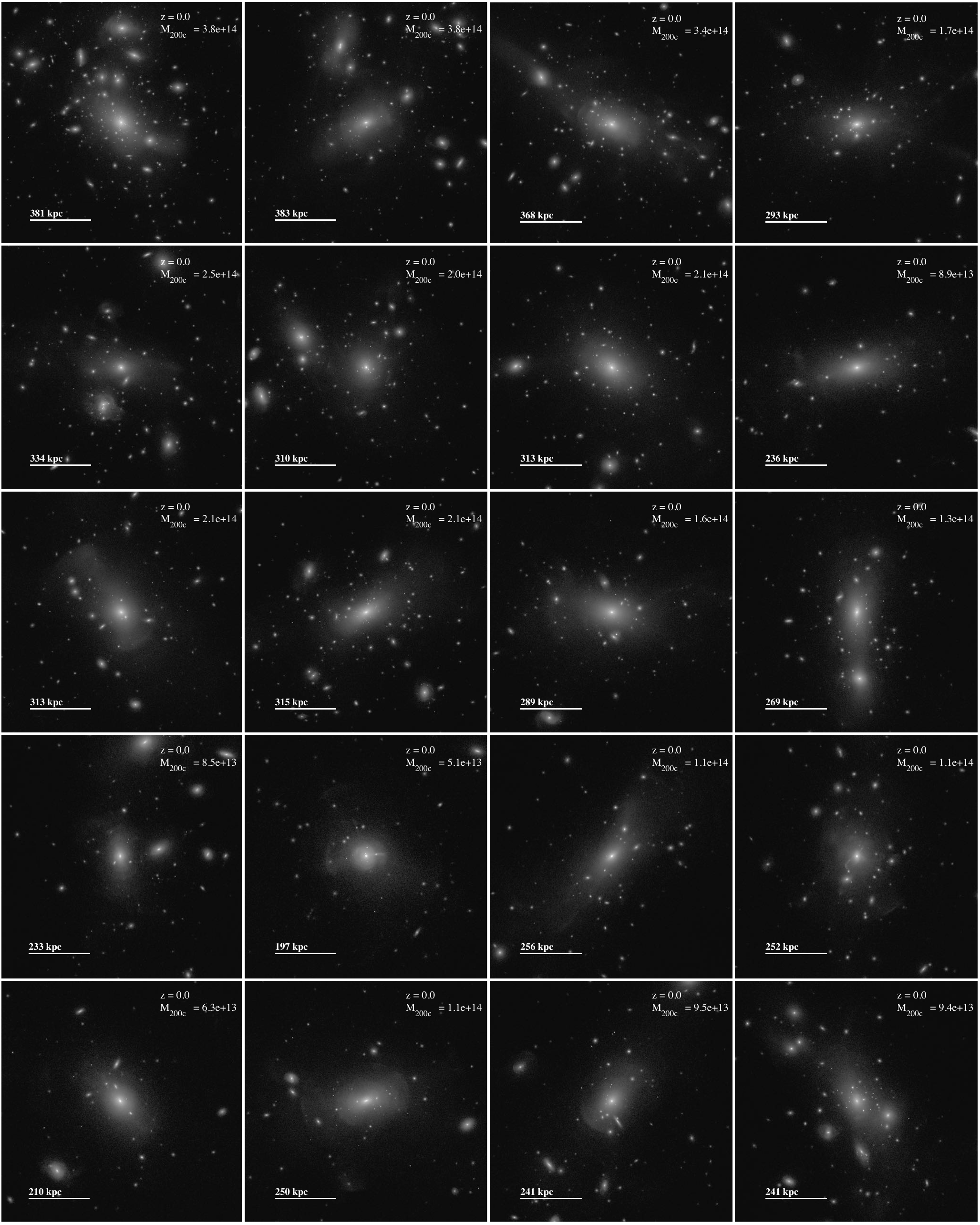}
\caption{\label{fig:stamps2} As in Figure \ref{fig:stamps1}, but for the 20 most massive objects in the TNG100 simulation: as the simulation volume is smaller, here we see a sample of slightly less massive and less rich clusters, with total halo masses between $9\times 10^{13}\MSUN$ and about $4\times 10^{14}\MSUN$. The diffuse stellar envelopes exhibit a diversity of shapes and morphologies, including subtle phase-space features like shells and stellar streams.} 
\end{figure*}

\section{The massive end of TNG galaxies and clusters}
\label{sec:sample}

In this paper we focus on the high mass end of the galaxy and halo mass functions over the last 8 Gyr of cosmic evolution ($0 \leq z \leq 1$).  In particular, we consider all (central) haloes in TNG100 and TNG300 with total halo mass $\MTC \geq 10^{13}\, \MSUN$, including their families of member galaxies down to stellar masses of about $10^8$ and $10^9\,\MSUN$ in TNG100 and TNG300, respectively.
Fig.~\ref{fig:stats} presents some basic statistics of the sample. The top panel gives the cumulative histogram of the number of simulated haloes above a given mass, for TNG100 (blue) and TNG300 (orange), both at $z=0$ (solid) as well as at $z=1$ (dotted). In every mass bin, the larger volume contains roughly twenty times more objects than TNG100, and has comparable or superior statistics to `zoom-in' cluster simulation projects with comparable numerical resolution, e.g. the Trieste cluster sample \citep{Ragone:2013, Planelles:2014}, C-Eagle \citep{Barnes:2017} and Rhapsody-G \citep{Hahn:2017}. In fact, while the most massive object in TNG300 has a mass of $\MTC=1.54\times10^{15}\MSUN$, the Trieste, C-Eagle, and Rhapsody-G samples extend to larger halo masses ($\MTC \sim 3.6, 2.4, 1.5 \times 10^{15} \MSUN$, respectively at $z=0$), with the latter (former ones) achieving approximately 6 times better (10 times worse) particle mass resolutions than TNG300. In practice, TNG300 contains numerous Virgo and Perseus mass analogs (i.e. $M \sim 1.4-4 \times 10^{14} \MSUN$ and $M \sim 4.6-6.7 \times 10^{14} \MSUN$, respectively), but only a few Coma-like clusters (i.e. $M \sim 1.3-2 \times 10^{15}\MSUN$, see \citealt{Weinmann:2011}).

In TNG100, the most massive halo at $z=0$ has a total mass of $3.8\times 10^{14}\,\MSUN$ ($\MTC$), its central galaxy hosting a stellar mass of $1.1 \times 10^{12}\,\MSUN$ (measured within 30 kpc). In the 300 Mpc box, on the other hand, 3 objects exceed $\MTC = 10^{15}\,\MSUN$ and 280 are more massive than $10^{14}\MSUN$. This gives us a sample of 182 and 3733 for $z=0$ groups and clusters ($\MTC \geq 10^{13}\,\MSUN$) from TNG100 and TNG300, respectively. At $z=1$ the number of groups and clusters is reduced to 118 and 2333 in TNG100 and TNG300, respectively. Partly to maintain good statistics at the high-mass end, we therefore restrict the analysis of the present paper to $z<1$. In doing so we capture the majority of cosmic evolution ($\sim$8 Gyr) and the period of massive cluster formation, after the dynamically rapid high redshift Universe begins to evolve more slowly.

\begin{figure*}
\centering
\includegraphics[width=16.8cm]{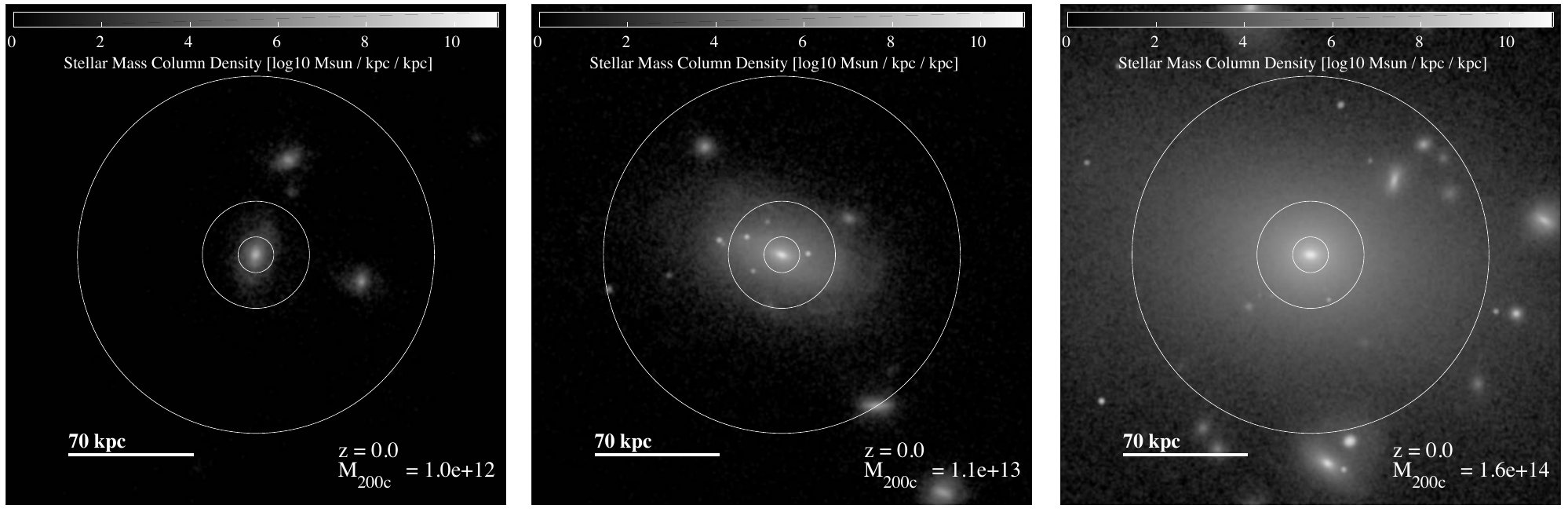}
\includegraphics[width=5.6cm]{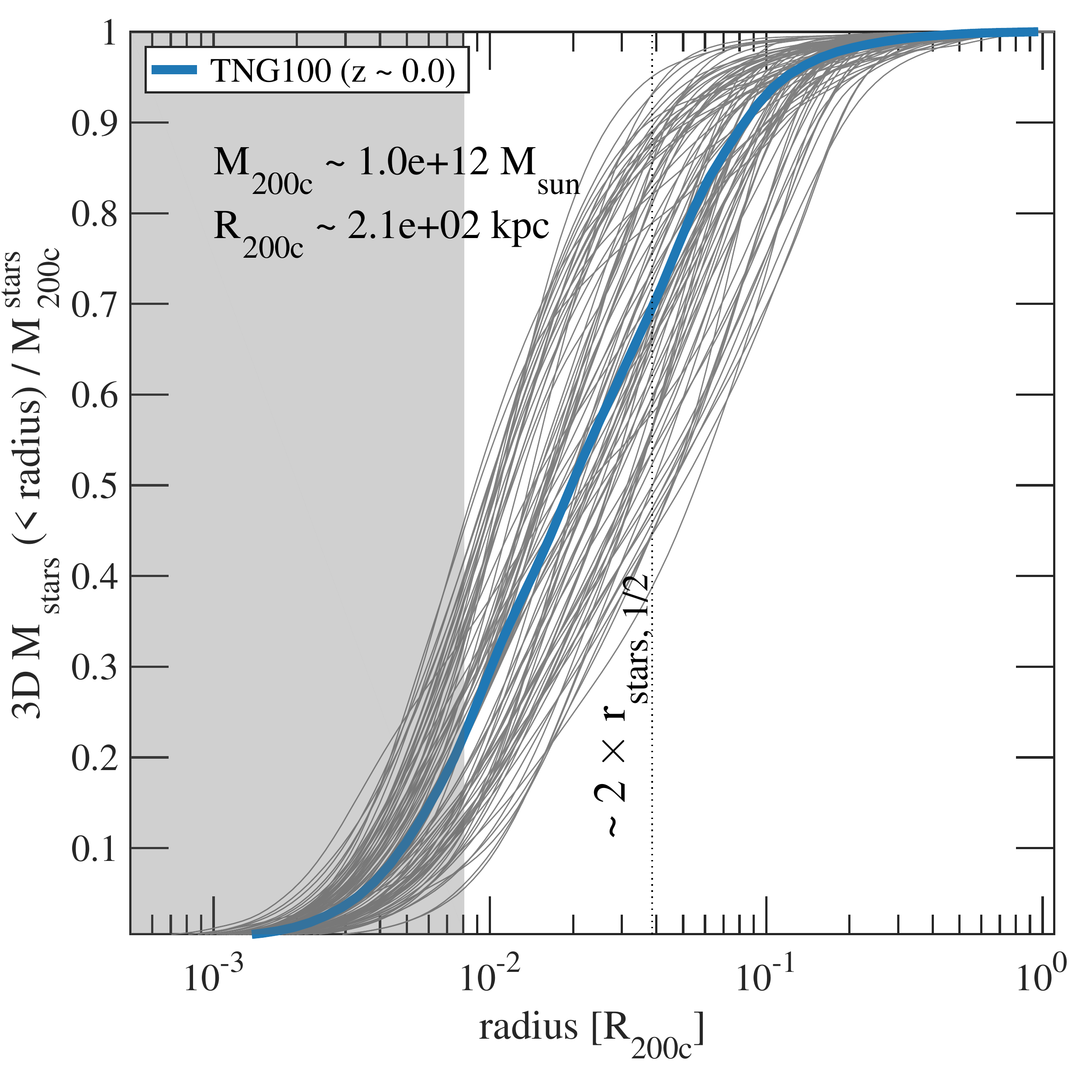}
\includegraphics[width=5.6cm]{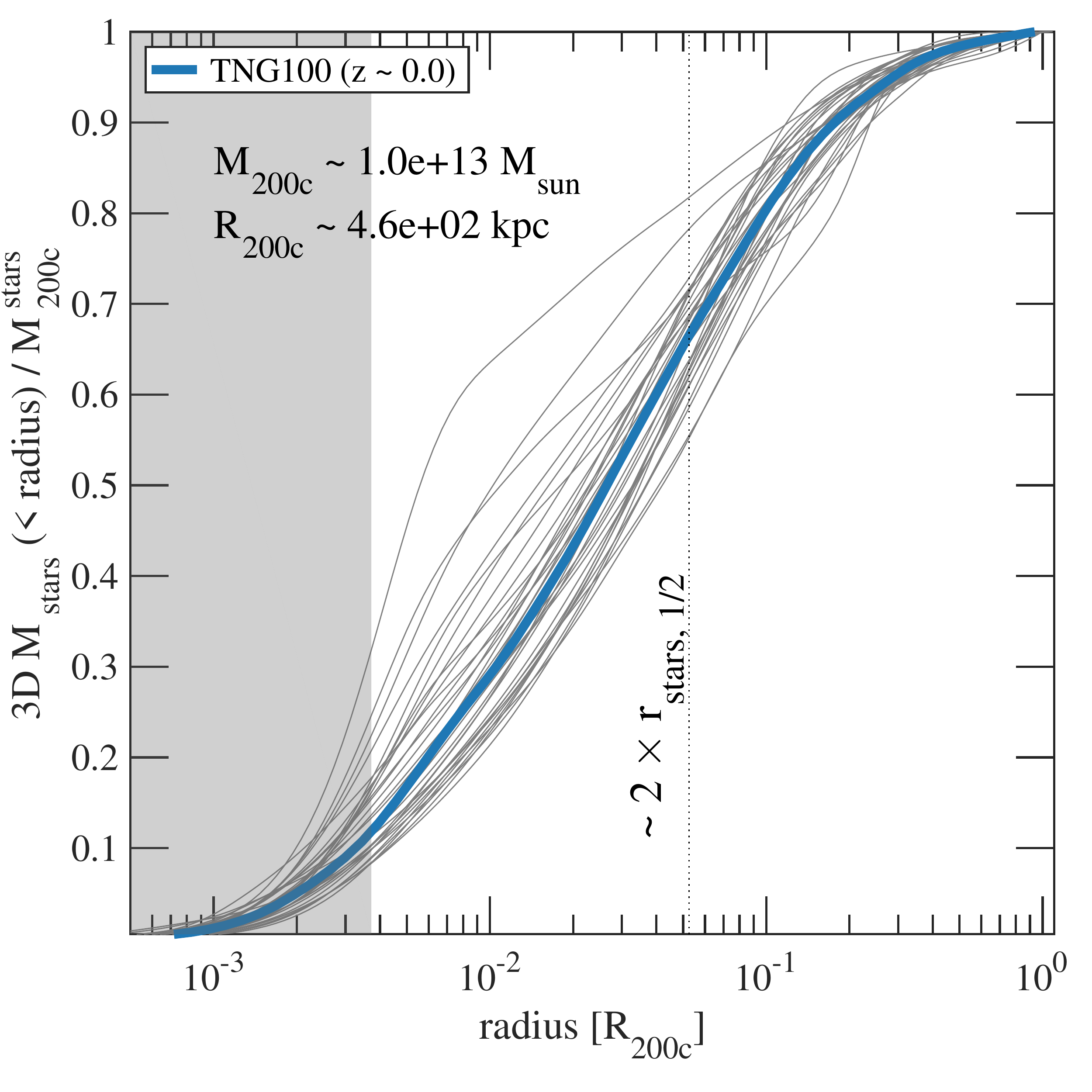}
\includegraphics[width=5.6cm]{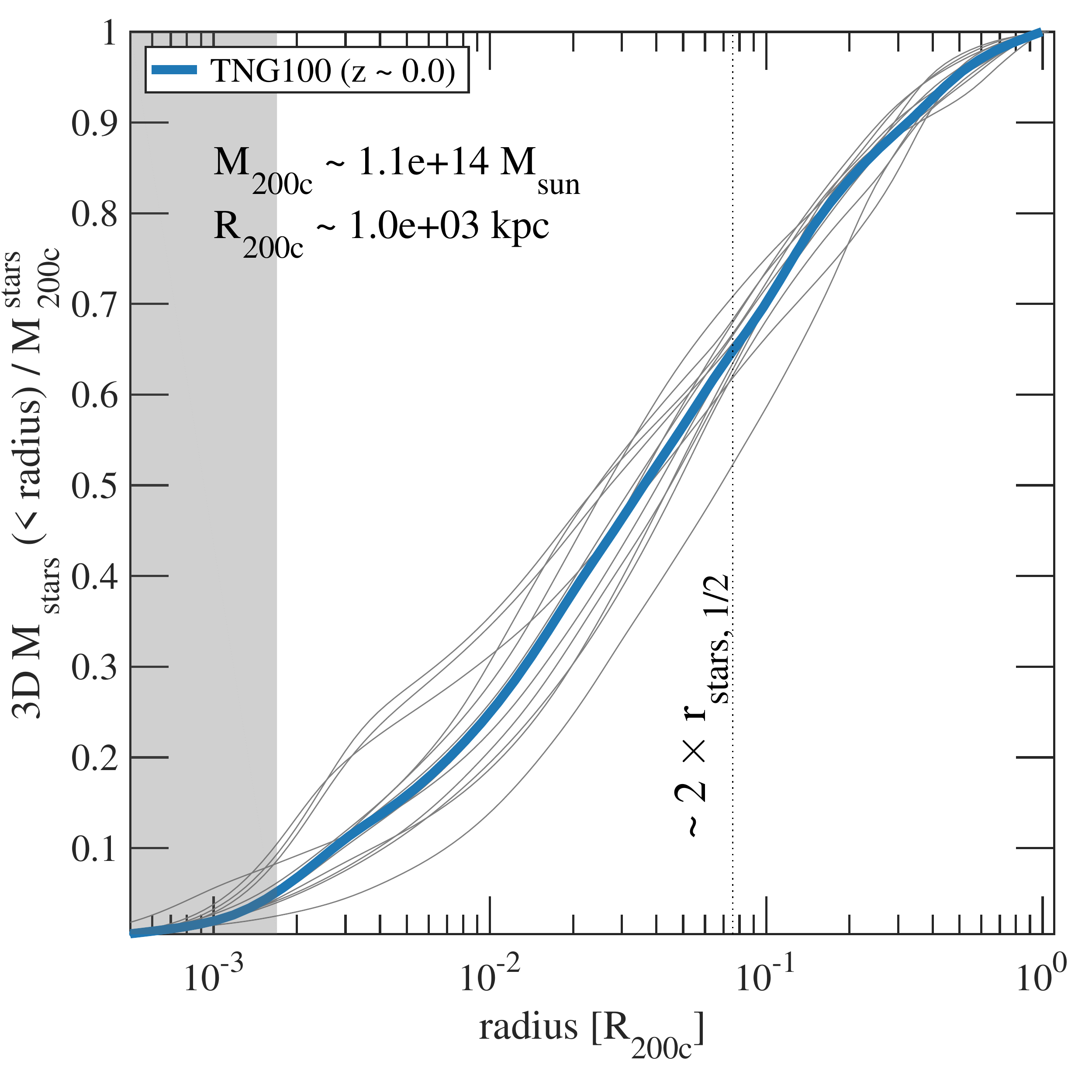}
\caption{\label{fig:profiles_prototypes} Top: stellar mass density projections of three typical central galaxies and their surrounding satellites, from low (left) to high (right) halo masses (in $\MSUN$). Bottom: thin grey curves depict profiles of the diffuse (i.e. excluding satellites) stellar mass of individual objects, thick blue curves are average stacked profiles in the labeled mass bin ($\pm0.01, 0.05, 0.2$ dex from left to right, in $\rm{log}_{\rm10}\MTC$). In the $y$-labels, $\MSTC = M_{\rm stars}(< \RTC)$. The stellar content of small haloes extends to much smaller distances than that of more massive haloes, even when the distances are renormalized by the virial radius.}
\end{figure*}

The resolution of TNG100 (TNG300) is such that $10^{13}\,\MSUN$ objects are on average resolved with 2-3 million (300-400 thousand) resolution elements each, among gas cells, stellar particles, DM particles and black holes. The most massive objects in the large TNG300 run, on the other hand, are sampled at the current time with about 50 million resolution elements each. The minimum satellite stellar mass values of $10^8$ and $10^9\,\MSUN$ correspond to about 100 stellar particles per galaxy in TNG100 and TNG300, respectively, and we have checked that they are suitable for the demographic statistics we are after in the following sections. The bottom panel of Fig.~\ref{fig:stats} shows the richness, or halo occupation number (HON), i.e.~the number of galaxies above a given stellar mass threshold in different mass haloes. The most massive simulated clusters host as many as $\sim 1000$ resolved galaxies with $M_{\rm stars} > 10^8 \MSUN$, while the smallest groups considered here contain about 10-20 such satellites. In practice, galaxies more massive than about $10^{11}\MSUN$ can only be found at the center of rich groups and clusters, i.e. surrounded by at least a few other galaxies more massive than $5\times 10^{10}\MSUN$ in stars.

In Figures~\ref{fig:stamps1} and \ref{fig:stamps2}, we show the 20 most massive galaxy groups and clusters of the TNG300 and TNG100 boxes, respectively: they are depicted in stellar mass density projections at $z=0$, each shown in a volume of $\RTC$ on a side, namely many hundreds of kpc across, in a randomly chosen projection. Because of the different simulation volumes, the two figures focus on two rather different mass ranges, mostly massive and rich clusters above a few $10^{14}\MSUN$ in Fig.~\ref{fig:stamps1}, and about one order of magnitude less massive galaxy groups in Fig.~\ref{fig:stamps2}. The inspection of these objects demonstrates the diversity and richness of the simulated sample, of which here we show a mere ten percent only. Our simulations naturally produce groups and clusters of galaxies, with one (or more) massive galaxies dominating the most luminous regions, surrounded by a spectrum of more or less numerous satellite galaxies, all amid a background of low surface brightness diffuse stellar material (also known as stellar halo, intra-cluster light or intra-halo light, hereafter used interchangeably). The latter can extend out to hundreds of kpc from the central object, exhibits a variety of different shapes in projection, and at times displays prominent configuration and phase-space features such as stellar shells and streams. This accreted stellar material encodes the hierarchical assembly history of the underlying dark matter halo, providing an observationally accessible witness to the process of hierarchical cosmic structure formation in $\Lambda$CDM.

\section{Stellar profiles to large radii}
\label{sec:profiles}

We are interested in quantifying the stellar mass content of groups and clusters of galaxies from the TNG simulations (see selection in Section \ref{sec:sample}). The aim is to provide a new comprehensive benchmark for comparison against observations as well as other theoretical models. To this end, instead of attempting to reproduce one or a few standard observational techniques usually adopted to recover the stellar mass of observed galaxies, we use the simulations to identify an optimal characterization of the total stellar mass budget and spatial distribution of galaxies and galaxy clusters. 
In the following, we occasionally extend our analysis to haloes smaller than $10^{13}\MSUN$ to gain insights on the trends across a larger mass range or to compare to existing results from the literature.

\subsection{Radial distribution of the diffuse stellar mass}
In Figure \ref{fig:profiles_prototypes}, we show the stellar mass radial distribution within a selection of objects from TNG100 at $z=0$, specifically looking at the three discrete halo mass bins centered at $10^{12}$, $10^{13}$, and $10^{14}\,\MSUN$. In the upper panels, the stellar mass density projection is shown for three typical central galaxies of comparable masses (and their surrounding satellites) across a fixed aperture of 280 kpc (this is a much smaller field of view than the stamps of Figures \ref{fig:stamps1} and \ref{fig:stamps2}). They are ordered from left to right in increasing host halo mass and thin circles denote fixed spherical apertures of 10, 30 and 100 kpc.

In the lower panels, we show the cumulative enclosed stellar mass as a function of radius, normalized to the total stellar mass within the virial radius ($\RTC$). Throughout this paper, stellar profiles only account for {\it diffuse} stellar mass, i.e. the latter does {\it not} include the mass in satellite galaxies or unbound stars, which are excised based on gravitational binding/unbinding criteria (see Section \ref{sec:haloes}). Profiles are measured in spheres evenly spaced in logarithmic radius ($\Delta{\rm  log}_{10} (r/[{1 \rm kpc}]) \approx 0.03$) between the minimum stellar distance and the virial radius, with uniform weighting. Thin grey curves denote profiles from individual objects; thick blue curves are average stacked profiles of all the objects in the labeled mass bins, also ordered from left to right for increasing host halo mass (bin sizes \ap{$\pm0.01, 0.05, 0.2$} dex, respectively, in ${\rm log}_{\rm 10}\MTC$). Vertical thin lines denote twice the stellar half mass radius.

From Figure \ref{fig:profiles_prototypes} it is clear that galaxies residing in less massive haloes have significantly {\it more centrally concentrated} stellar mass distributions when compared to more massive galaxies, {\it even} when the radial apertures are renormalized to the virial radius of the host haloes. This behaviour is similar, at least qualitatively, to that of DM, for which less massive haloes exhibit more centrally concentrated DM density profiles than more massive objects. While more than 90\% of the total stellar mass of a $10^{12}\,\MSUN$ halo is within 10\% of its virial radius (bottom left panel), there is at least another 30\% of stellar mass beyond 10\% of the virial radius of a $10^{14}\,\MSUN$ group (bottom right panel). Similarly, the 3D stellar half-mass radius moves from a few kpc for a Milky Way-like halo -- equivalent to a few per cent of the virial radius -- to tens of kpc for a galaxy residing at the centers of a group \citep[see][for an analysis of galaxy sizes in IllustrisTNG]{Genel:2017}. Finally, Milky Way-like galaxies have more than 95\% of their stellar mass contained within a 3D spherical aperture of 30 kpc -- hence the standard use of 30 kpc as a natural boundary for galaxy stellar mass \citep[see e.g.][who demonstrate that this physical aperture provides stellar mass estimates in good agreement to those within Petrosian radii in observations]{Schaye:2015}. However, this fraction decreases to less than 50\% within the same aperture for the most massive galaxies at the centers of groups and clusters. It is also apparent that, while there are some variations in the stellar profile shapes across different objects (thin grey curves), the averaged profiles in narrow halo mass bins are remarkably regular and share a similar {\it sigmoidal} form across halo and galaxy masses. 

\begin{figure*}
\centering
\includegraphics[width=8.6cm]{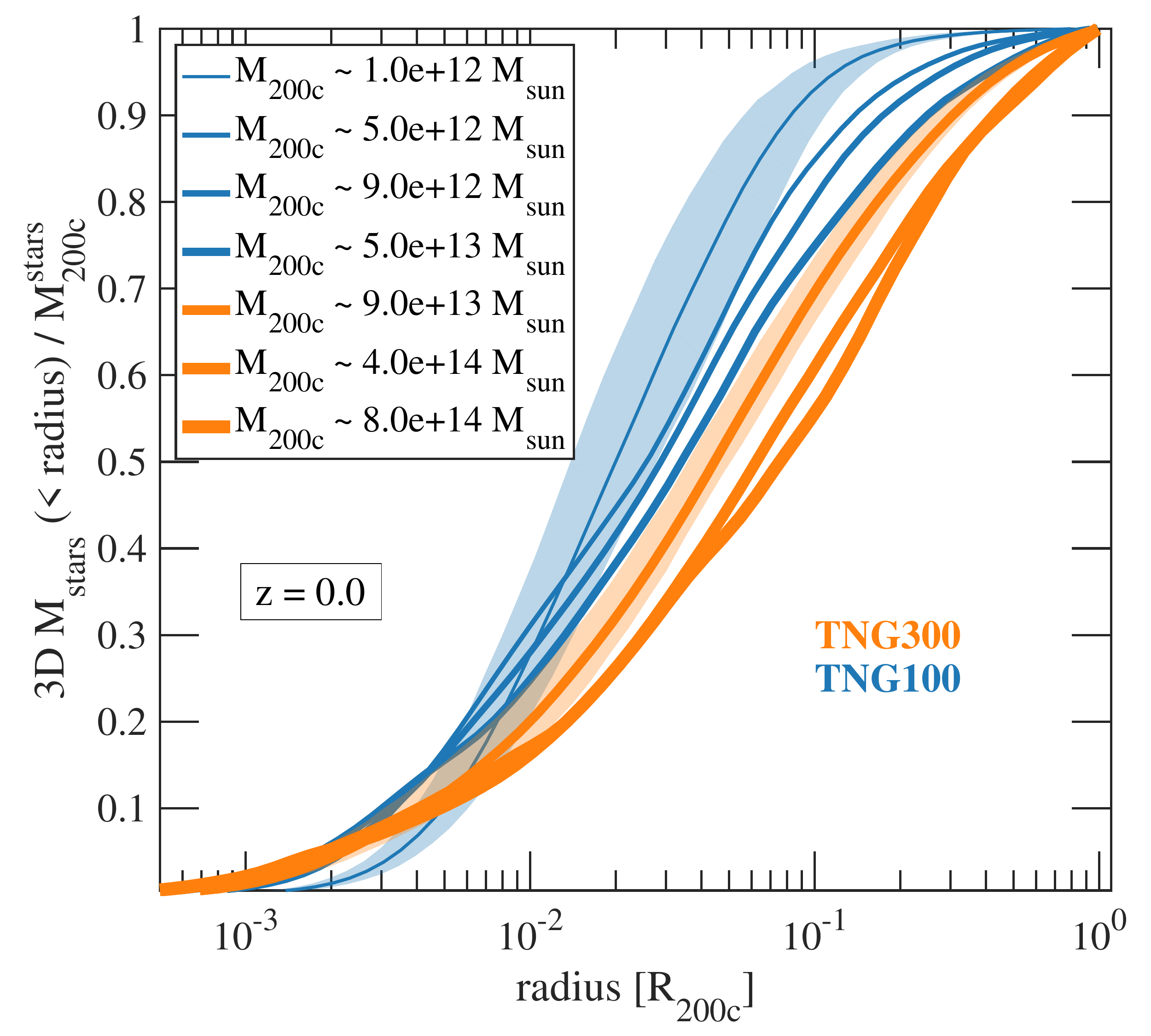}
\includegraphics[width=8.6cm]{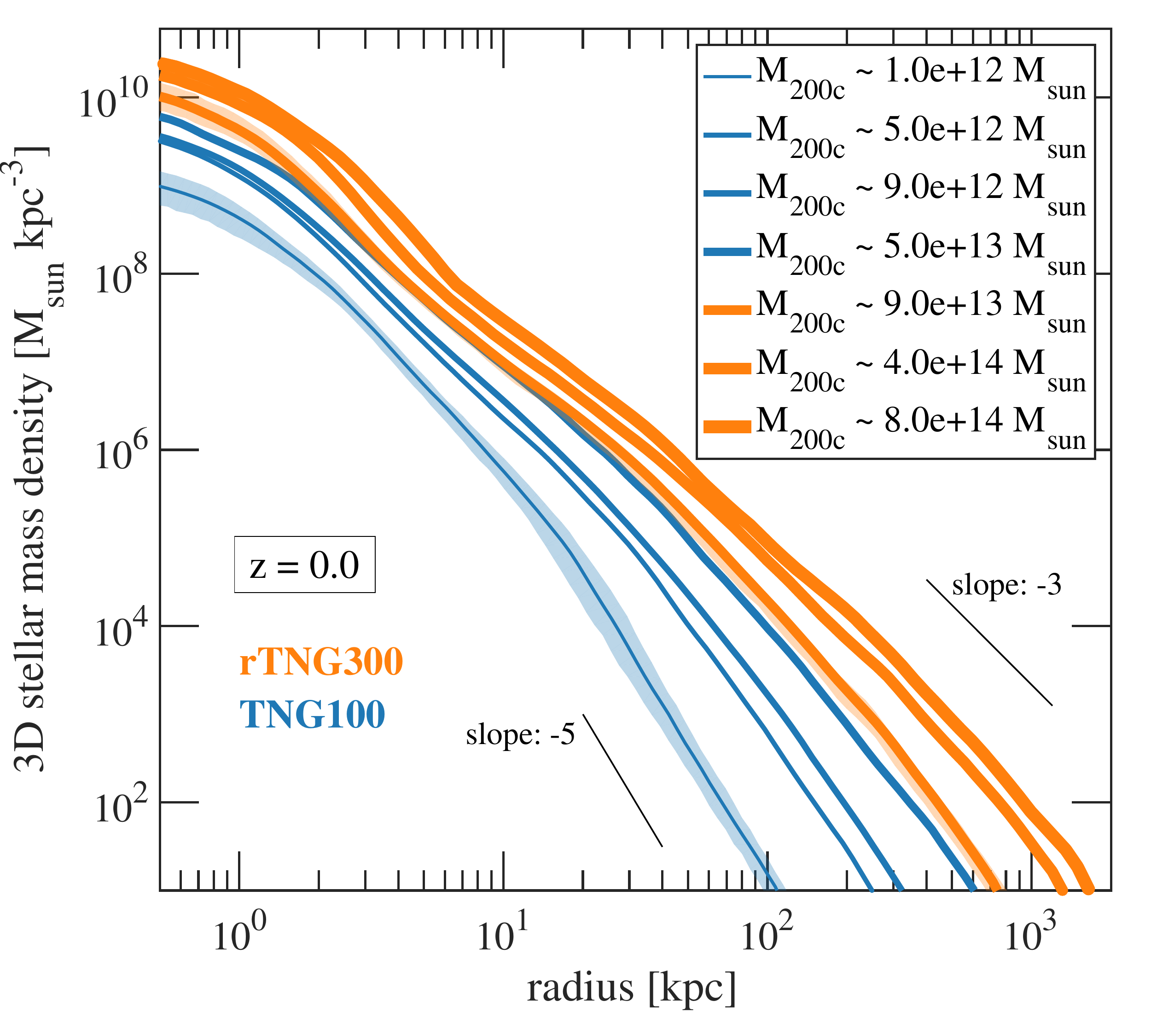}
\includegraphics[width=8.5cm]{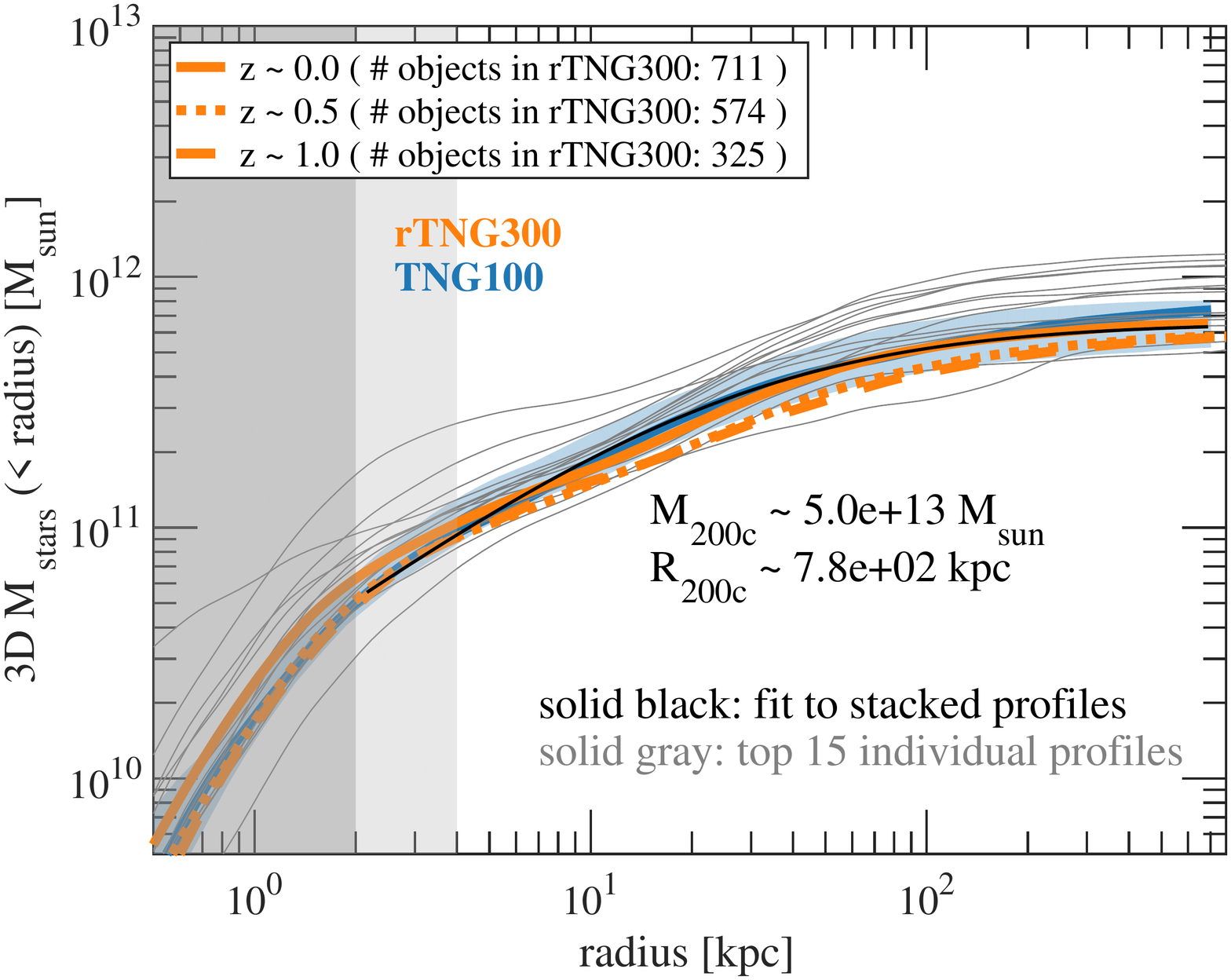}
\includegraphics[width=8.5cm]{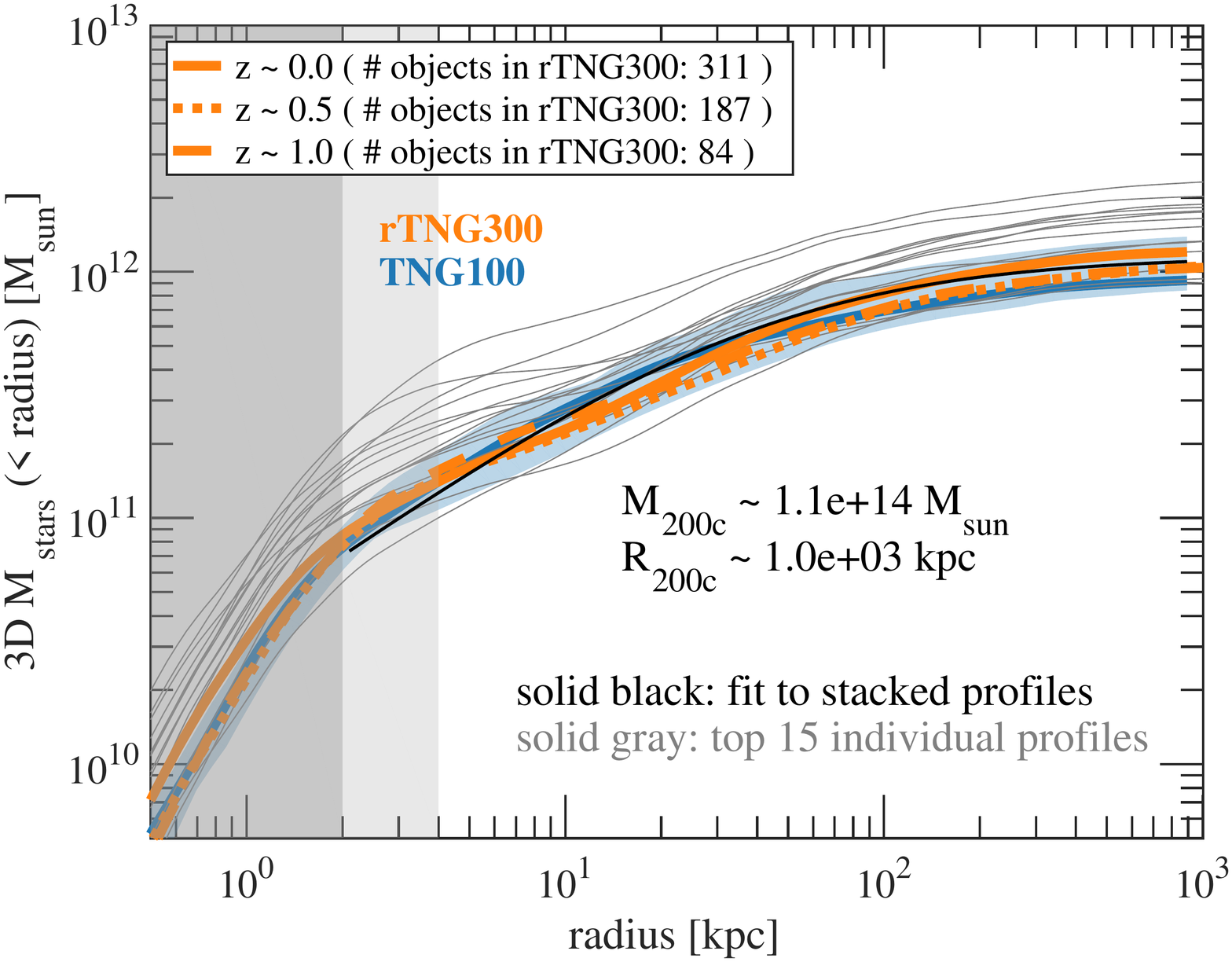}
\includegraphics[width=8.5cm]{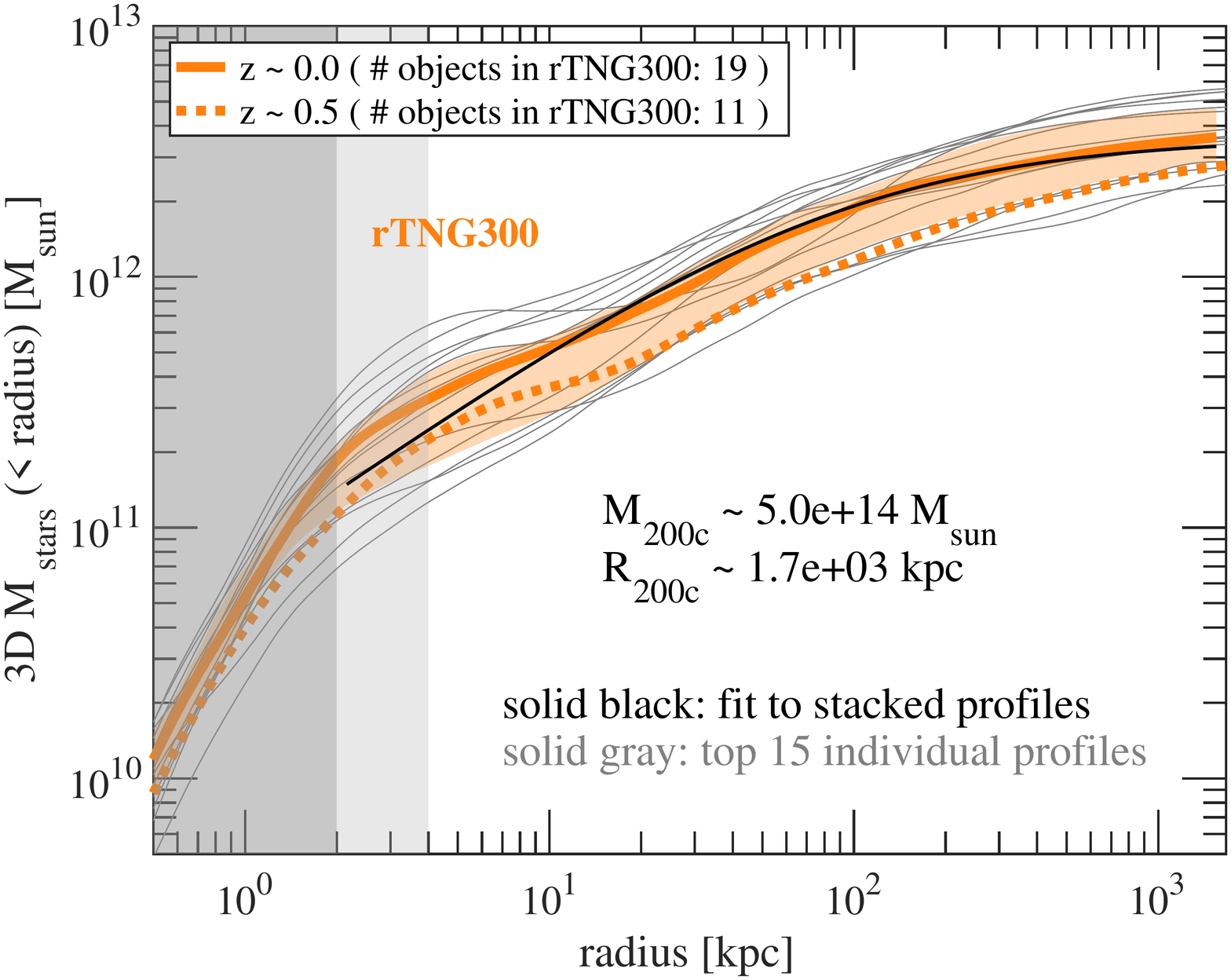}
\includegraphics[width=8.5cm]{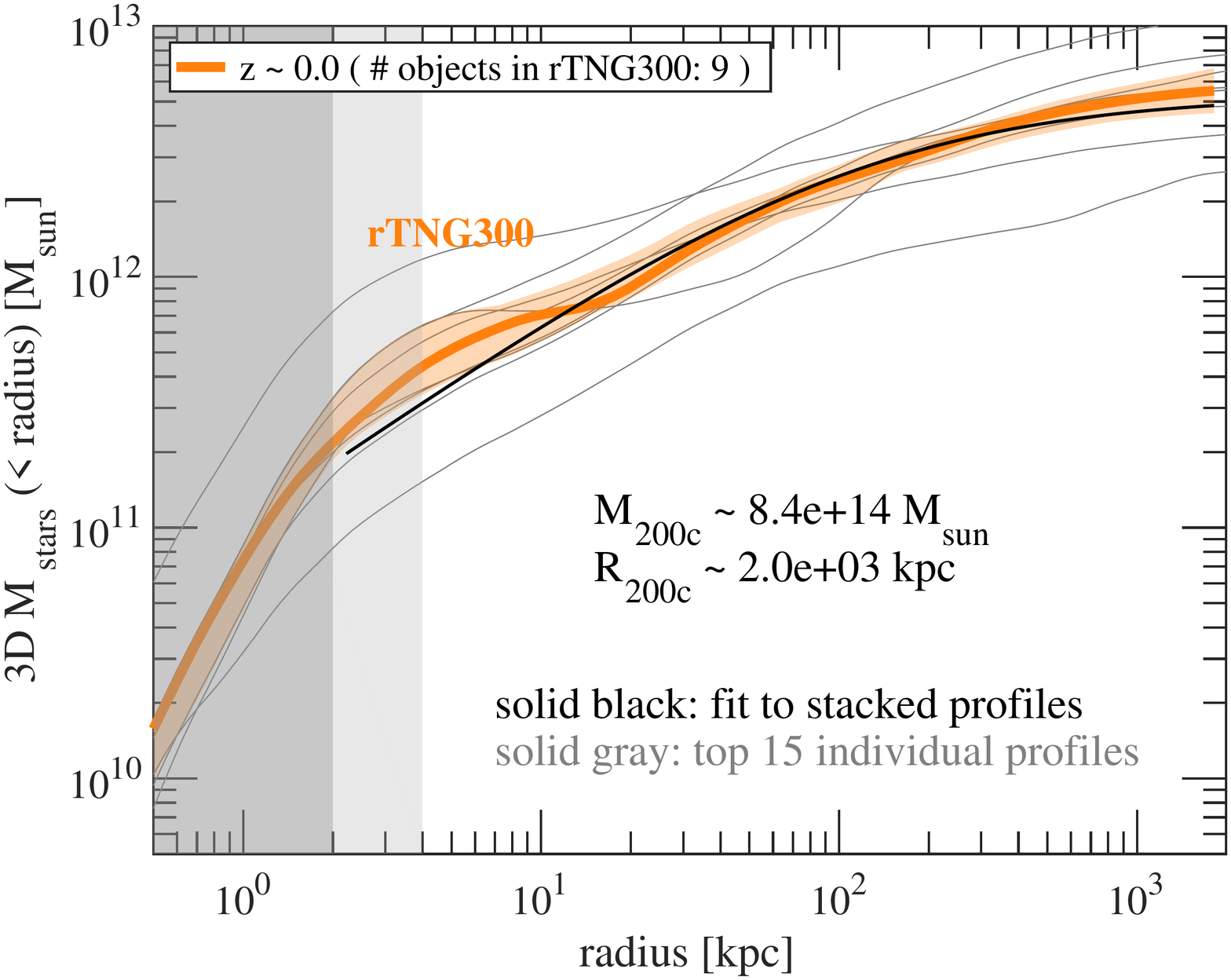}

\caption{\label{fig:stellarprofiles} 3D stellar radial profiles of massive haloes in the TNG simulations: mass enclosed (top left and bottom four panels) and mass density (top right) as a function of clustercentric distance at various redshifts, $z\la 1$. No distinction or separation is applied between stars in the central galaxy and stars in the diffuse, low-surface brightness ICL. Any contribution from satellites is excluded. In the top left panel, the 3D mass enclosed is normalized by the virial radius and the total mass within the virial radius, otherwise profiles are given in solar masses, and spatial scales are in comoving units. In the bottom four panels, profiles are given in the labeled bins of halo mass ($0.5, 1.1, 5.0, 8.4 \times 10^{14}\MSUN \pm 0.2$ dex in $\rm{log}_{\rm10}\MTC$). In all panels, thick color curves show the average stacked profiles of haloes (if at least 3 objects lie in the bin). Color shaded area denote the 25th and 75th per centiles. Thin grey curves are up to 15 most massive individual haloes in each bin - hence possibly biased high compared to the stacks if the bins contain many more individual haloes. All annotations and symbols are $z=0$ measurements, except for the dotted and dashed curves, which show stacked profiles for rTNG300 in the indicated mass bin at $z\sim 0.5$ and $z\sim1$, respectively. 
Grey radial regions indicate where interpretations require care, given our numerical resolution. Solid thin black curves are the results of the procedure described in Section \ref{sec:thefit}: the recovery of the full large-scale profiles given the total halo mass only.}
\end{figure*}

We expand this analysis in Figure \ref{fig:stellarprofiles} focusing on the mass trend at $z=0$ of the stacked 3D enclosed stellar mass and stellar mass density profiles (top panels) and on the diversity of the former as a function of radius for objects in different mass bins at different times (bottom four panels, one mass bin per panel). Radial profiles are given in terms of comoving kpc and include outcomes from TNG100, TNG300 and/or rTNG300, when possible. The curves for rTNG300 are obtained as described in Appendix \ref{sec:app_res}. 

From Figure \ref{fig:stellarprofiles}, top left panel, we confirm with TNG100 and TNG300 (not rescaled) that the slope of the enclosed mass profiles becomes progressively shallower for larger galaxy/halo masses, as can be seen in the flattening of the slope at mid point of the enclosed mass; and the stellar half mass radius moves to larger fractional radii. The changes to the enclosed mass profiles with halo mass appear remarkably regular. In the radial mass density (top right), the central value rises monotonically with halo mass in reflection of its approximate rank ordering with galaxy stellar mass. Increasingly massive central galaxies are accompanied by more extended stellar mass at large distances -- the mass density of stars at 100 kpc for a $10^{13}\, \MSUN$ ($10^{15}\, \MSUN$) halo is roughly 100 (10'000) times larger than for a $10^{12}\, \MSUN$ halo at the same radius. Indeed, the power-law slope of the 3D stellar mass density profiles at a few tens kpc from the centers clearly decreases from about $-3$ to about $-5$ in the mass range between $\sim 10^{15}\MSUN$ to $10^{12}\MSUN$ (see Section \ref{sec:powerlaw}).

In the lower four panels of Figure \ref{fig:stellarprofiles} we show both individual (thin grey curves, only the most massive 15 or less in each panel, to avoid overcrowding) and stacked (thick orange or blue curves) profiles of the 3D stellar mass enclosed out to the virial radius, in bins of halo mass ($\pm 0.2$ dex). Solid curves refer to $z=0$, dotted and dashed to redshifts $z=0.5$ and 1, respectively, where possible. Note that, because of its volume, TNG100 does not contain any halo larger than a few $\times 10^{14}\,\MSUN$ at any time. These profiles correspond to those in the top left panel, plotted in actual physical units and without normalization, to highlight possible features and provide quantitative references for comparison to observations. In all panels, vertical lines denote specific spherical apertures. As a reminder, 0.7 (1.4) kpc is about the stellar softening length in our TNG100 (TNG300) simulation. No strong claims regarding the shapes of the profiles can be made below radial distances of about 2.8 times the softening length: this corresponds to about 2 kpc and 4 kpc for TNG100 and (r)TNG300, respectively, and two grey zones in each panel remind us where to be cautious in any quantitative or qualitative interpretation (see also Appendix~\ref{sec:app_res} and Fig.~\ref{fig:resolution_2}). 
Beyond a few kpc from the centers, the stacked enclosed mass profiles exhibit radial trends at different halo or galaxy masses which are similar in shape but change substantially in normalization. In fact, the distributions of enclosed stellar mass at fixed physical aperture across galaxies can be very broad, with the full scatter across the individual gray lines being possibly much larger than the 25th-75th percentile gap denoted by the shaded areas around the median.

In contrast to the mass trends, the stellar mass distribution at fixed halo mass evolves surprisingly little from redshift $z=1$ to today. We shall return to this point in the context of Fig.~\ref{fig:stellarmasses}, and here only point out that the enclosed mass profiles show marginal if any change for the $\sim 5 \times 10^{13}$ and $\sim 10^{14}\,\MSUN$ halo mass bins (lower two panels). At $\sim 5 \times 10^{14}\,\MSUN$, there is a hint that the enclosed mass density increases as much as 50\%, depending on radius, between $z=0.5$ and $z=0$. However, even with the TNG300 volume the statistics become too poor at high redshift to make a population generalized statement: the apparent trend in the fixed mass bin reflects a mass trend within the few clusters in the bin itself, i.e. the samples at high redshift are biased towards slightly lower halo masses that the $z=0$ sample. Similarly, without larger simulated volumes, we cannot yet probe the evolution of haloes in the most massive bin beyond $z=0$.

\begin{figure}
\centering
\includegraphics[width=8.3cm]{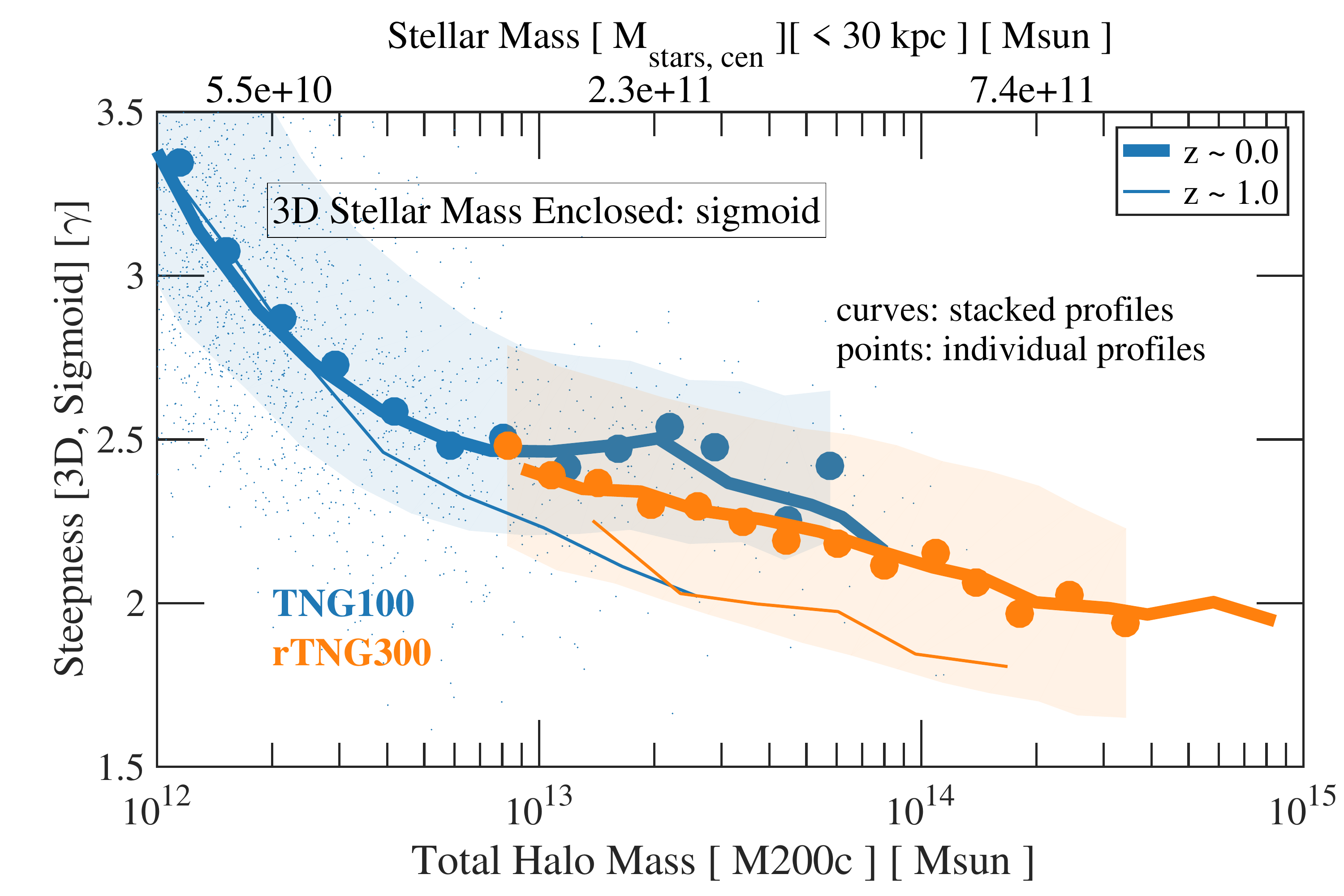}
\includegraphics[width=8.3cm]{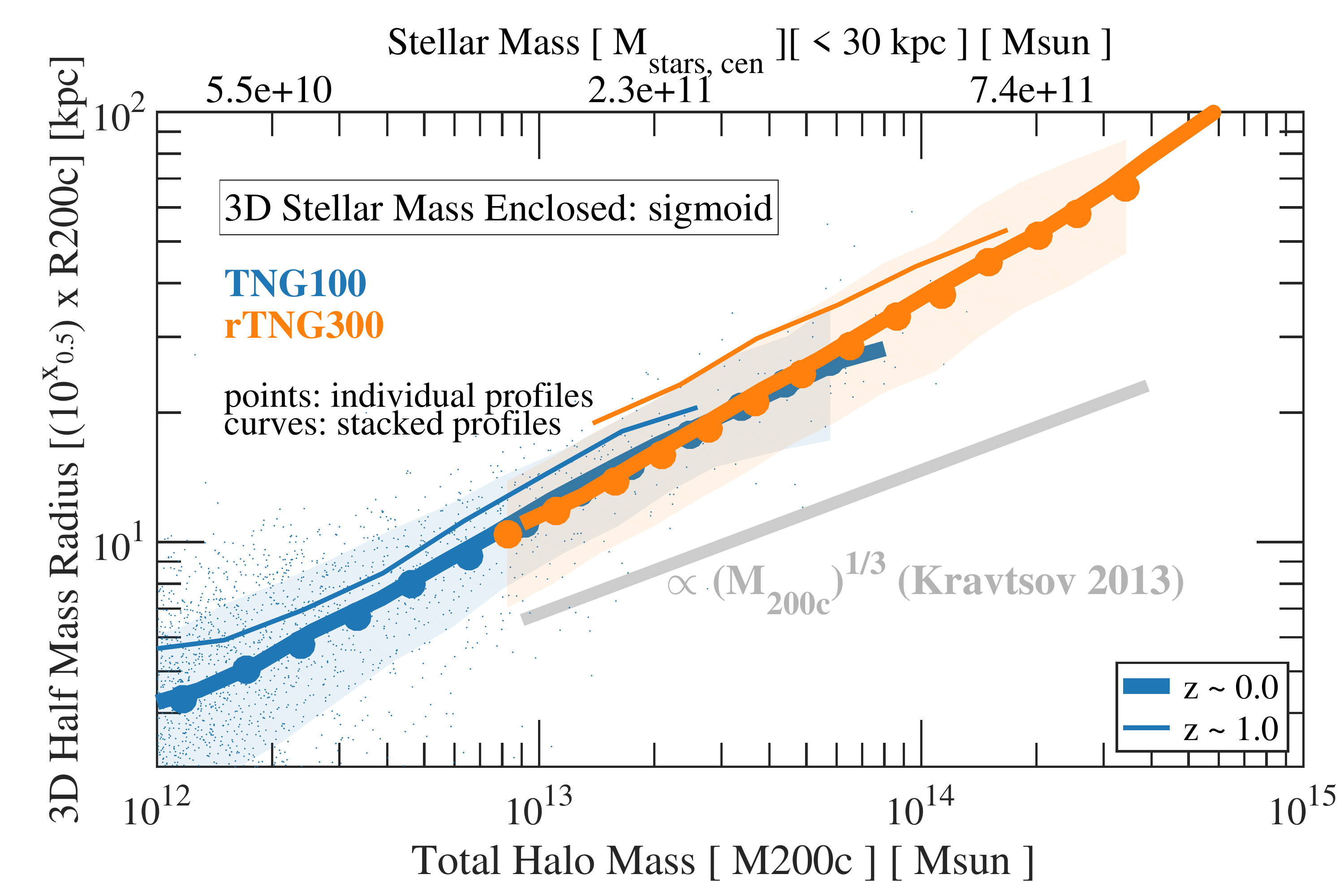}
\caption{\label{fig:stellarprofilesfits_1} Best-fit parameters to the normalized enclosed stellar mass profiles of TNG galaxies (Fig.~\ref{fig:stellarprofiles}, top left) via the sigmoid functional form of Eq.~(\ref{eq:logistic}) -- for radii larger than 1 kpc. We show the trends with halo mass of the best-fit parameters at $z=0$ and $z\sim1$ of TNG100 and rTNG300 haloes. In the upper panel, the steepness of the sigmoid function: the larger the slope, the steeper the profile at the mid point. In the lower panel, the stellar half mass radius obtained by fitting for the pivot parameter x$_{0.5}$. Solid curves are the results from stacked profiles in bins of 0.2 dex in $\MTC$; dots and shaded areas denote results from fits to individual profiles, with dots denoting best fit parameters to individual galaxies (only for TNG100 at $z=0$ to avoid overcrowding) and filled circles indicating the medians to the individual best fit parameters in bins of 0.2 dex in $\MTC$; shaded area denote 1-sigma galaxy-to-galaxy variation results. Corresponding stellar masses are indicated in the top x-axes, their values being taken from the average TNG100 stellar mass to halo mass relation.}
\end{figure}

\subsection{Functional forms and fits}

The regularity of the average stellar mass profiles demonstrated in Figure \ref{fig:stellarprofiles} allows us to proceed and provide a quantitative description of the spatial distribution of stellar mass in massive TNG haloes. 
We fit a functional form to the radial stellar mass content predicted in the TNG simulations as a function of halo mass or galaxy stellar mass, and redshift. In considering the three-dimensional mass distributions we deliberately use spherical instead of ellipsoidal apertures: the former are comparatively less sensitive to details of the measurement procedure and therefore more appropriate for comparison across analysis works.
If the shapes of the extended stellar components are systematically non-spherical, then the spherically symmetric radial profiles are clearly a simplification. However, we focus on the group and cluster populations as a whole and across two orders of magnitude in mass, and so neglect a study of the detailed morphologies of individual haloes. 

\subsubsection{Mass-enclosed profiles, beyond a few kpc from the centers}
\label{sec:sigmoid}

The 3D enclosed stellar mass profiles of massive TNG groups and clusters can be well described by the following sigmoid function:
\begin{equation}\label{eq:logistic}
M_{\rm stars} (< r, z) = \frac{M^{\rm stars}_{200c} }{1+ {\rm exp}[-\gamma (x - {\rm x}_{0.5})] \vphantom{\sum^A}},
\end{equation}
where $ x = {\rm log}(r/\RTC)$ and $M^{\rm stars}_{200c}$ is the total amount of stellar mass (diffuse, no satellites) enclosed within the virial radius $\RTC$. Both the steepness parameter $\gamma$ and the pivot, mid-point x$_{0.5}$ are a function of mass and redshift. In practice, we fit the normalized, stacked profiles of enclosed stellar mass in bins of halo mass ($\MTC$) 0.2 dex wide, across the whole available mass range. We also fit the profiles of individual objects, to get a sense of the halo-to-halo variation. In both cases, the profiles are measured in spheres evenly spaced in logarithmic radius ($\Delta{\rm  log}_{10} (r/[{1 \rm kpc}]) \approx 0.03$) between the minimum stellar distance and the virial radius, with uniform weighting, and the fit is performed only outwards of 1 kpc. For halo masses up to $10^{14}\MSUN$ ($10^{15}\MSUN$) at $z=0$, the sigmoid of Eq.~(\ref{eq:logistic}) represents the actual normalized stacked profiles to better than a 2-3 per cent (5-6 per cent) accuracy, and equally so when averaging over all radii larger than either one or ten per cent of the virial radius. In fact, 90 per cent of the {\it individual} profiles in the halo mass range $10^{13}\MSUN \leq \MTC \leq 10^{15}\MSUN$ are captured by their {\it individual} sigmoid fit to better than 5 per cent accuracy over the same radial ranges.

The best-fit results are given in Figure \ref{fig:stellarprofilesfits_1}, with the steepness parameter $\gamma(\MTC)$ in the top panel and the radial pivot parameter $x(\MTC)$ in the bottom panel. 
The latter effectively corresponds to the stellar half mass radius of the galaxy in units of the virial radius ($r_{\rm stars, 0.5} = 10^{{\rm x}_{0.5}} \times \RTC$). For the $z=0$ cases, we show both results from the stacks and the individual fits, the latter averaged too in bins of halo mass: in each case the agreement is good. 

Both parameters exhibit remarkably clear trends with total halo mass (or galaxy stellar mass measured e.g. within 30 kpc), with $\gamma$ slowly declining as the radial distribution of stellar mass flattens for larger haloes. At the same time, x$_{0.5}$ increases with larger halo mass. Above halo masses of $10^{13}\MSUN$, both parameters can be expressed as linear or power-law functions of halo (or stellar) mass, which we provide in Table \ref{tab:profilefits}. Little evolution in redshift appears in the shapes of the normalized stellar enclosed mass profiles between $z\sim 1$ (thin curves) and today (thick curves): up to a maximum of 15-20\% reduction in the profile steepness at fixed mass.

In the bottom panel we recover the empirical result that the 3D half mass radius of the galaxy and the virial radius (or virial mass) of the underlying host halo are tightly correlated  \citep{Kravtsov:2013}. From the fits to the individual profiles, we find a galaxy-to-galaxy variation for the stellar half mass radius at fixed halo mass of $\sim$ 0.16 dex: this is slightly smaller than the 0.2 dex found by \cite{Kravtsov:2013} based on the abundance matching ansatz in concert with a diverse ensemble of observational datasets. 

Yet, the TNG relation does not agree with the one by \cite{Kravtsov:2013} in terms of normalization nor slope, with the latter scaling essentially like $\propto \MTC^{1/3}$ (grey annotations in Figure \ref{fig:stellarprofilesfits_1}, bottom panel), while ours is steeper, with a power-law slope in the range $0.41-0.53$ depending on resolution and exact mass range  \citep[see however][for a more direct comparison to galaxy size observations]{Genel:2017}. It is also important to note that this power-law relation breaks at lower masses and hence is applicable only to TNG haloes above a few $10^{12}\MSUN$: at $z\sim1$, galaxies deviate from the scaling relation already at $\sim 10^{12}\MSUN$.

{\renewcommand{\arraystretch}{1.2}
\begin{table}
\centering

  3D $M_{\rm stars} (< r)$ as a function of $\MTC$ at $z=0$
  (Eq.~ \ref{eq:logistic})\\
  Here: $m = {\rm log}_{10}(\MTC) - 14$
  
  \begin{tabular}{l  c c c}
    \hline
    Parameter &  slope $a$ & norm $b$ & 1$-\sigma$ scatter\\
    \hline
    
       ${\rm log}_{10}(\RTC)$& 0.33 & 2.99  & 0.05 dex\\
       ${\rm log}_{10}(\MSTC)$ & 0.74 & 12.04 & 0.18 dex\\
       $\gamma$ & -0.25 & 2.14 & 0.34 \\
       x$_{0.5}$ & 0.19 & -1.42 & 0.15\\
  
  \hline
  \end{tabular}

  \vspace{0.5cm}
  3D $M_{\rm stars} (< r)$ as a function of $M_{\rm stars}(<$ 30 kpc) at $z=0$\\
  Here: $m = {\rm log}_{10}(M_{\rm stars}(<$ 30 kpc$)) - 11.5$

  \begin{tabular}{l c c c }
    \hline
    Parameter & slope $a$ & norm $b$ \\
    \hline
       ${\rm log}_{10}(\RTC)$& 0.70 & 2.88 \\
       ${\rm log}_{10}(\MSTC)$ & 1.56 & 11.80 \\
       $\gamma$ & -0.52 & 2.22 \\
       x$_{0.5}$ & 0.41 & -1.48\\
    \hline
  \end{tabular}

\caption{\label{tab:profilefits} Instructions to recover {\it average} 3D enclosed stellar mass profiles from the TNG galaxies at $z=0$. These average stellar profiles follow from Eq.~(\ref{eq:logistic}), which depends on four quantities: $\RTC$, the stellar mass within this radius $\MSTC$, the sigmoid slope $\gamma$, and the sigmoid midpoint \mbox{x$_{0.5}=\,$log$_{10}(r_{\rm stars,0.5}/\RTC)$}. Each of these four parameters lies on tight linear or power-law relations with halo mass or stellar mass within a given aperture, at fixed redshift. The tables above give such relations: $y = am + b$, where $m$ is the logarithm of the input (see top of the sub tables) and $y$ is any of the parameters in the first columns. Range of validity: $10^{13}< \MTC/\MSUN < 10^{15}$; $z\sim 0$. In the upper panel, we also give the one sigma scatter in the parameters at fixed halo mass, as an indication of the halo-to-halo (or galaxy-to-galaxy) variation of the parameters. All masses are in $\MSUN$ and assume a Chabrier IMF; all lengths are in kpc.}
\end{table}}
\subsubsection{From halo (galaxy) mass to the whole stellar mass profile: a practical tool}
\label{sec:thefit}

In order to develop a useful and practical tool, we give formulas for the four parameters needed to recover the full 3D $M_{\rm stars}(< r)$ profile as a function of halo mass and redshift, or galaxy stellar mass and redshift. Table \ref{tab:profilefits} gives simple fit relations for the virial radius $\RTC$, the total diffuse stellar mass included within this radius $\MSTC$, the sigmoid steepness $\gamma$, and the sigmoid midpoint x$_{0.5}$ as a function of total halo mass or galaxy mass. In practice, the third and fourth rows in Table  \ref{tab:profilefits} (the sigmoid steepness $\gamma$ and midpoint x$_{0.5}$, respectively) provide fits to the average relations depicted in Figure \ref{fig:stellarprofilesfits_1}, at $z=0$, as a function of halo or galaxy mass.

By plugging in a halo mass, it is possible to easily recover the full stellar mass profile of the halo (outwards of a few kpc from the center), and so by definition also the stellar mass of the corresponding galaxy restricted to any given aperture. In fact, the total halo mass $\MTC$ and the total diffuse stellar mass within $\RTC$ also lie on a very tight scaling relation at the group and cluster mass scales (we show this explicitly in the next Section), so in practice, from the observational side, if instead of halo mass a particular stellar mass is known, e.g. the mass within 30 kpc, this can still be used to determine the entire stellar mass profile of the galaxy. For any other galaxy stellar mass measurement, this process can be easily inverted, by calculating a grid of stellar mass profiles for different halo masses, then selecting the one which contains the matching stellar mass within the observationally measured aperture. In this way, the full `unseen' stellar mass distribution of the observed system is predicted, as well as its total dark matter halo mass. 

The parameterization given in Table \ref{tab:profilefits} in combination with Eq.~(\ref{eq:logistic}) returns the TNG {\it average} stellar enclosed mass profiles (not normalized) to an accuracy of 10 per cent or better, across the whole mass range $10^{13}< \MTC/\MSUN < 10^{15}$ at $z\sim 0$ and at all radii larger than 2 kpc. This is demonstrated with the black solid curves in Fig.~\ref{fig:stellarprofiles}, bottom four panels, which have been obtained by solely knowing the total halo mass of the reported bin. By construction, this tool cannot reproduce the special features of certain high mass haloes within a few kpc in clustercentric distance nor the fast drop within the innermost cores of the simulated galaxies. Yet, the large scale regularity of the spherically-averaged stellar mass profiles as a function of total halo or galaxy mass is striking.

\begin{figure}
\centering
\includegraphics[width=8.3cm]{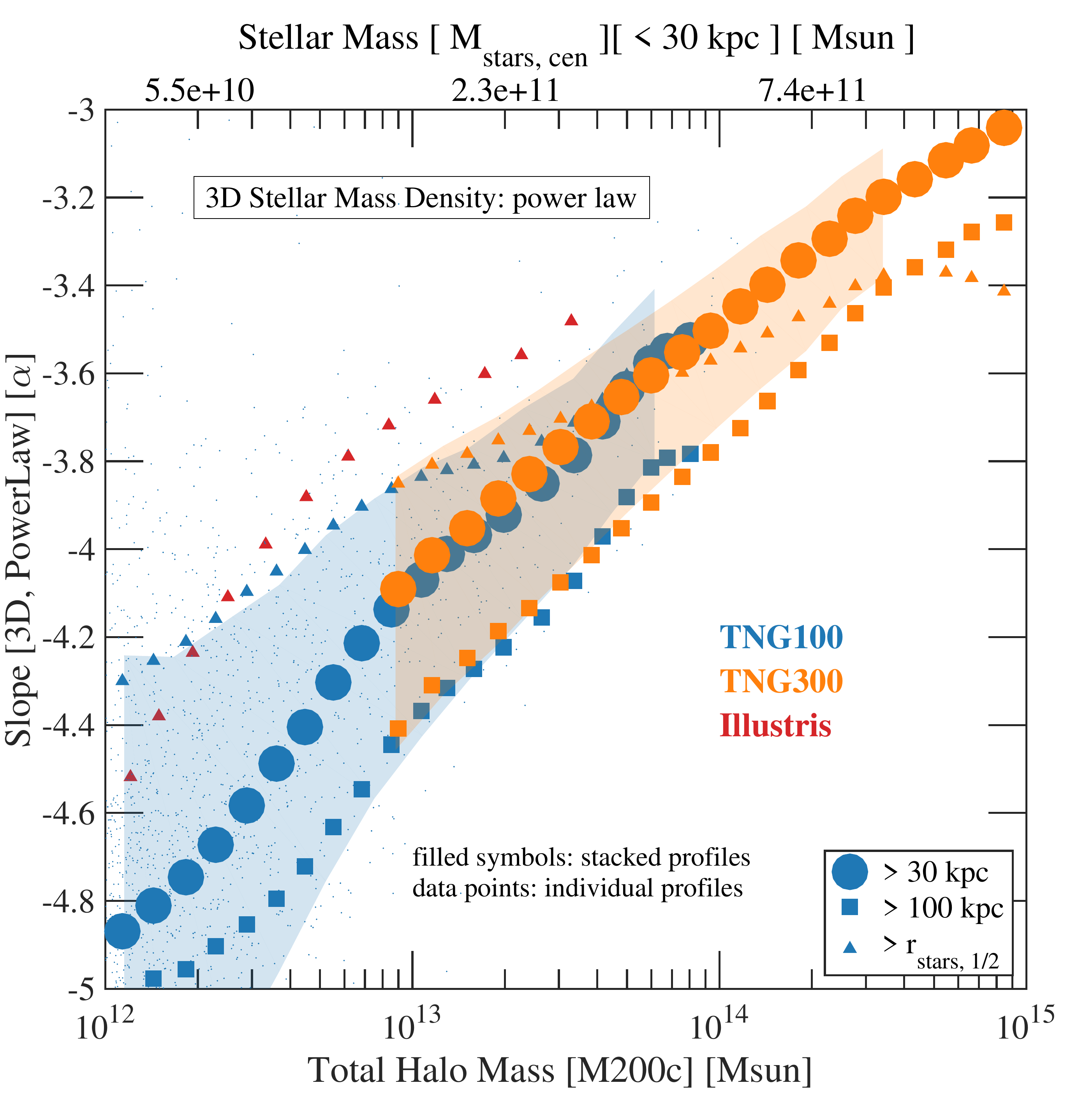}
\caption{\label{fig:stellarprofilesfits_2} Best-fit power-law slope to the 3D stellar mass density in the stellar haloes or intra-cluster light, i.e. at large distance from the central galaxy. We show trends as a function of total halo or stellar mass at $z=0$. The measurement is made on all stars between a given radial inner aperture and the virial radius: $>$ 30 kpc (circles), $>$ 100 kpc (squares), or $> r_{\rm stars,0.5}$ (triangles). Results are shown from TNG100 (blue), the not rescaled TNG300 (orange), and original Illustris (red). Results from stacked profiles (filled symbols) are in very good agreement with the averages from the individual profile fits (not shown). Small data points are the slopes of individual TNG100 galaxies and shaded areas denote the 1-sigma halo-to-halo variation (both for $>$ 30 kpc). Over 3 dex in halo mass the power-law slope of the stellar halo drops from $-3$ to $-5$, reflecting the flattening of the extended stellar mass profile.}
\end{figure}

\subsubsection{Stellar mass density at large distances}
\label{sec:powerlaw}

To specifically characterise the outer stellar slopes, we follow \cite{Pillepich:2014bb} and express the 3D diffuse stellar mass profiles at large radii with a power-law function, 
\begin{equation}
\rho_{\rm stars} (< r, z) = \rho_0 ~ r^{\alpha}, \,{\rm for }~ r \gg {\rm 1 ~ kpc}, 
\end{equation}
where the slope parameter $\alpha$ and the central stellar density $\rho_{0}$ are both functions of mass and redshift. The profiles, measured in spherical shells evenly spaced in logarithmic radius ($\Delta{\rm  log}_{10} (r/[{1 \rm kpc}]) \approx 0.03$), are obtained as before by stacking the density profiles of individual objects in running bins of halo mass ($\MTC$) 0.2 dex wide. In practice, the fitting procedure is performed in logarithmic space, both for stacked and individual profiles, by minimizing the summed squares of the residuals from a first-degree polynomial fitting function, equally weighting all bins containing at least one particle each. These {\it ICL} power-law slopes are then evaluated for different radial apertures, with the inner boundary ranging between fixed 30 or 100 kpc to the 3D stellar half mass radius (with an outer boundary fixed to the virial radius in all cases). 

The results of the power-law fits are shown in Figure \ref{fig:stellarprofilesfits_2}, where the shallower stellar halo profiles towards larger halo masses are immediately recovered. Note that TNG300 and the rescaled rTNG300 simulations return the same median stellar halo slopes, and hence in Figure \ref{fig:stellarprofilesfits_2} we show results for the unprocessed TNG300 output. 
Considering all gravitationally-bound stars outside of a fixed aperture, either 30 kpc or 100 kpc, but excising satellites, the $\alpha$ slopes increase from $-5$ to $-3$ between halo masses of $10^{12}$ and $10^{15}\,\MSUN$, in broad agreement with the original findings of \cite{Pillepich:2014bb} based on the Illustris simulation. The adaptive $r_{\rm stars,1/2}$ aperture includes different amounts of the central galaxy at different masses, making the mass trend of $\alpha$ correspondingly weaker than with the fixed 30 and 100 kpc boundaries. The power-law slopes from the original Illustris simulation, outward of the 3D stellar half mass radius, are also shown (red triangles): they are steeper than the corresponding TNG averages (blue triangles) below a halo mass of $4 \times 10^{12}\, \MSUN$, and shallower above. Indeed, the 3D stellar half mass radii of Illustris and TNG galaxies are different \citep{Pillepich:2017}, implying overall rather different radial distributions of the stellar material. However, the TNG model has been shown to reproduce galaxy sizes in better agreement with observations \citep{Genel:2017} than Illustris \citep{Bottrell:2017}, lending more credibility to the outcome of the TNG stellar mass distributions also at highest mass end.  
Regardless of the exact definition for the inner boundary of the stellar envelopes, the stellar mass in the outskirts of the most massive galaxy clusters is almost always as shallow as the dark matter's, with average density slopes in the range $[-3.5, -3]$.


\section{The mass budget in groups and clusters between \texorpdfstring{$z \sim 1$}{z=1} and today}
\label{sec:budget}

\begin{figure*}
\centering
\includegraphics[width=8.5cm]{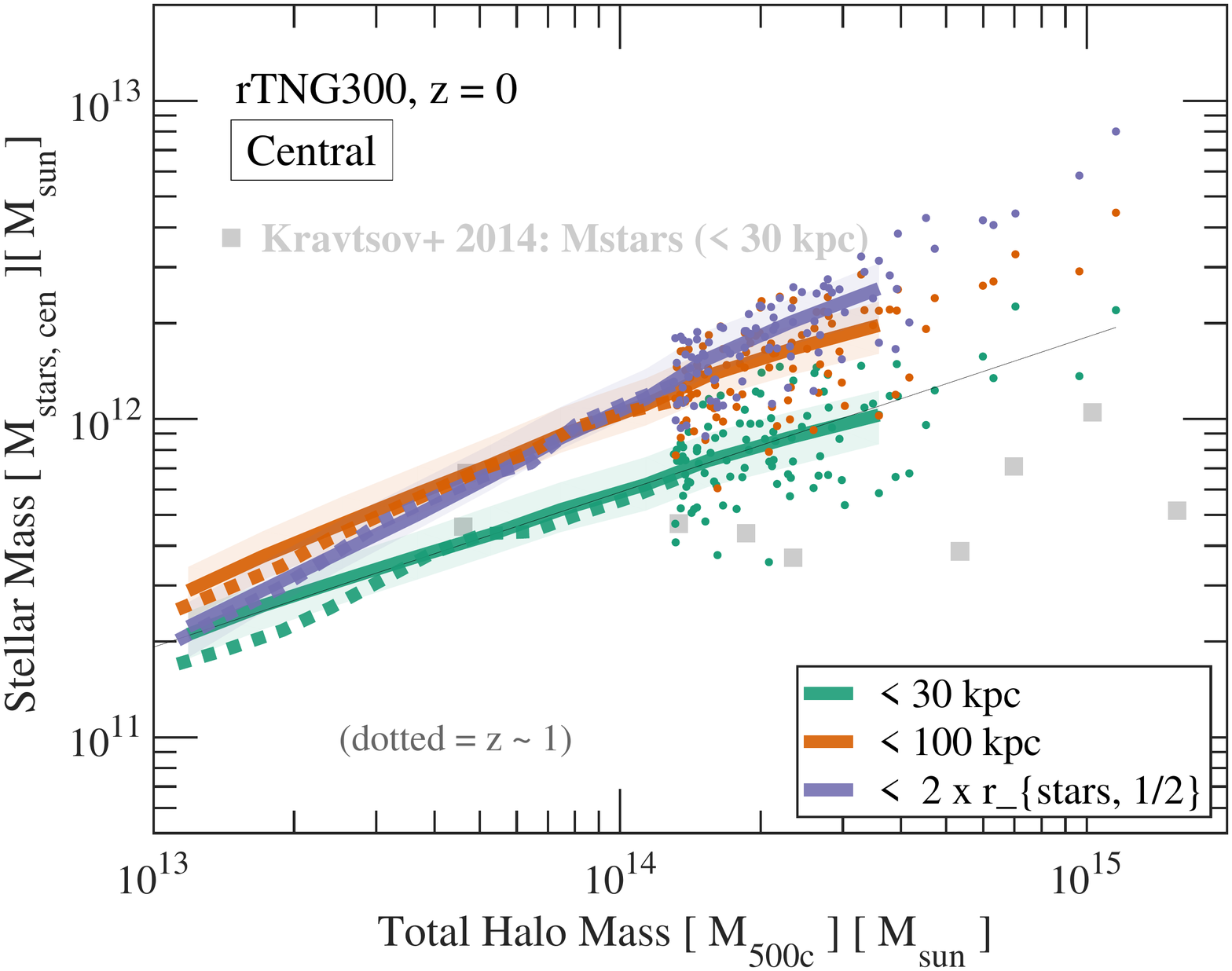}
\includegraphics[width=8.5cm]{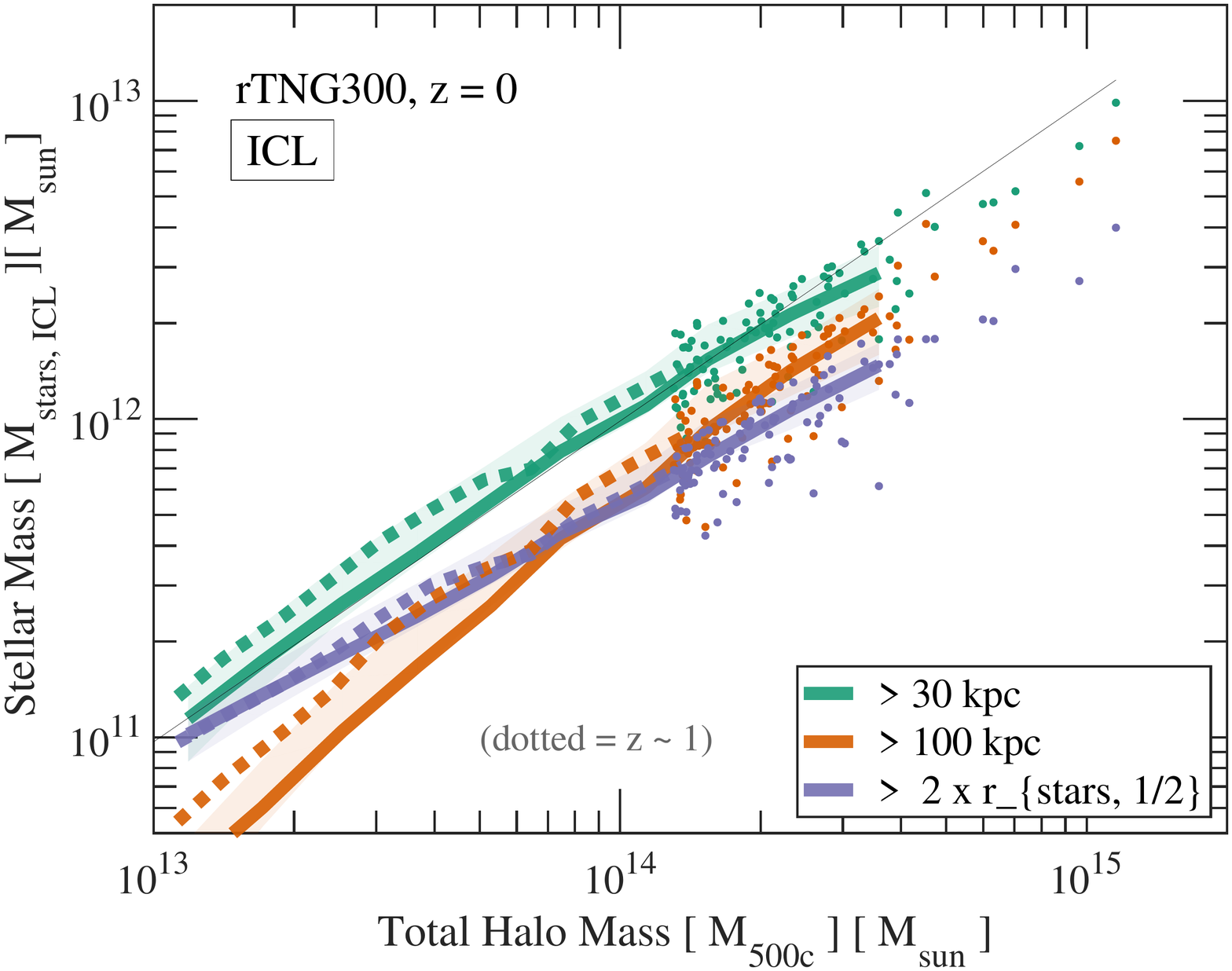}
\includegraphics[width=8.5cm]{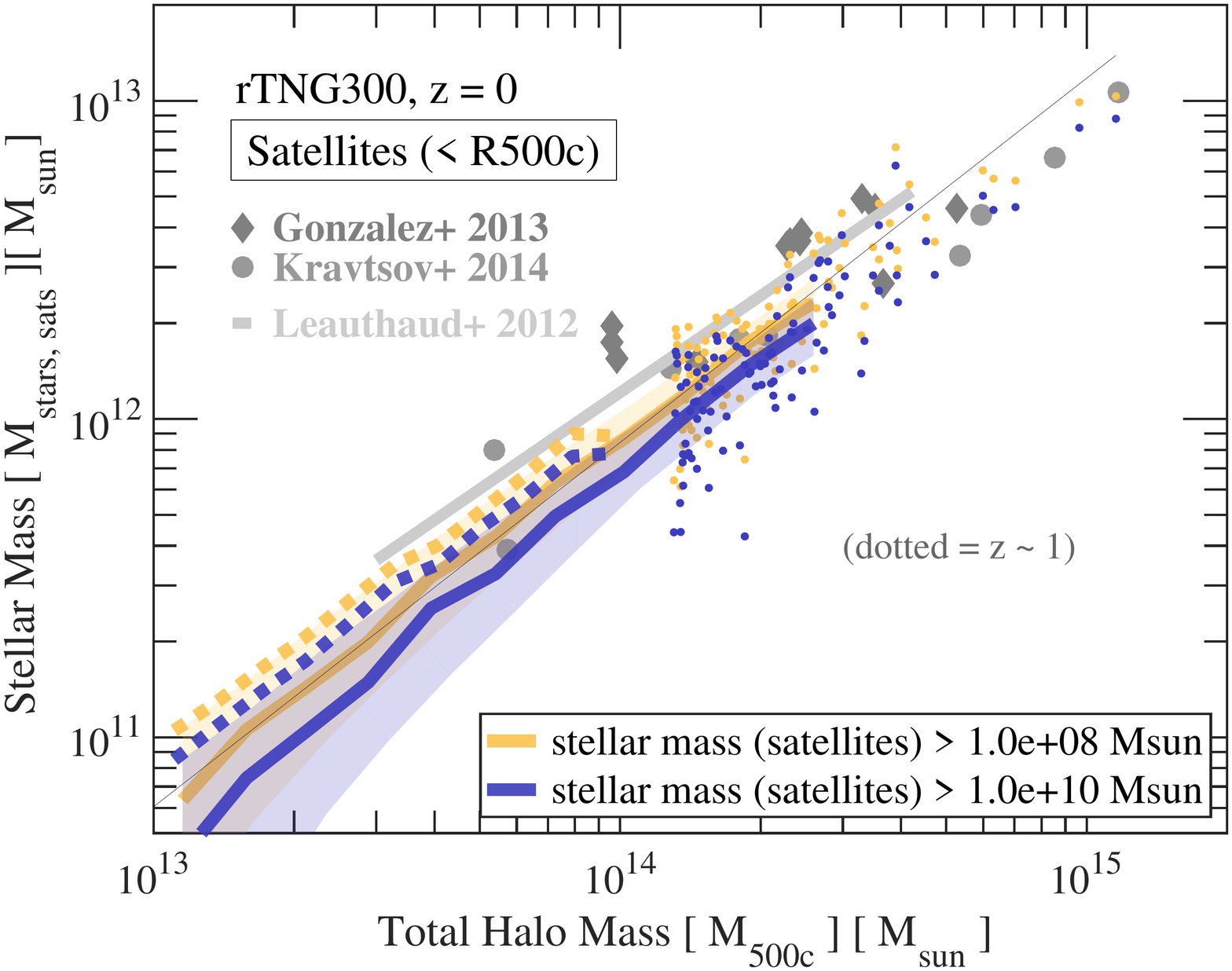}
\includegraphics[width=8.5cm]{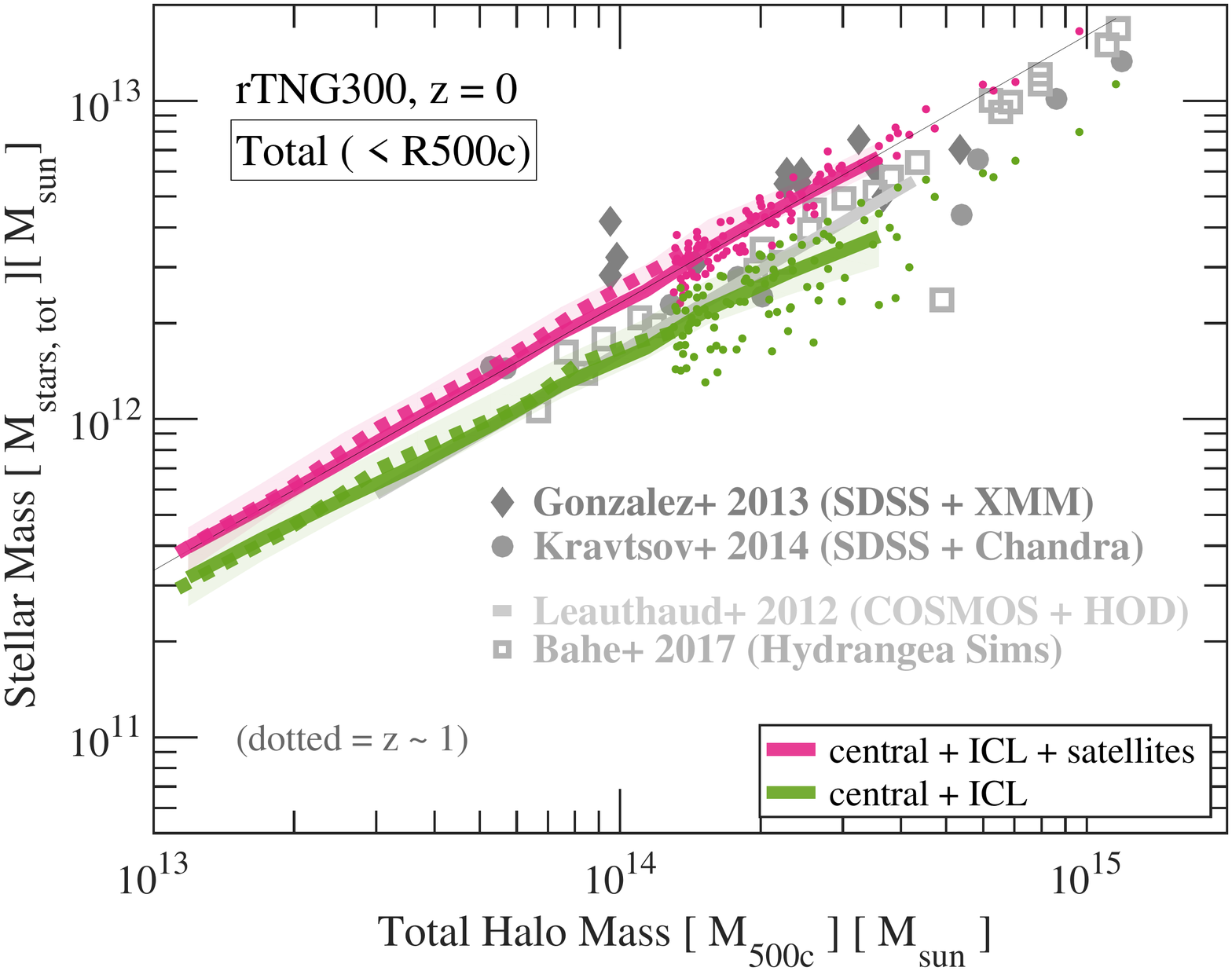}

\caption{\label{fig:stellarmasses} The stellar mass budget of galaxy groups and clusters from the TNG simulations, in the central galaxy (top left), in the ICL (top right), cumulatively in satellites (bottom left), and in the whole halo: central+ICL+satellites vs central+ICL (bottom right). For the operational definitions of the cluster components, see Table \ref{tab:defs}. Fixed apertures are all 3D, for both observations and simulations, and in comoving units. Solid lines denote medians in bins (0.2 dex in $\rm{log}_{\rm10}\MFC$) with at least 5 objects; individual dots represent the top 100 most massive objects in the simulation; shaded area indicate the 1-sigma halo-to-halo variation. All solid curves, symbols and annotations refer to $z=0$ results, but for the dotted median curves, denoting $z=1$ results. Constraints from the literature, mostly observational, are indicated in grey curves and large symbols, with the Bahe+ 2017 data points being at $z=0.1$. In the lower left panel, observational data are nominally for all the stellar mass locked in satellites with no threshold or selection bias on the cluster members. In the lower right panel, the literature data points are all to be intended for the `central+ICL+satellites' case. Fitting functions for the stellar mass profiles of TNG objects are provided in Section \ref{sec:profiles} and Table \ref{tab:profilefits} and can be used to reproduce these TNG predictions. Basic linear fits to the scaling relations of this Figure can be found in Table \ref{tab:massfits}, and are indicated in selected cases with solid thin black lines.}
\end{figure*}

From the inspection of the stellar profiles in the previous Section, we see that individual haloes can exhibit changes of steepness across their radial extent. In our simulated massive clusters there are hints of possibly different profile components (e.g. bottom panels of Figure \ref{fig:stellarprofiles}). However, our limited spatial and numerical resolution -- precisely at the highest mass end and at small clustercentric distances -- does not allow us to support population-wide quantitative reasons to prefer a description of the mass profiles in Figures~\ref{fig:profiles_prototypes} and \ref{fig:stellarprofiles} in terms of a superposition of multiple profiles of different shapes rather than a unique function. In practice, we cannot identify any optimal, qualitative, or generalized boundary between the innermost, brightest regions of galaxies and their lower surface brightness envelopes. 
We hence proceed in this Section to quantify the stellar mass census of the cluster components by adopting the definitions given in Section \ref{sec:components} and Table \ref{tab:defs}, all based on arbitrary fixed-apertures decompositions.

\subsection{Central galaxy, intracluster light, and cluster satellites}
\label{sec:budget_1}

In Fig.~\ref{fig:stellarmasses}, we give the amount of mass in different cluster components in rTNG300 as a function of total halo mass. Here we use $\MFC$ to facilitate observational comparisons with X-ray derived mass estimates. We quantify: (1) the stellar mass in the central galaxy within fixed spherical apertures of 30 kpc, 100 kpc or twice the stellar half mass radius, excising satellites (top left); (2) the ICL, namely the stellar mass outside fixed spherical apertures of 30~kpc, 100~kpc or twice the stellar half mass radius, excising satellites (top right); (3) the total satellite stellar mass component, with the sum carried out for different thresholds in satellite stellar mass (bottom left); and (4) the total stellar mass in all components across the whole total mass overdensity (bottom right). Solid lines denote medians in total halo mass bins (0.2 dex in ${ \rm log}_{\rm10}\MTC$) with at least 5 objects. Individual dots represent the 100 most massive simulated objects, while shaded areas indicate the 1-sigma halo-to-halo variation. All solid curves and symbols refer to $z=0$ results, except for the dotted median curves which show the same measurements at $z=1$. 

For all components in the mass range studied here ($10^{13}\MSUN \leq \MTC \leq 10^{15}\MSUN $), the stellar versus total halo masses are accurately described by power law relations, with the most massive clusters containing the largest amount of stellar mass, regardless of type. The trends with total mass, however, are different for different components: the stellar mass enclosed in the outskirts (top right) and in satellites (bottom left) is a much steeper function of total mass than the stellar mass of the innermost regions (top left). In other words, Fig.~\ref{fig:stellarmasses} is a quantitative demonstration that the TNG  galaxy formation model, coupled with the hierarchical growth of structure, naturally produces massive clusters of galaxies which are surrounded by relatively more numerous and more luminous satellites, and by larger amounts of stellar mass in the far outskirts, than less massive groups. 

In more detail, the slope and normalization of the relations of the diffuse mass depend on the aperture within or beyond which the central galaxy and ICL are defined (top panels). The total stellar mass (central+ICL+satellites, magenta curve in lower right panel) is a steeper function of halo mass when satellites are included, in comparison to the case when satellites are excluded (total diffuse mass = central + ICL, green curves in lower right panel). The stellar mass bound to satellites (lower left panel) is also a strong function of the stellar mass of the satellites themselves, due to the steep stellar mass function of satellites and the steep relation between richness and total halo mass (see Fig.~\ref{fig:stats}, lower panel).

We can also characterize the time evolution of the relations and their scatter. In Figure \ref{fig:stellarmasses}, dotted curves denote the analog $z\sim1$ average stellar-to-total mass relations, albeit truncated to lower masses because of the lack of massive objects sampled at early cosmic times. These should be compared to the solid curves at $z=0$. Interestingly, there appears to be very little redshift evolution in the scaling relations of Fig.~\ref{fig:stellarmasses}, in the sense that $z=1$ relations lie within the present day 1-sigma scatter. Moreover, differences among different definitions can be much larger at any fixed time than the evolution of the relations between $z \sim 1$ and today. Still, a few interesting trends are evident. First, at earlier times the scaling relations are slightly tilted compared to those at $z=0$, steeper or shallower and by varying amounts for the different components. Secondly, the amount of stellar mass contained in satellite galaxies is larger (up to a factor of 2) at earlier times than today. In the last few Gyrs of cosmic evolution, stars are stripped from orbiting and incoming satellites and mergers, thereby contributing to the diffuse components labeled here as central galaxies and ICL. Yet, interestingly, the stellar mass in the outskirts at fixed halo mass was larger at earlier times (Figure \ref{fig:stellarmasses}, top right panel), although only beyond {\it fixed comoving} clustercentric distances. 

Overall, massive haloes build up their stellar mass at roughly the same pace as they assemble their total dark matter mass -- the total stellar mass at a fixed halo mass scale is essentially invariant from $z=1$ to $z=0$ (as seen in the bottom right panel), while the satellite contribution is higher at $z = 1$ than today and the central (ICL) contribution is at most (at least) as high. One needs to keep in mind, however, that the total measured mass of each component depends on definition, and particularly if reference is made to an evolving mean or critical overdensity. Moreover, the different redshifts comparison of Fig.~\ref{fig:stellarmasses} does not represent the evolution of individual systems. In Section \ref{sec:theory} we explore the origin of the stellar mass growth further, breaking down its predominantly accretion origin at late times.

\begin{table}
 \begin{tabular}{l l c c c}
 
    \hline
    Relation & definition & slope & norm & scatter\\
    \hline
    &&&\\
    $M_{\rm stars, cen} - \MFC$& $< 30$ kpc & 0.49 & 11.77& 0.12\\
    $M_{\rm stars, cen} - \MFC$& $< 100$ kpc & 0.59 & 12.00& 0.13\\
    $M_{\rm stars, cen} - \MFC$& $< 2\times r^{\rm stars}_{0.5} $ &0.74&12.02&0.12\\
    &&&\\
    $M_{\rm stars, ICL} - \MFC$& $> 30$ kpc & 1.01&12.01&0.13\\
    $M_{\rm stars, ICL} - \MFC$& $> 100$ kpc & 1.25&11.74&0.14\\
    $M_{\rm stars, ICL} - \MFC$& $> 2\times r^{\rm stars}_{0.5} $ & 0.77&11.72&0.11\\
    &&&\\
    $M_{\rm stars, sat} - \MFC$& $>10^8 \MSUN$& 1.14&11.93&0.22\\
    $M_{\rm stars, diff} - \MFC$& cen+ICL& 0.75&12.19&0.11\\
    $M_{\rm stars, tot} - \MFC$& cen+ICL+sats& 0.84&12.36 & 0.07\\
    &&&\\
    \hline
    
\end{tabular}
\caption{\label{tab:massfits} The stellar mass budget in TNG groups and cluster at $z=0$: best fit parameters to the relations presented in Figure \ref{fig:stellarmasses}. The adopted fitting function reads: $y = am + b$, where $m={\rm log}_{10}(\MFC / \MSUN) -14$ and $y = {\rm log}_{10}(M_{\rm stars} / \MSUN$). Important: the reported scatter (in dex) is simply the median 1-sigma halo-to-halo variation in bins of ${\rm log}_{10}\MFC$.  Range of validity: $10^{13}< \MTC/\MSUN < 10^{15}$; $z\sim 0$. All masses are in $\MSUN$ and assume a Chabrier IMF; all apertures are 3D.}
\end{table}

With respect to other findings, the TNG stellar masses across the various components are in the ball park identified by \citet{Gonzales:2013} combining SDSS and XMM data, \cite{Kravtsov:2014} combining SDSS and Chandra data, as well as those from \citet{Leauthaud:2012} from COSMOS (lower two panels). All adopt a Chabrier IMF, consistently with the TNG choice, and we report here their measures within 3D apertures in all cases.

In more detail, the mass in simulated central galaxies (top left panel, 3D 30 kpc aperture) appears still up to a factor of 3 (or $0.4-0.5$ dex) larger than observational constraints by \citealt{Kravtsov:2014} (their Table 4, third column, for 3D 30 kpc aperture). In fact, the TNG and \citealt{Kravtsov:2014}'s relations between stellar mass ($<30$ kpc$)$ and halo mass differ substantially in {\it slope}, but their normalizations are largely consistent at the $10^{14}\MSUN$ halo scale. The mismatch among simulated and observed stellar masses at the highest mass end is indeed still a recurring issue for models, with e.g. the simulated clusters by \cite{Ragone:2013, Hahn:2017, Bahe:2017} all returning central's stellar masses that are up to $\sim 0.6$ dex larger than observational constraints, all taken at face value and measured within a few tens of kpc from the centers. Instead, the overall predicted {\it total} amounts of stellar mass in clusters appears in better agreement with the available observational values.
%
The steep increase of the stellar mass locked up in orbiting satellites (bottom left) is commonly recovered, although our simulations lie about 30-50 per cent lower than data constraints. In TNG, we find a power law slope of order unity above a satellite threshold mass of $10^8 \MSUN$, total satellite mass and total halo mass rising in concert. Generally, the slope of the total stellar mass trends (bottom right, central + ICL + satellites) is slightly larger than that favored by \citet{Gonzales:2013}, while it is quite similar to e.g. those from the Hydrangea simulations \citep{Bahe:2017} and the semi-empirical constraints by \cite{Leauthaud:2012}, yet TNG is about 30\% higher than the observational data by \cite{Kravtsov:2014} in normalization. 

To facilitate future comparisons between the TNG model predictions and the observations, in Table \ref{tab:massfits} we provide fitting formulas to these particular relations of the various stellar components versus halo mass. We note that the same quantities can be obtained by adopting the more general fitting functions provided previously which describe the enclosed stellar mass profiles of TNG galaxies (Section \ref{sec:sigmoid}): we argue that conclusive remarks for the comparison between models and observations should be determined by contrasting the whole stellar mass profiles, as we will do in future works. Interestingly, all scaling relations between stellar mass and total halo mass at the massive end exhibit very little scatter, in the range 0.1-0.2 dex (see Table \ref{tab:massfits}). For example, for the total stellar mass we find an intrinsic scatter at fixed halo mass of 0.07 dex, in between the findings of \citealt{Andreon:2012} ($\la 0.06$ dex) and e.g. \citealt{Kravtsov:2014} (0.1 dex).

\begin{figure}
\centering
\includegraphics[width=8.4cm]{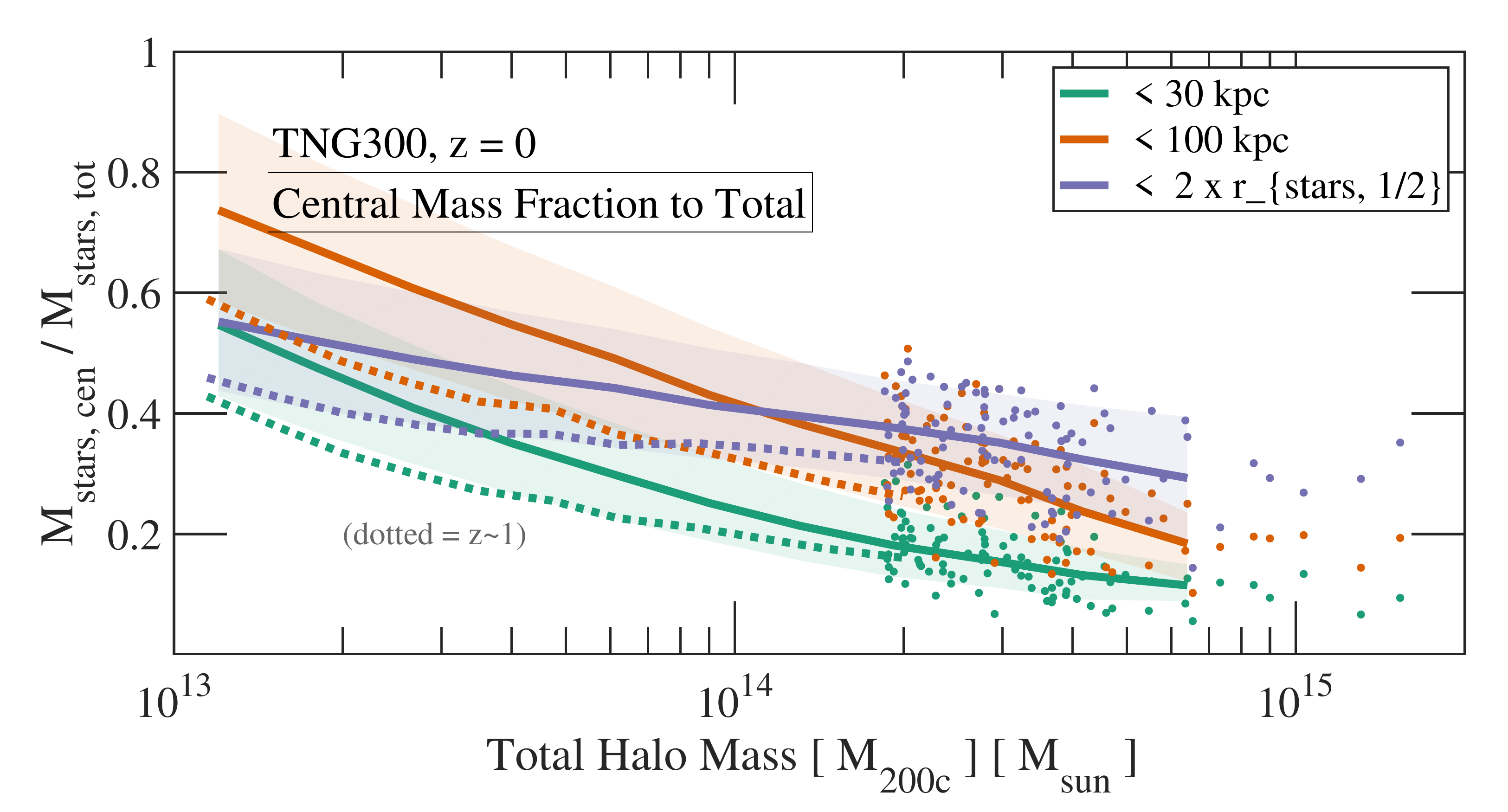}
\includegraphics[width=8.4cm]{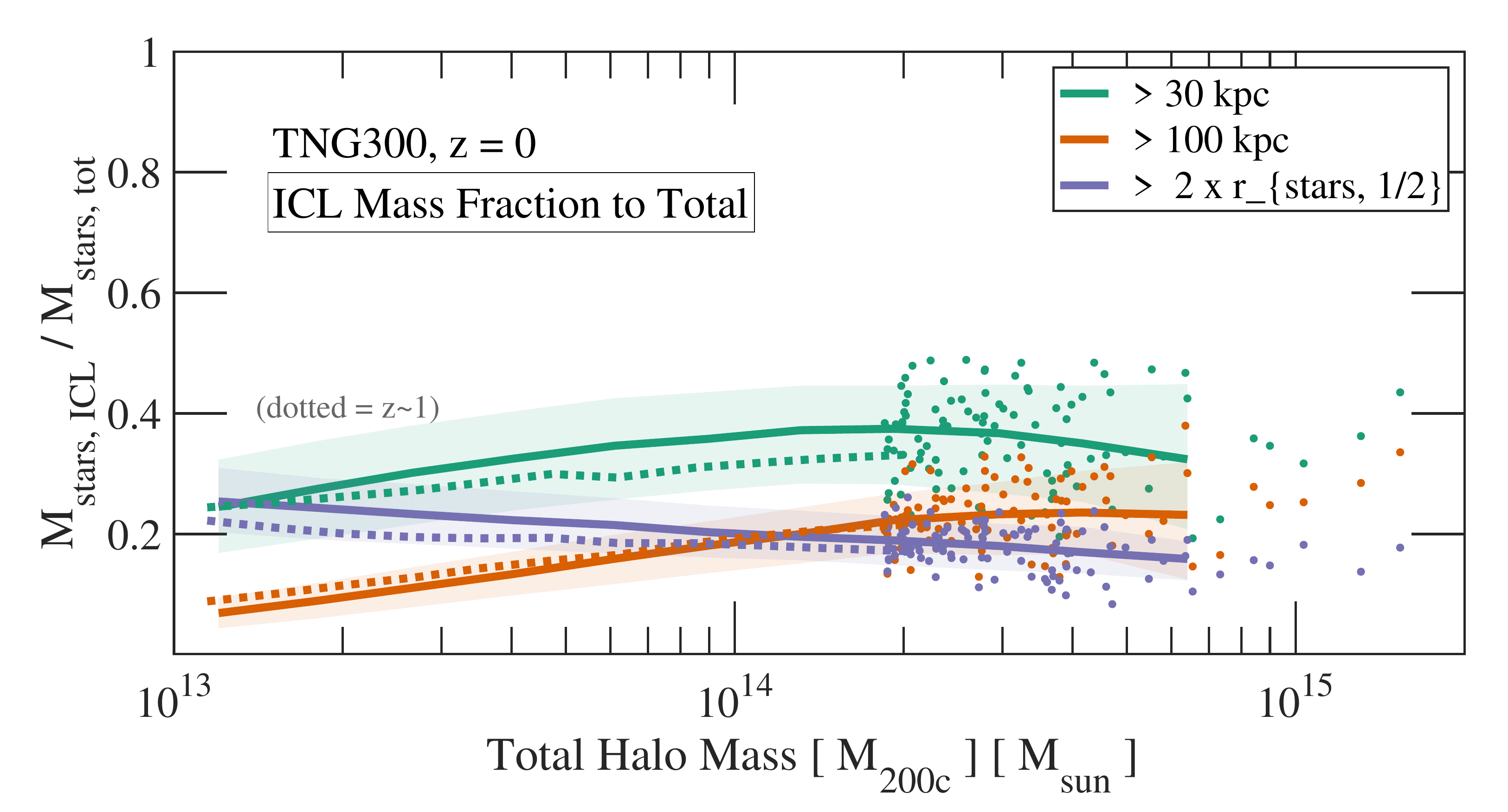}
\includegraphics[width=8.4cm]{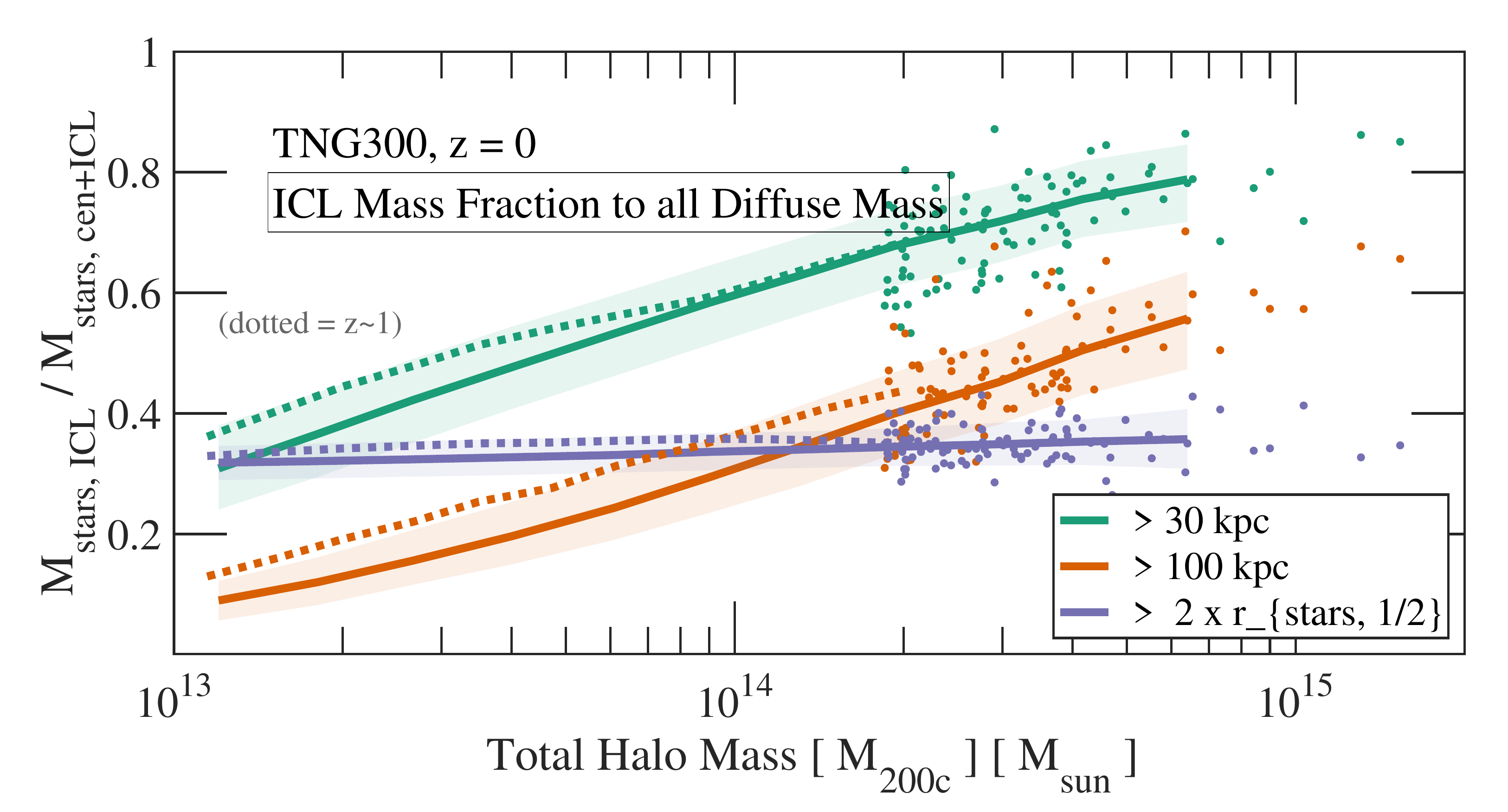}
\includegraphics[width=8.4cm]{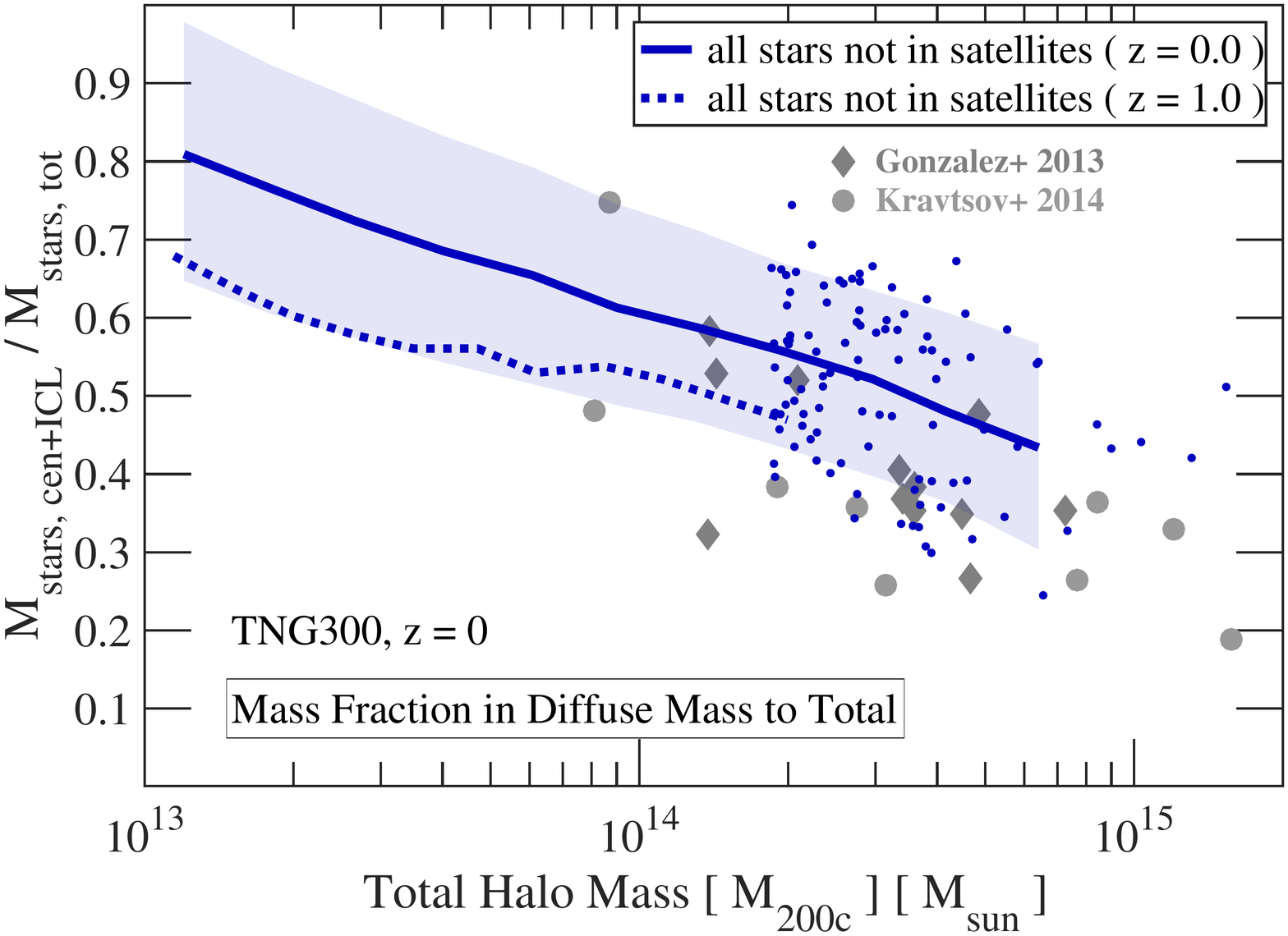}
\caption{\label{fig:fractions} Stellar mass fractions in the various luminous components of groups and clusters, as a function of total halo mass. From top to bottom: stellar mass in the central galaxy to the total stellar mass out to the virial radius; stellar mass in the ICL or outskirts to the total stellar mass; mass in the ICL or outskirts to the stellar mass in the diffuse components (central + ICL); stellar mass in the diffuse component (central + ICL) to total stellar mass out to the virial radius. Annotations are as in Figure \ref{fig:stellarmasses}, with all solid curves, symbols and annotations referring to $z=0$ relations, but for the dotted median curves, denoting $z=1$ results. }
\end{figure}

\subsection{Stellar mass fractions across halo mass and cosmic time}
\label{sec:budget_2}

To highlight the relative importance of different elements of the stellar mass budget, Fig.~\ref{fig:fractions} gives the contribution of individual components to the total {\it stellar} mass enclosed within the virial radius, as a function of total halo mass. From top to bottom we show: (1) the fraction of stellar mass in the central galaxy relative to the total stellar mass out to the virial radius; (2) the stellar mass in the ICL to the total stellar mass, both out to the virial radius; (3) the stellar mass in the diffuse components (central + ICL) to the total stellar mass out to the virial radius; and (4) the mass in the ICL to the stellar mass in the diffuse components (central + ICL). As in Fig.~\ref{fig:stellarmasses}, the stellar mass associated to the ICL is obtained by integrating from a given spherical aperture all the way out to the virial radius (here $\RTC$, see Table \ref{tab:defs}).

The results of Fig.~\ref{fig:fractions} are shown for TNG300 systems, but we note that all trends and quantities are unchanged in TNG100 at its improved numerical resolution, in the overlapping mass range $10^{13}-10^{14}\MSUN$. This is because the effects of resolution largely vanish when taking ratios of stellar masses. 

At the current epoch, the relative contribution of the innermost regions to the total stellar mass (top panel) is smaller for more massive objects: for $10^{15}\MSUN$ clusters, only about 10\% (20\%) of the total stellar mass is within 30 (100) kpc of the cluster centers (top panel). The relative amount beyond large, fixed cluster-centric distances (ICL) is a larger fraction of the total diffuse stellar mass (central + ICL) for more massive haloes (bottom panel). In $10^{15}\MSUN$ clusters, more than 80\% (60\%) of the cluster stellar mass {\it excluding satellites} is found beyond 30 (100) kpc distance from the centers (third panel from the top). And finally, the contribution of the diffuse stellar mass to the total (cen+ICL to cen+ICL+satellites) decreases towards larger masses. Conversely, the amount of mass locked up in surviving satellites is larger for more massive objects (Figure \ref{fig:fractions} bottom panel): the richest clusters have more than  half of their stars bound to satellite galaxies. In the bottom panel of Figure \ref{fig:fractions} grey data points represent the findings of \cite{Kravtsov:2014} and \cite{Gonzales:2013} for what they call their BCG fraction to the total mass, with the former being measured by a triple-Sersic fit to the central's profile extrapolated to infinity. Here we have converted the observational data from $\MFC$ to $\MTC$, adopting the simulation-based relation. As in the case of the stellar mass within 30 kpc (top left panel of Figure \ref{fig:stellarmasses}), the contribution to the total budget of our TNG massive galaxies is shifted upwards by 10-15 per cent in comparison to observational constraints, namely it is 30-40 per cent relatively larger than the values inferred from observations.

We draw attention to the relatively shallow trends found when the adaptive $2 \times r_{\rm stars,0.5}$ boundary is used (purple lines). In particular, the fraction of ICL mass relative to the total diffuse stellar mass is nearly constant with halo mass, reflecting a degree of invariance in the shapes of the radial stellar profiles as previously seen in Figure \ref{fig:stellarprofiles}.

Similar trends hold between redshift $z=0$ and $z=1$ (solid versus dotted lines): the relations at these two times are nearly always parallel. Interestingly, the 1-$\sigma$ halo-to-halo variation in the fractional contribution of the central mass is larger for less massive haloes, with $10^{13}\,\MSUN$ groups exhibiting central galaxies contributing anywhere between 55 and 95\% to the total within a 100 kpc aperture. At a few times $10^{14}\,\MSUN$ this variation decreases to of order 10-20\%. On the other hand, the scatter in the extended components is roughly constant with halo mass, implying that for groups and clusters merger histories with $\gg 1$ event may suppress the stochastic diversity of accreted mass.

\begin{figure*}
\centering
\includegraphics[width=8.6cm]{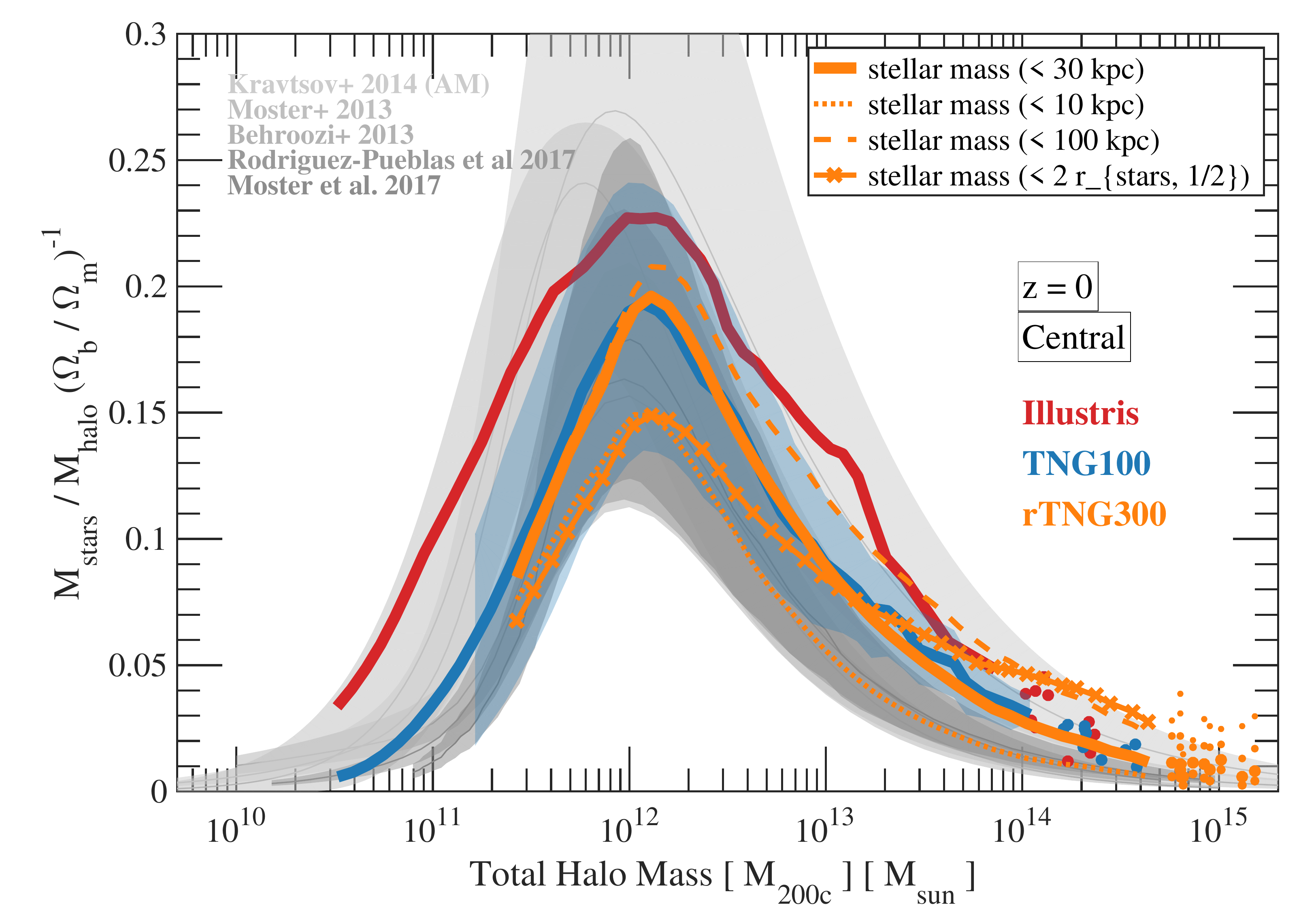}
\includegraphics[width=8.6cm]{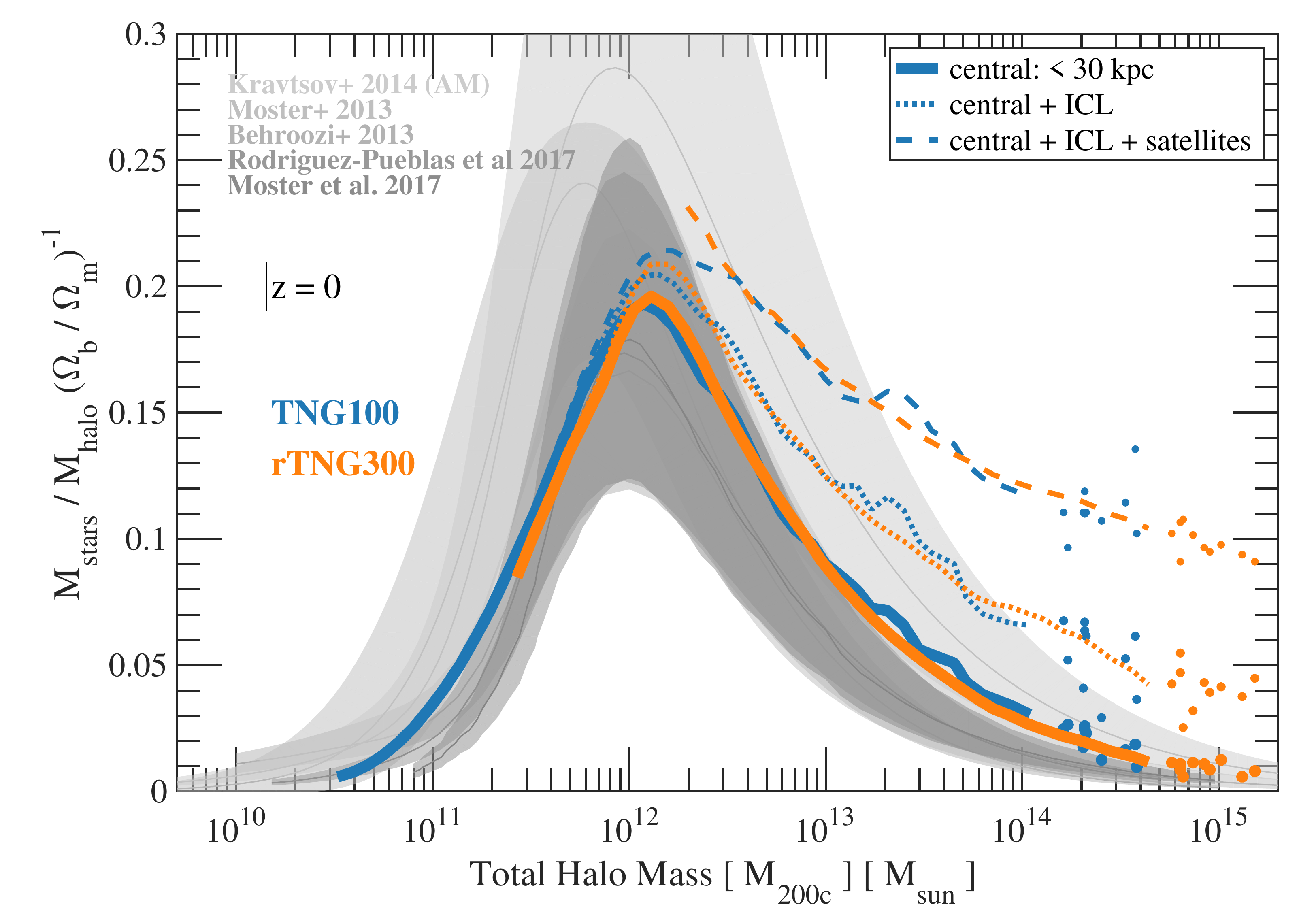}
\includegraphics[width=8.6cm]{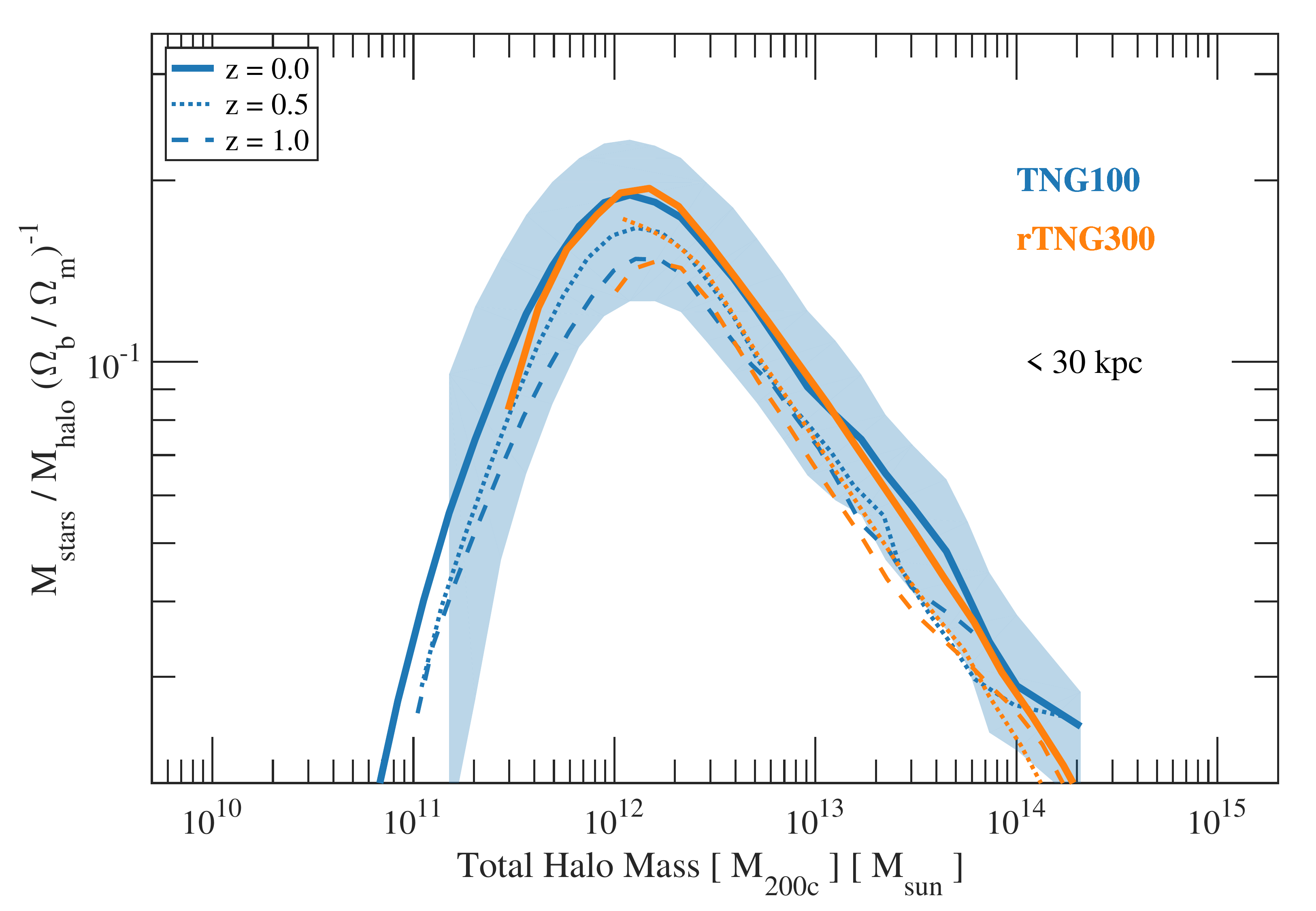}
\includegraphics[width=8.6cm]{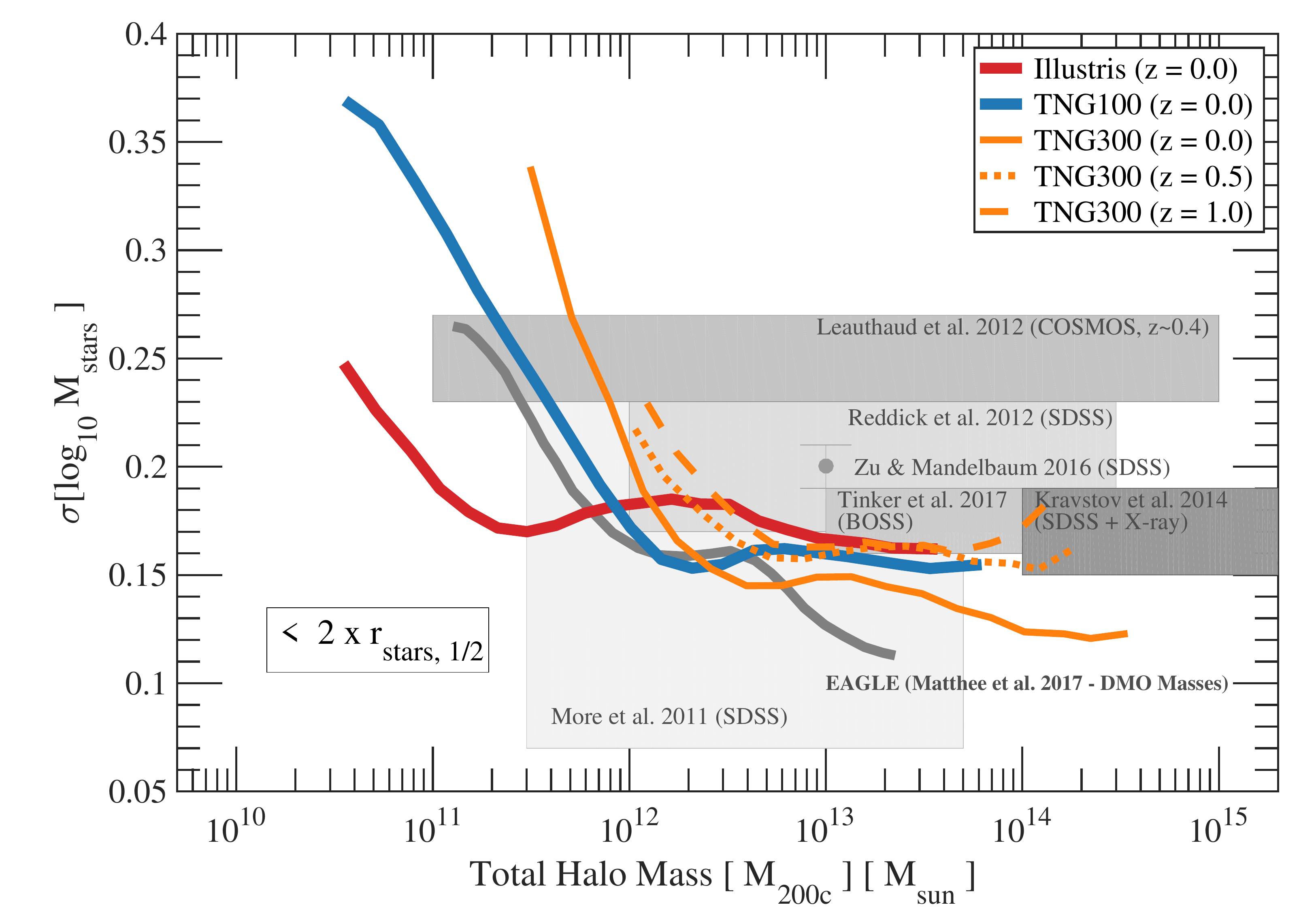}
\caption{\label{fig:sm2hm} The stellar-mass to halo-mass (SMHM) relation and its scatter. Top left: SMHM relation at $z=0$ for central galaxies, for different simulations (Illustris, TNG100, TNG300 or rTNG300 in red, blue, and orange thick curves, respectively) and for different definitions of a galaxy's stellar mass: within fixed spherical apertures of 30 kpc (solid thick), 10 kpc (dotted), 100 kpc (dashed), and twice the stellar half mass radius (crossed). Top right: the SMHM in TNG galaxies at $z=0$ within the central galaxy (solid thick), for the whole diffuse stellar mass (central + ICL, excluding satellites: dotted), for the whole stellar content within a halo (total stellar mass = central + ICL + satellites, out to the virial radius of the halo: dashed). Bottom left: redshift evolution of the SMHM relation (central galaxy mass) from $z \sim 1$ to the present day. Bottom right: 1-sigma scatter in the logarithmic galaxy stellar mass (within twice the half-mass radius) at fixed halo mass, as a function of halo mass, compared to EAGLE and observational estimates (see boxes). }
\end{figure*}

\subsection{The stellar mass -- halo mass relation}
\label{sec:budget_3}

Until now we have focused on the relative contributions of different stellar components to the total stellar budget within groups and clusters. We now turn to the relative contribution of luminous mass to the total halo mass, measured in terms of a baryonic conversion efficiency, $M_{\rm stars} / M_{\rm halo} \,(\Omega_b/\Omega_m)^{-1}$.

Although more massive haloes contain more stellar mass than their low mass counterparts, it has been long recognised that the baryonic conversion efficiency is maximal at the Milky Way mass scale \citep[e.g.][and references therein]{Conroy:2009,Leauthaud:2012}. Towards both higher and lower masses, dark matter haloes become progressively more inefficient at converting their baryons into stars. However, measurement of this efficiency as a function of halo mass is a difficult observational task, and different model and observational constraints have not yet reached a quantitative consensus.

In Fig.~\ref{fig:sm2hm} we therefore show several aspects of this stellar mass - halo mass (SMHM) relation for TNG haloes.
The most effective way to highlight subtle discrepancies across models, observations, and definitions is with the linear ratio (upper left panel). In such a view, we are interested in tensions on the order of ten per cent or less; differences that would be invisible in a logarithmic plot of stellar vs. total halo mass. Here we include the $z=0$ SMHM relations from Illustris, TNG100 and rTNG300 (red, blue and orange curves, respectively). Thick solid curves give the median relations, while the 10 most massive haloes are represented as individual points. For visual clarity, the 1-$\sigma$ scatter at fixed halo mass is shown with shaded areas for TNG100 only. Note that a version of the upper left panel Fig.~\ref{fig:sm2hm} has been already quantified in \citealt{Pillepich:2017} (e.g. their Fig. 4), for a selection of the apertures adopted here and using the outcome of smaller simulated volumes, hence possibly affected by sample variance. There the focus was a relative comparison between the Illustris and TNG model.

We contrast our TNG results to the outcome of abundance matching and other semi-empirical models \cite[][the latter two denoted as SHARC and EMERGE, respectively]{Behroozi:2013, Moster:2013, Kravtsov:2014, Rodriguez:2017, Moster:2017}. The comparison is favourable, as indeed the TNG model has been devised to reproduce the general features of the semi-empirical constraints on the SMHM relation \citep[see][]{Weinberger:2017, Pillepich:2017}. The TNG model produces a strongly peaked SMHM function, with galaxies residing in $\simeq 10^{12}\,\MSUN$ haloes being the most effective at forming stars, with a conversion efficiency of $\sim$ 20\%. This efficiency halves rapidly, reaching $\sim$ 10\% for $2 \times 10^{11} \MSUN$ haloes (towards lower masses) and by $8 \times 10^{12}\,\MSUN$ (towards higher masses). In comparison to Illustris, the TNG model results in a much more strongly peaked SMHM, albeit at a slightly lower peak value. The differential suppression of star formation to either side of the peak is stronger in TNG, steepening the slope of the SMHM. With a canonical galaxy stellar mass definition ($<$\,30 kpc; solid lines) both TNG realizations produce a simulated SMHM whose shape and amplitude as a function of halo mass is in good agreement with the various constraints, over the full resolved halo mass range, although with the location of the peak possibly up to a few fractions of a dex higher than the locus from semi-empirical models. 

Unfortunately, a detailed evaluation of the level of quantitative agreement proves more difficult. While it has become common to compare the results of semi-empirical models to the outcomes of hydrodynamical simulations as a fundamental benchmark, no quantitative SMHM relation can at present be directly compared between the two. In a semi-empirical model, the exact values of $M_{\rm stars} / M_{\rm halo}$ are a non-trivial convolution of the different observational datasets adopted for its calibration, each of which may invoke a different definition of galaxy stellar mass, possibly varying as a function of mass. Unless a single observational dataset has been used to calibrate a given model and the galaxy stellar mass definition from that dataset is made explicit, it is impossible to make a well-posed comparison with simulated stellar mass content which is, by construction, sensitive to definition.

We show this effect for rTNG300 galaxies in Fig.~\ref{fig:sm2hm} (upper left panel; orange curves) by accounting for the stellar mass within different spherical apertures: 30 kpc (solid), 10 kpc (dotted), 100 kpc (dashed), and twice the stellar half mass radius (crossed). The different choices not only modify the SMHM relation at the high mass end, but also the height of the peak itself. For example, the baryonic conversion efficiency of Milky Way-like galaxies can differ by 30\% if the stellar mass accounted for is within 10 or 100 kpc. For $10^{13}\,\MSUN$ haloes, the baryonic conversion efficiency can differ by more than a factor of 2 for different operational definitions of central galaxy stellar mass -- this discrepancy increases to a factor of a few at larger masses, modifying the overall slope of the SMHM relation at the high mass end.

A similar consideration holds when extended stellar mass outside the central galaxy is included (top right panel). For both TNG100 and rTNG300, we show how the stellar to total mass fraction increases when accounting for the central galaxy only (thick solid curves), all the diffuse stellar mass (central + ICL, dotted), or all the stellar mass within the virial radius (+ satellites, dashed curves). Surprisingly, the SMHM recast in terms of the total stellar mass of the halo is much shallower on the high-mass end of the peak, and decreases by only a factor of two between Milky Way mass haloes and the most massive clusters in the Universe. As we have already seen, the satellite contribution at the high mass end is significant and a clear signpost of hierarchical assembly.

Considering the amount of evolution of the stellar mass content at a given halo mass scale, we contrast the $z=0$, $z=0.5$, and $z=1$ SMHM relations of the two TNG simulations (lower left panel), showing the $<$\,30 kpc aperture only. Broadly speaking, the whole relation shifts to lower normalization across this redshift range, without any significant change in its shape. At $z=1$, the relation is slightly less peaked, implying that it is shallower to both sides of the peak, although this is a small effect. The location of the peak in $M_{\rm halo}$ remains remarkably constant -- the shift with redshift is within $<$\,50\% at most. This is in contrast to the SMHM of Illustris \citep{Genel:2014}, where the peak shifted more than half a dex towards higher masses from $z=0$ to $z=1$. The value of $M_{\rm stars} / M_{\rm halo}$ at the high-mass end is definitely larger at $z=0$ than at $z=1$ in TNG, qualitatively regardless of aperture, and the same is true also at the low-mass end. Finally, although we only show the evolution for the $<$\,30 kpc definition, we note that the trends at $z < 1$ are similar for the other aperture definitions considered in this paper.

Different semi-empirical models disagree on the redshift evolution of the SMHM in both the details and in the broad, overall behaviour, making any comprehensive comparison difficult. We note that \cite{Moster:2017} favour a \textit{higher} peak efficiency at $z=1$ as compared to $z=0$, as well as a larger efficiency at high masses for the central galaxy, and we differ in both conclusions. The dependence of the peak mass in TNG is similar to \cite{Behroozi:2013}, although they favor a near constant efficiency over this redshift range. In apparent agreement with that work is a largely unchanged slope, at the low-mass and high-mass ends, although their high-mass end amplitude is consistent with being constant or even slightly decreasing from $z=1$ to $z=0$. With respect to \cite{Moster:2013}, TNG finds qualitatively different behavior in the degree of shift in the peak mass, and the change of both slopes across this redshift range, agreeing only on the \textit{sign} of the overall normalization increase. 

As a rather orthogonal constraint on the stellar assembly histories of dark matter haloes, we consider the amount of allowed variation in the stellar mass content at a given halo mass scale. The final panel of Figure \ref{fig:sm2hm} shows the 1-$\sigma$ scatter in the logarithmic galaxy stellar mass at fixed halo mass (lower right panel). In addition to TNG100 (blue) and the redshift evolution of TNG300 (orange), we have also included original Illustris (red) results and the finding from EAGLE \citep[][grey]{Matthee:2017}. Combined with the results of N-body simulations, several observational estimates of the scatter have been made \citep{More:2011,Leauthaud:2012,Reddick:2013,Zu:2016,Tinker:2017}. Halo abundance or subhalo abundance matching models typically place the value of the intrinsic scatter in the range 0.1 to 0.26 between $z\sim1$ and today, in broad agreement with direct observational constraints that are available at the high-mass end \citep{Kravtsov:2014}. Hydrodynamical simulations (Illustris, TNG, as well as EAGLE) commonly find $\sigma = 0.15 \pm 0.05$, for haloes above a few $10^{12}\MSUN$, with no dependence on resolution. Our estimates are based on stellar masses measured within twice the stellar half mass radius, but do vary somewhat with the other apertures studied thus far. They can be compared to the 1-sigma logarithmic scatters reported in Table~\ref{tab:massfits} for the different clusters components, all in the $\sigma = 0.15 \pm 0.05$ range, with the total (satellites) stellar mass exhibiting the smallest (larger) halo-to-halo variations.

The redshift evolution of the scatter is mild (Figure \ref{fig:sm2hm}, lower right panel, solid vs dotted vs dashed curves), less than 0.05 dex from $z=1$ to $z=0$. Towards lower mass, on the other hand ($M_{\rm halo} \la 10^{12}\,\MSUN$), the simulations predict a sharp increase in logarithmic scatter, with a non-negligible dependence on physical model. This regime is now becoming accessible with deep lensing surveys \citep[e.g. in the HSC-SSP;][]{Harikane:2016} and motivates an in-depth study of the SMHM scatter and its redshift evolution in the near future.


\section{Stellar Mass Assembly and Stellar Mass Function}
\label{sec:theory}

So far we have seen that for $\MTC \ga 10^{13}\MSUN$, in the group and cluster regime, halo mass is a good predictor of stellar mass measured within any fixed spherical aperture. We can use the simulations to assess how the stellar mass in massive haloes is assembled across time, and across cluster components. We therefore measure the importance of the accreted `ex-situ' stellar component, determine its origin through the progenitor satellite mass spectrum, and connect the ICL of the most massive central galaxies to the numerous, low-mass end of the galaxy stellar mass function.

\subsection{In-situ vs ex-situ material}

\begin{figure}
\centering
\includegraphics[width=8cm]{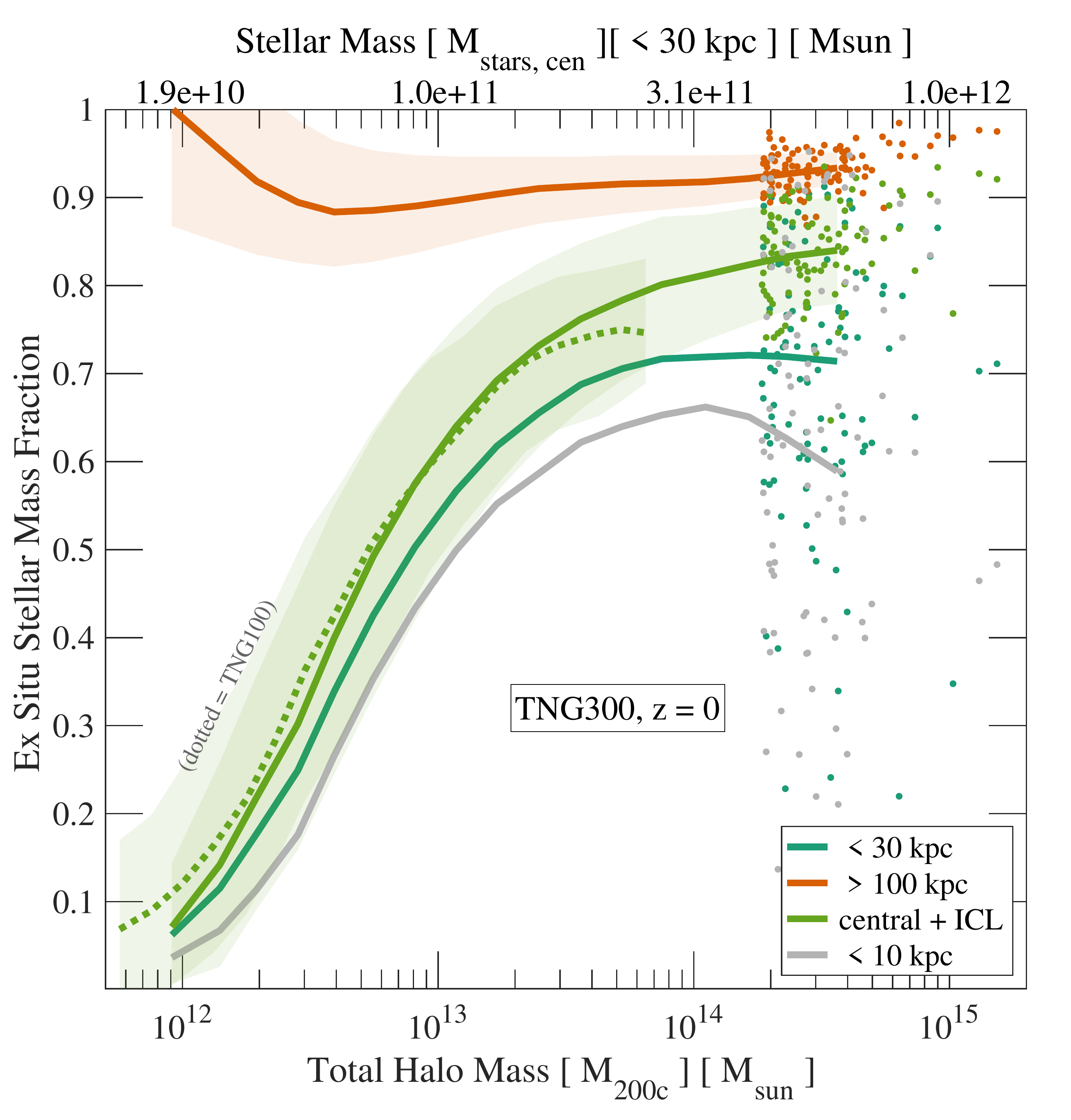}
\caption{\label{fig:exsitu} The fraction of ex-situ (accreted) stellar mass to the total (diffuse) stellar mass (i.e. in-situ+ex-situ) within different apertures, as a function of halo mass at $z=0$. We consider four apertures which examine different regions of the halo: the innermost region of groups and clusters, i.e. within the massive central galaxy ($<$ 10 kpc, in grey, and $<$ 30 kpc, in dark green); the outskirts, i.e. the outer stellar mass excluding satellites ($>$ 100 kpc, in orange); and the stellar diffuse component of the whole halo ($\MSTC$ = central + ICL, excluding satellites, in light green). Solid lines show TNG300 (bands and points are the same as in previous figures), while the dotted line from TNG100 demonstrates that the ex-situ fractions are reasonably well converged.}
\end{figure}

Cosmological simulations have demonstrated that the most massive galaxies in the Universe grow both in mass and sizes by the merging with smaller units. More quantitatively, the fractional contribution of {\it accreted} stars to the total stellar mass of a galaxy is a steep function of stellar or halo mass \citep{Oser:2010,Lackner:2012,Rodriguez-Gomez:2016}. Using the original Illustris simulation, in \cite{Rodriguez-Gomez:2016} we measured the ex-situ fraction with respect to the total stellar mass of the halo (central + ICL, excluding satellites, effectively), which rises from less than 10\% to more than 50\% for galaxies in the $10^9$ to $10^{11.5} \MSUN$ stellar mass range. With the TNG simulations, we extend this analysis to the most massive objects, adopting the in-situ vs ex-situ classification of \cite{Pillepich:2014bb} and \cite{Rodriguez-Gomez:2016}, and by using the baryonic merger trees of \cite{Rodriguez-Gomez:2015}. As a reminder, an individual stellar particle is classified as ex-situ if it forms outside of the main progenitor branch of its $z=0$ host; otherwise, a star is in-situ.

In Figure \ref{fig:exsitu} we measure the ex-situ stellar mass fraction as a function of halo mass for TNG300 at $z=0$. Four results for four different operational definitions of the stellar component are given: within 10 kpc (grey), within 30 kpc (dark green), the total central galaxy + ICL, excluding satellites (light green), and stars outside 100 kpc only, excluding satellites (orange). The first three cases have similar behavior: a steep rise from $\la$ 10\% at $M_{\rm halo} < 10^{12} \MSUN$ through an equal 50\% contribution at $M_{\rm halo} \simeq 10^{13} \MSUN$ to asymptotic values between 60\% and 90\% for $M_{\rm halo} > 10^{14} \MSUN$, depending on definition, independent of resolution, and with about 1-sigma 10\% galaxy-to-galaxy variation. In practice, we demonstrate that, while low-mass galaxies form almost all of their stars in-situ, the central galaxies of the most massive haloes ($\MTC\ga10^{14}\MSUN$) acquire more than 80\% of their mass through cosmological assembly and mergers. Remarkably, even within the innermost regions of the most massive galaxies (e.g. within just 10 kpc), the contribution of ex-situ stars is as a high as 60\%.

The behavior of the stellar component at large physical radii, considering stars beyond 100 kpc only (orange curve), is markedly different. Regardless of the halo mass scale, stellar mass at these distances originates almost exclusively from accreted material, with a $\ga$ 90\% ex-situ fraction. This is because star formation is not expected to occur at such large radii, and other mechanisms need to be invoked to move in-situ stars to large distances from their original birth sites. The ex-situ contribution also exceeds 50\% for a smaller physical inner boundary of $>$ 30 kpc (not shown). This bounds the contribution from in-situ stars to the extended stellar halo. In particular, relatively few stars can be stripped from a central galaxy and ejected to large radii as a result of e.g. dynamical interactions. The properties of the extended stellar component, beyond a few tens of kpc from the centers, are therefore a largely pristine record of the baryonic assembly history of a given dark matter halo.

\subsection{The mass spectrum of accreted satellites}

\begin{figure}
\centering
\includegraphics[width=7.8cm]{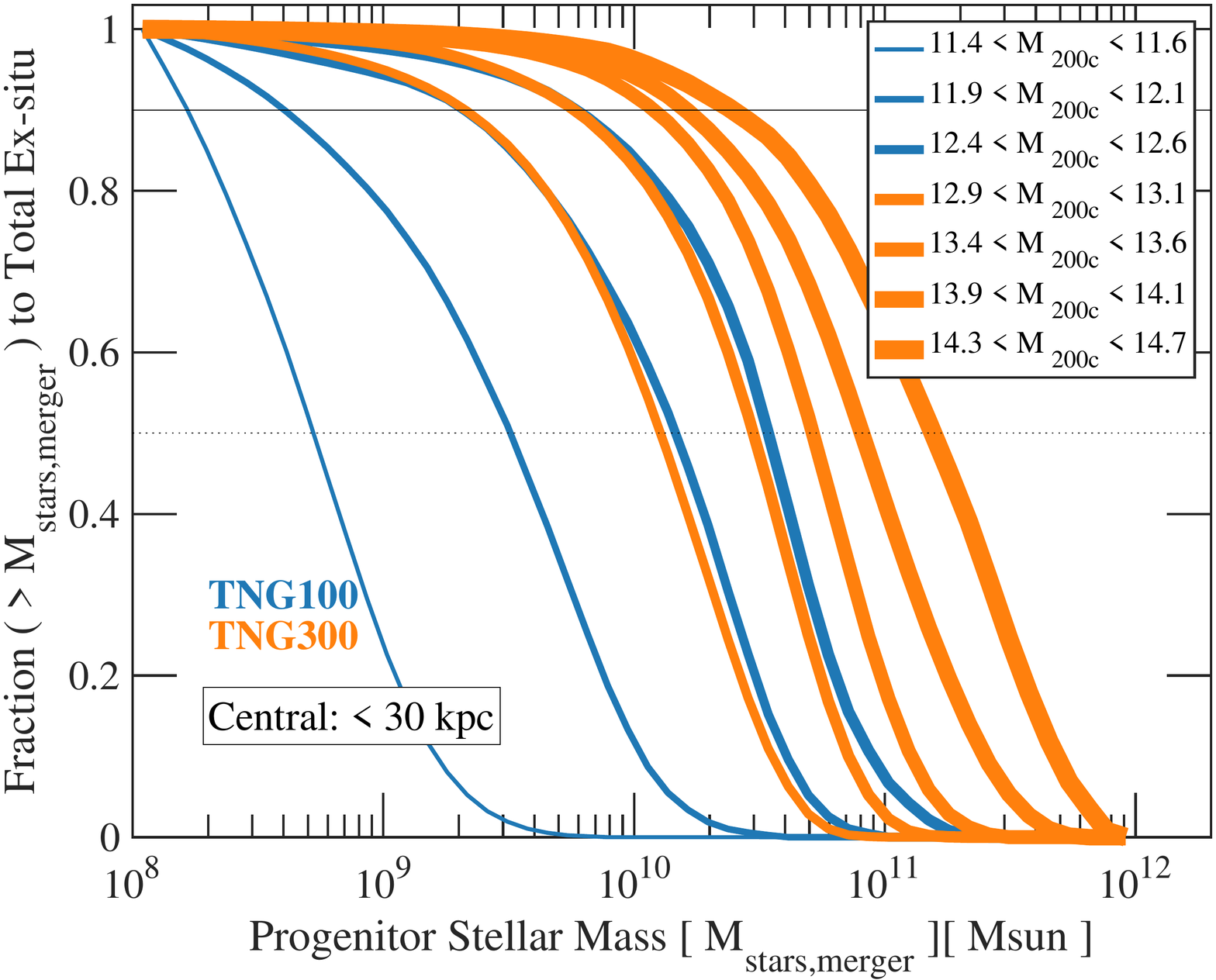}
\includegraphics[width=7.78cm]{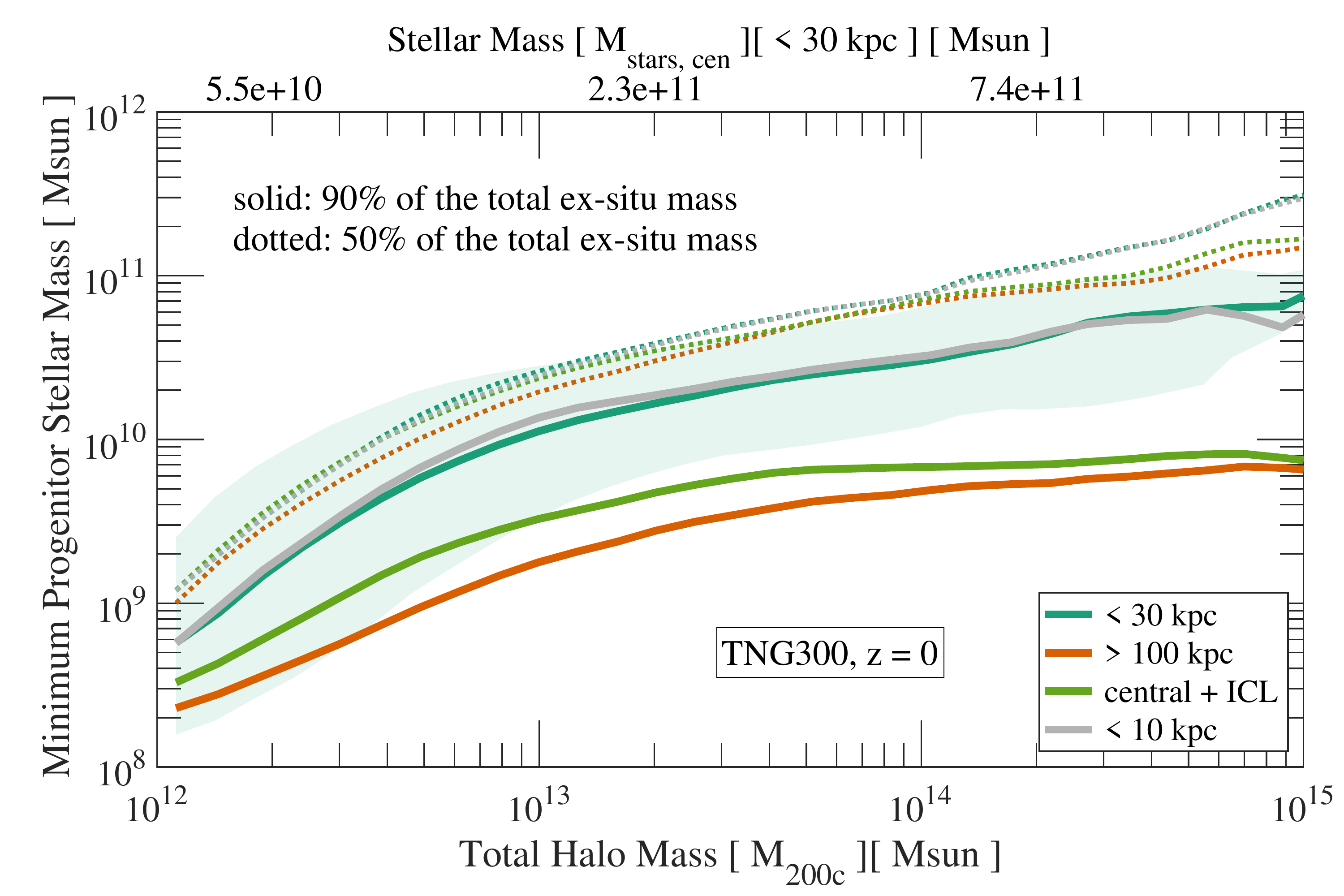}
\caption{\label{fig:progmass} The relative contribution from accreted satellites of different masses to the ex-situ stellar component of massive central galaxies at $z=0$. In the top panel we stack haloes in seven mass bins, indicated in the legend. For all stars within the centrals ($<$ 30 kpc), we show the cumulative fraction of the total ex-situ mass contributed by satellites of mass $M_{\rm stars,merger}$ or greater. Horizontal lines indicate threshold contributions of 50\% and 90\%. In the bottom panel we plot these thresholds as a function of halo mass, such that satellites of this mass and higher contribute half (dotted curves) or 90 per cent (solid curves) of the total ex-situ stellar mass. Here we consider the same four apertures as in Figure \ref{fig:exsitu}, showing the degree to which accreted satellites of different masses contribute to the various stellar components of the halo.}
\end{figure}

Not surprisingly, given that a significant fraction of the stars of galaxies above the Milky Way mass scale are ex-situ in origin, the bulk of such material for very massive galaxies has been brought in by rather massive galaxies. We demonstrate this in Figure \ref{fig:progmass}.

Figure \ref{fig:progmass} deconstructs the accreted (diffuse) stellar component into the spectrum of progenitor satellites from which it arose.
The top panel shows the relative contribution of ex-situ stars from infalling satellites of different stellar masses.\footnote{As initiated and motivated  by \cite{Rodriguez-Gomez:2015}, we follow the common practice in merger analyses to take, for the mass of a merging satellite, its maximum mass at any time along its main progenitor branch, as opposed to its mass at infall or its mass at the time of the merger, the latter of which are sensitive to both physical and numerical issues.} In particular, we show the fraction of the ex-situ stars within 30 kpc contributed by progenitors \textit{above} a given threshold in stellar mass, stacking the results for seven different halo mass bins (colored lines) spanning $ 10^{11.5} \MSUN < M_{\rm halo} < 10^{14.5} \MSUN$. The intersection of each curve with the horizontal 50\% ex-situ line indicates, for example, that satellites of this mass and greater contribute half of the total ex-situ stellar mass found within the galaxy. This threshold mass increases with increasing halo mass -- the centrals of more massive haloes are built up from larger satellites, with correspondingly older and more metal-rich stellar components \citep[e.g.][]{Gallazzi:2005, Bernardi:2010}. For $M_{\rm halo} > 10^{13} \MSUN$, the satellites which contribute the dominant part of the ex-situ stars ($<$ 30 kpc) exceed stellar masses of $10^{10.5} \MSUN$. This is an important threshold mass in the TNG model above which central galaxies are predominantly quiescent, gas-poor, often spheroidal-morphology systems, and so likely to distribute a different physical mix of stars differently throughout the halo of their descendant host.

In the lower panel we measure two particular satellite threshold masses, those contributing 50\% and 90\% of the total ex-situ M$_{\rm stars}$, as a function of halo mass and as a function of four different aperture definitions. These are the same as in Figure \ref{fig:exsitu}, probing within the massive central galaxy ($<$ 10 kpc, in grey, and $<$ 30 kpc, in dark green) as well as the outskirts alone ($>$ 100 kpc, in orange) and the stellar component of the whole halo (central + ICL, excluding satellites, in light green). We find that the progenitor stellar mass needed to account for half of the accreted stars is nearly invariant to aperture, i.e. the bulk of the stellar component of a host, in any given radial extent, is contributed by a similar mass spectrum of satellites. This minimum contributing mass increases sharply from $M_{\rm merger} \simeq 10^9 \,\MSUN$ for host haloes of total mass $10^{12} \,\MSUN$, to $M_{\rm merger} \simeq 10^{10.5} \,\MSUN$ for $10^{13} \,\MSUN$ haloes. It then plateaus and rises to $M_{\rm merger} \simeq 10^{11.2} \,\MSUN$, such that half of the extended stellar components of the most massive clusters is contributed by accreted satellites with stellar mass at least as massive as a few $10^{11}\MSUN$. For Milky Way haloes, this corresponds to LMC-like satellites, while for $10^{15} \MSUN$ clusters these are mergers of already large ($\sim 5 \times$ MW mass) elliptical satellites.

If we consider instead the 90\% threshold, i.e. down to which satellite stellar mass must we go to recover essentially all of the stellar component of a host halo, the situation differs. Namely, a strong radial dependence becomes apparent. For the central galaxy ($<$ 10, or 30, kpc) the minimum contributing satellite mass $M_{\rm merger}$ is roughly 0.5 dex less massive than the 50\% threshold value, with the same dependence on host halo mass. However for the outskirts (central + ICL, and even more so, $>$ 100 kpc), the minimum satellite mass is significantly lower, and becomes halo mass independent at a nearly constant value of $\simeq 10^{9.8} \MSUN$. Therefore, accreted galaxies with stellar masses above this value contribute 90\% of the extended stellar mass of all groups and clusters. 

In conclusion, within the innermost 30 kpc of galaxies in $10^{15}\MSUN$ haloes (about $10^{12}\MSUN$ in stars), 70 per cent of the stars have been assembled by merging and 90 per cent of those stars have been brought by galaxies at least as massive as $4-6\times 10^{10} \MSUN$.

\subsection{The TNG stellar mass function from z=0 to z=4}
\label{sec:gsmf}

\begin{figure*}
\centering
\includegraphics[width=8.8cm]{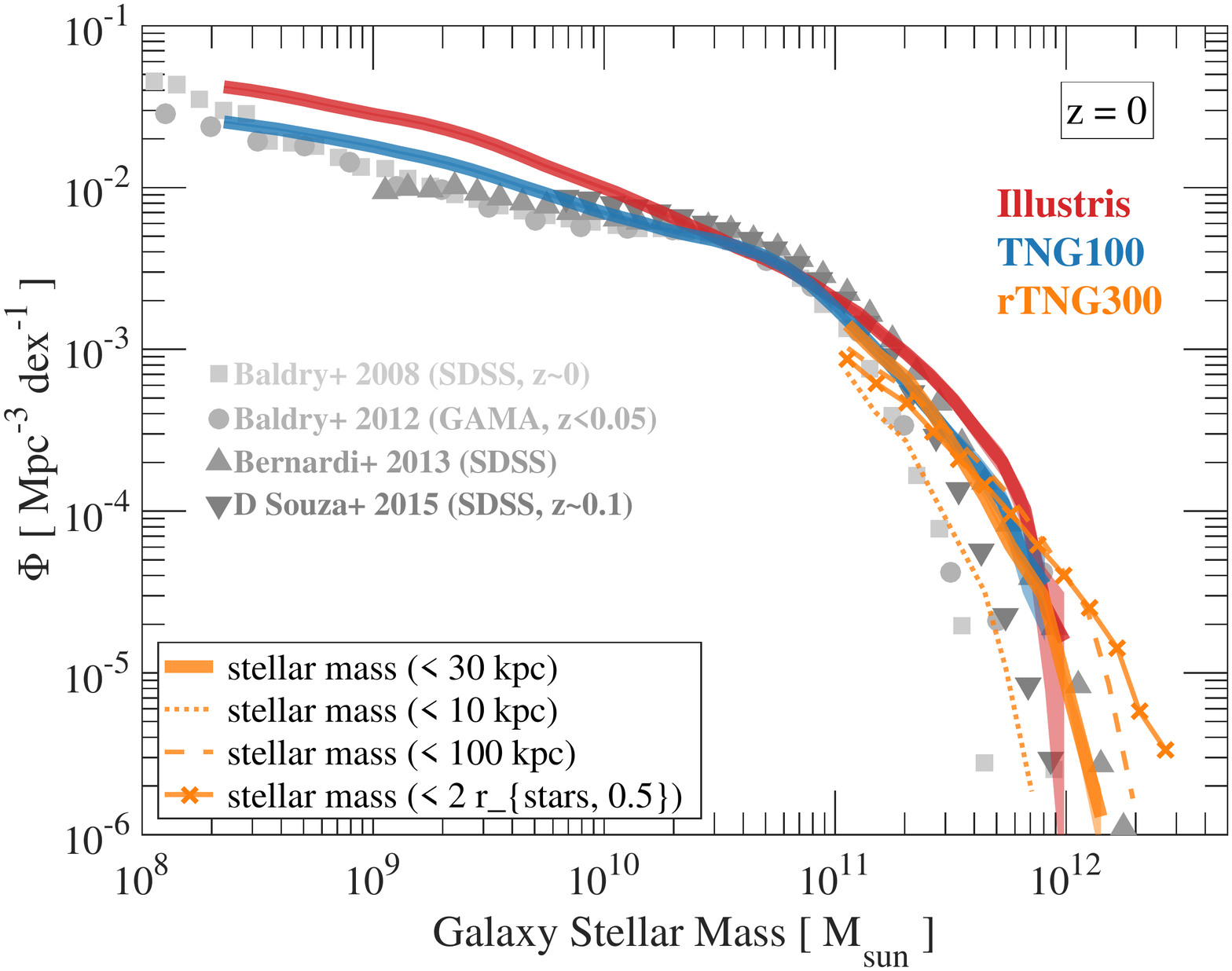}
\includegraphics[width=8.8cm]{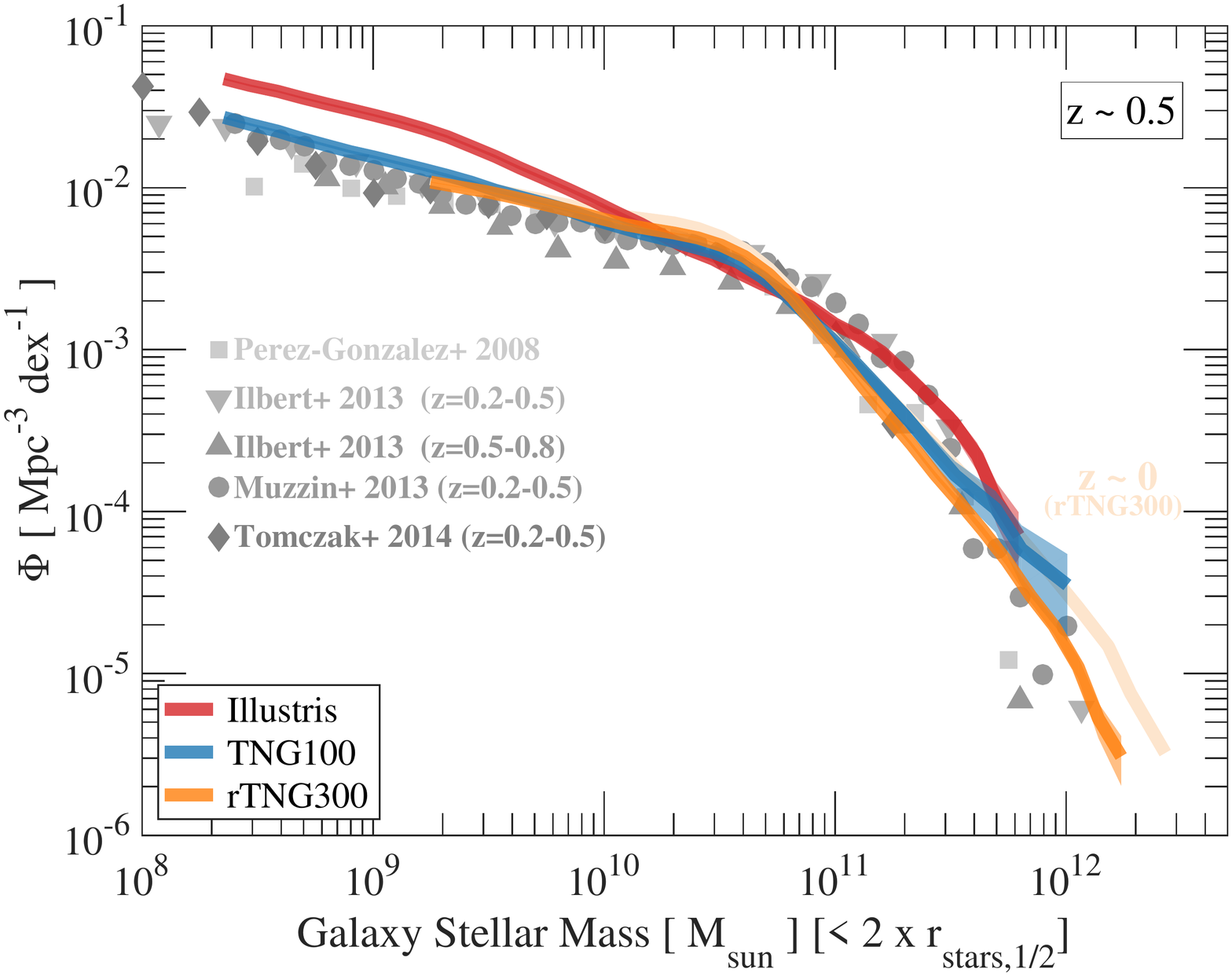}
\includegraphics[width=8.8cm]{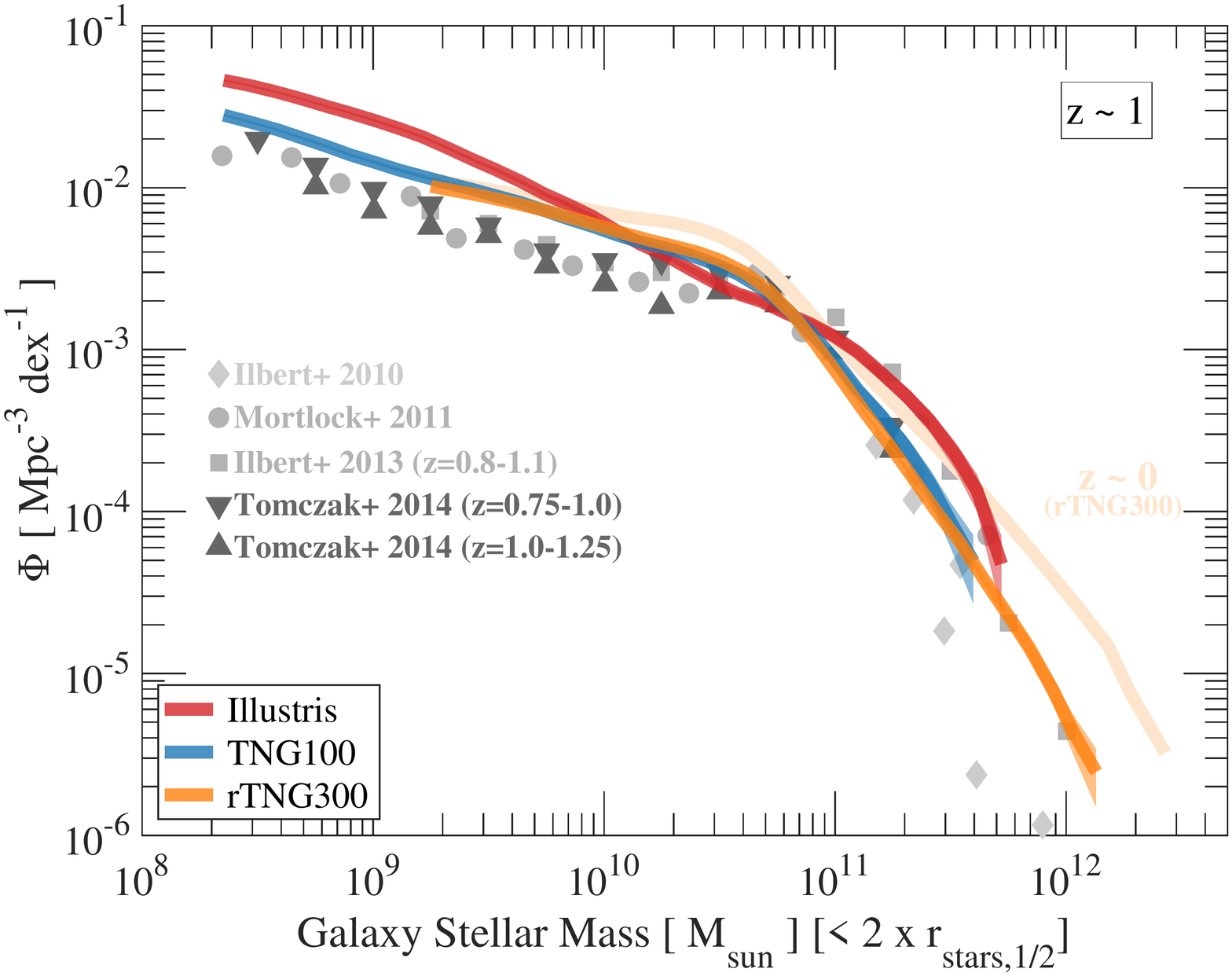}
\includegraphics[width=8.8cm]{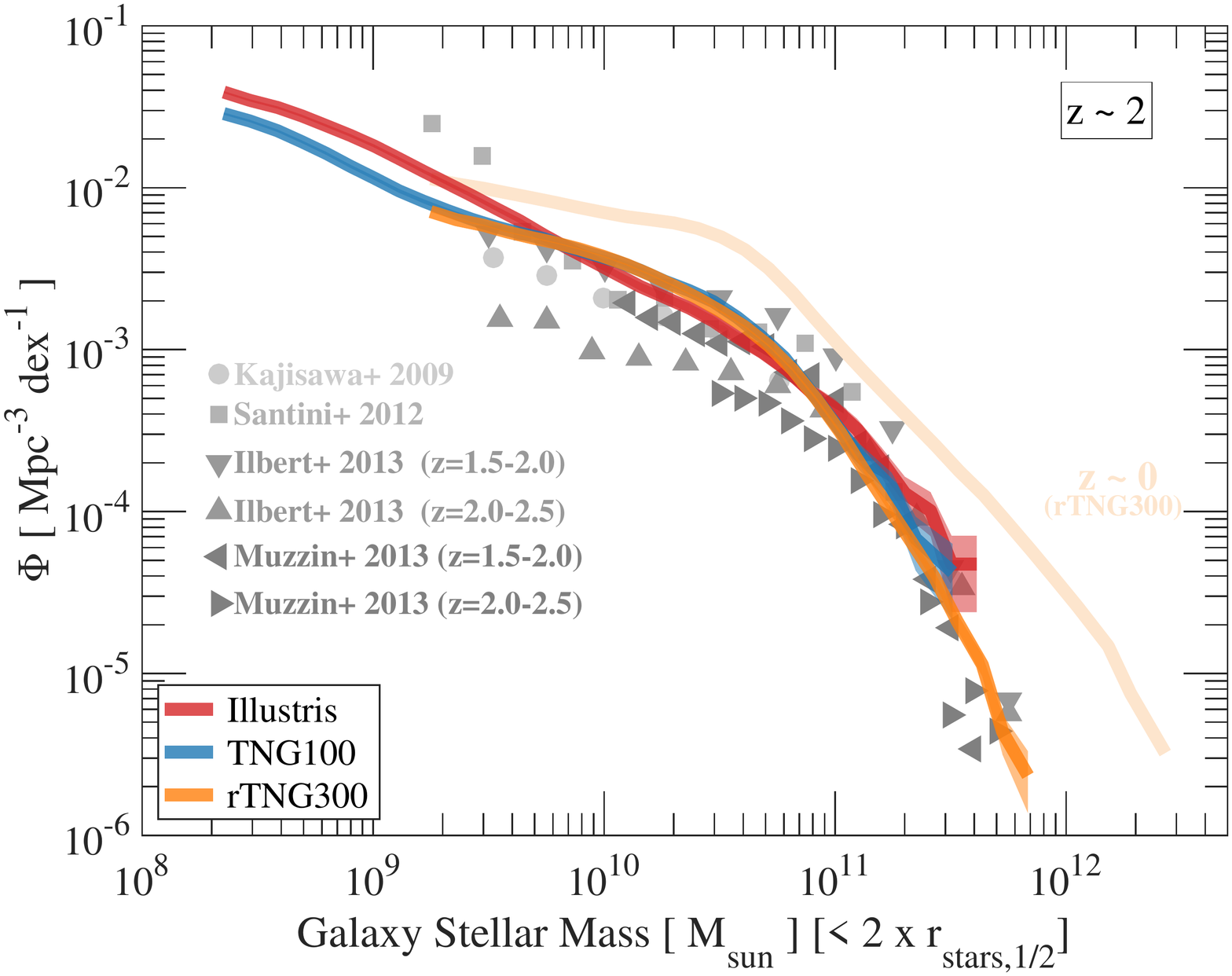}
\includegraphics[width=8.8cm]{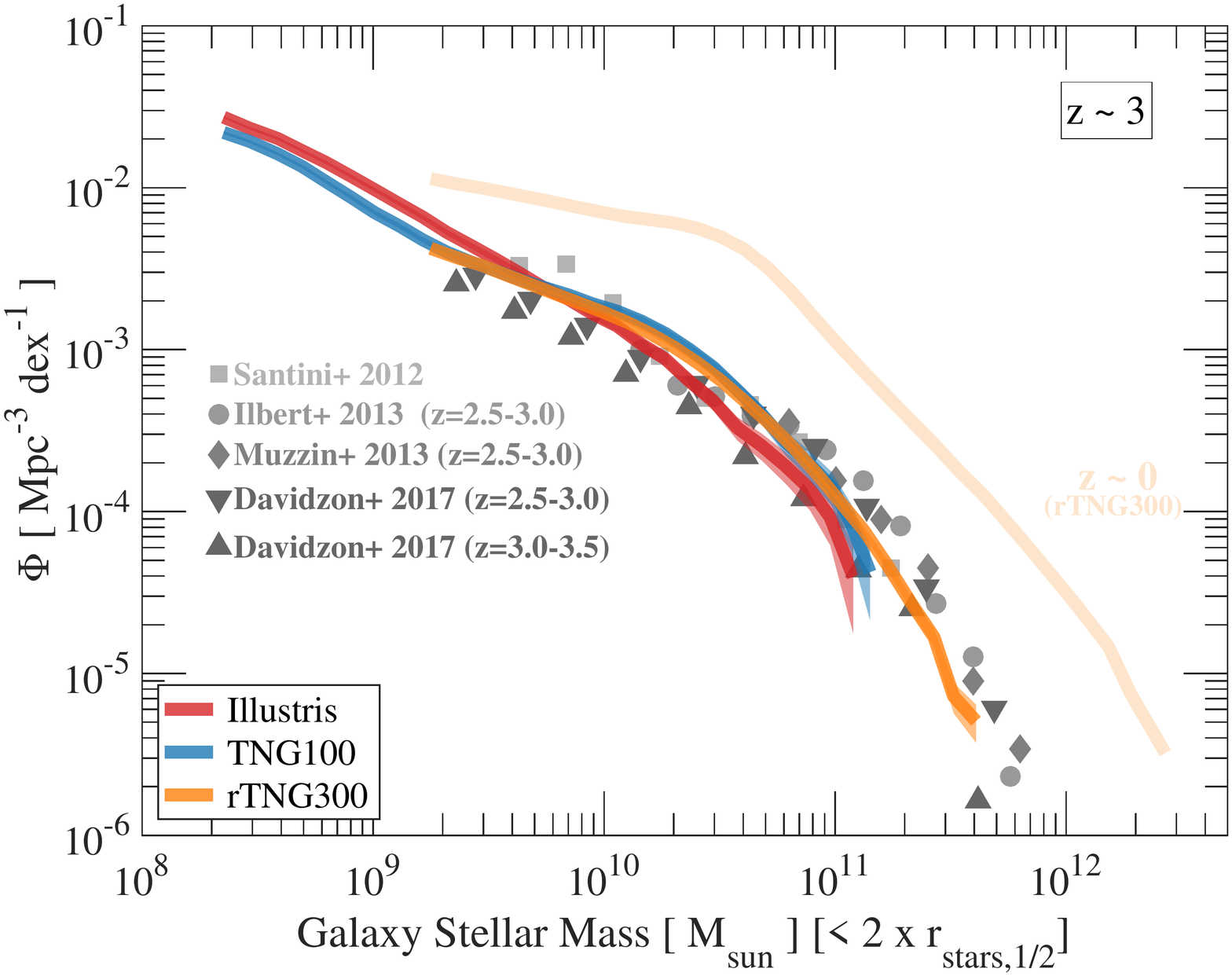}
\includegraphics[width=8.8cm]{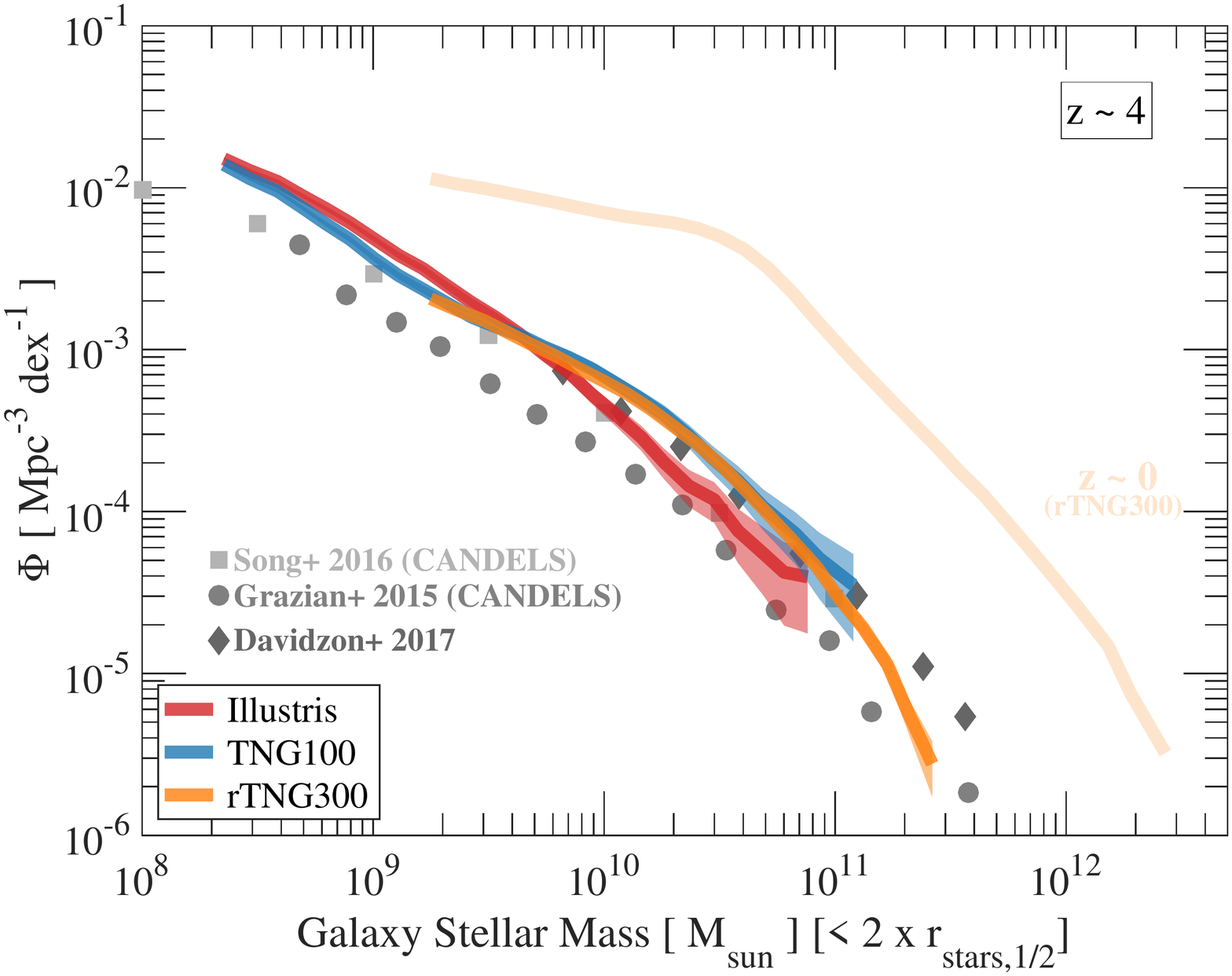}
\caption{\label{fig:mf} The TNG galaxy stellar mass functions after the Epoch of the Reionization, from $z\sim4$ to today. Unless otherwise specified, we show results from the simulations by accounting for all the stellar mass within twice the stellar half mass radius (thick colored curves from $z=0.5$ to 4). At $z=0$, we emphasize the importance of the galaxy mass definition by providing the predictions from TNG300 for different aperture measurements: 30 kpc (for all runs, thick curves), 10kpc (orange dotted), 100 kpc (orange dashed) and twice the stellar half mass radius (orange crosses). At $z>0$, we report in light orange the rTNG300 mass function (within twice the stellar half mass radius), for reference.
A selection of observational data points is included for comparison in grey symbols, all converted to Chabrier IMF \citep{Baldry:2008,Baldry:2012,Bernardi:2013,DSouza:2015,PerezGonzalez:2008,Mortlock:2011,Ilbert:2013,Muzzin:2013,Tomczak:2014,Kajisawa:2009,Santini:2012,Davidzon:2017,Grazian:2015}. 
}
\end{figure*}

The contributions of smaller merging companions only make physical sense if the properties of these satellite galaxies are realistically recovered in the simulations -- not only at $z=0$, but also backwards in time, as the stellar envelopes of galaxy groups and clusters assemble. We hence conclude the paper by demonstrating in Fig.~\ref{fig:mf} that the TNG model returns plausible galaxy populations across the whole mass spectrum and across redshift, and therefore that the non-trivial predictions of our numerical model presented throughout are dependable. 

Here we show the galaxy stellar mass function (GSMF) from the TNG simulations at six discrete redshifts from the present day to $z=4$, comparing to several observational datasets. We include TNG100, rTNG300, and Illustris (blue, orange, and red, respectively). Overall, the TNG simulations satisfy the available observational constraints, with the abundance of galaxies at the knee of the GSMF growing by more than an order of magnitude between $z=4$ and today.
We can also demonstrate here, across a larger mass range and with better statistical sampling than what was achieved in \cite{Pillepich:2017}, that with respect to Illustris, the amplitude of the low-mass end ( $\la 10^{11}\MSUN$) is significantly suppressed, particularly at $z \le 1$, resulting in much improved agreement. As discussed in \cite{Pillepich:2017}, this is mostly due to the changes in the implementation of the stellar-driven winds. At the higher mass end, the modifications to the black hole feedback (identically implemented in TNG100 and rTNG300) suppress, in comparison to Illustris, the GSMF at recent times in the critical mass regime above a few $10^{10}\MSUN$, with a more pronounced knee and a sharply falling abundance at stellar masses above a few times $10^{11}\MSUN$.

In the redshift zero comparison (upper left panel) we elaborate on the high-mass end shape of the galaxy abundance by implementing four distinct measurements  from the simulation perspective. Namely, we show the resulting curves for four different definitions of the galaxy stellar mass: $<$ 30 kpc (solid), $<$ 10 kpc (dotted), $<$ 100 kpc (dashed), and $<$ twice the stellar half mass radius (crossed). The value of $\Phi$ at $M_{\rm stars} > 10^{10.5} \MSUN$ sensitively depends on definition, with differences rising to more than one order of magnitude at stellar masses of $10^{12}\MSUN$. Our findings on the large effect of aperture at the high mass end confirm previous studies, both in the context of observations \citep[e.g.][]{DSouza:2015} and simulations \citep[e.g.][]{Schaye:2015}.
 In fact, in these four choices we have replicated in precise detail none of the observational definitions used in \citet{Baldry:2012,Bernardi:2013,DSouza:2015}. Therefore we point out only that, depending on definition, the high-mass end of the simulated galaxy stellar mass function spans the range of $z=0$ observational measurements without difficulty.

The situation at high redshift is much less constraining, with all three simulations showing similar levels of agreement with the observed $z \ge 2$ GSMF, despite the rather different mechanisms and energetics of the most important feedback processes. Certainly, a more quantitative comparison to the individual observational datasets would require more sophisticated and specifically tailored mock stellar mass modelings of the simulated galaxies.  Yet, it is reassuring to witness the fundamental progress this represents in such models, which had basic difficulties at recovering the zeroth-order abundances of galaxies just a few years ago \citep[e.g.][]{Keres:2009}.

As the main takeaway of this GSMF comparison, we conclude that the TNG galaxy formation model results in a realistic distribution of stellar masses across the bulk of cosmic time. Therefore, as these galaxies merge and accrete onto forming groups and clusters at $z < 1$, we have confidence that the deposition of stars into both the innermost and extended stellar components of the most massive objects in the Universe is well captured in our simulations.


\section{Summary and Discussion}
\label{sec:summary}

This study is one of five papers presenting the new IllustrisTNG ({\it The Next Generation}) project, a follow-up of the Illustris simulation. TNG is an ambitious set of cosmological magneto-hydrodynamical simulations aimed at studying the formation and evolution of galaxies across an unprecedented range of masses and environments. They are performed with the moving-mesh code \textsc{Arepo} and the underlying galaxy physics model includes prescriptions for star formation, stellar evolution, chemical pollution, primordial and metal-line cooling of the gas, galactic winds, and black hole formation, growth and feedback, the latter featuring a novel black-hole driven wind which acts at low accretion rates.

In this paper, we have given an overview of the stellar mass content at the massive end of IllustrisTNG galaxies at recent times ($z\leq 1$). We have focused on massive galaxy groups and clusters ($10^{13} \leq \MTC/\MSUN \leq 10^{15}$) and provided a census of the stellar mass content in their central galaxies, diffuse intra cluster light, satellite populations, and total stellar mass out to the virial radius. Our simulations naturally form large groups and clusters of galaxies (Figures \ref{fig:stamps1} and \ref{fig:stamps2}), with one (or more) massive galaxies dominating the most luminous regions, surrounded by a significant number of lower-mass satellite galaxies (Figure \ref{fig:stats}, lower panel), all amid an extended diffuse halo of stellar material. 

Here we have used the first two simulations of the IllustrisTNG series, TNG100 and TNG300 (Table \ref{tab:sims}). The large volumes encompassed by these runs (about 100 and 300 Mpc on a side, respectively) in combination with 2$\times$1820$^3$ and 2$\times$2500$^3$ resolution elements each, provide us with an unprecedented statistical power over a large mass range. We resolve tens of thousands of galaxies with stellar masses as small as $10^{8.5}-10^9 \MSUN$ as well as $\sim$ 300 central galaxies residing in galaxy clusters more massive than $10^{14}\MSUN$ and featuring stellar masses $\ga 5\times 10^{11}\MSUN$ (see Fig.~\ref{fig:stats}).

This is the first time that a theoretical, population-wide study of the stellar mass distribution in massive galaxies is accessible in a statistically-significant sense via numerical simulations that are also capable of resolving the structural details of smaller satellite galaxies. 
%

Throughout, we have characterized the halo to halo variation in all derived quantities (e.g. Figure \ref{fig:sm2hm}, bottom left, and Table \ref{tab:massfits}) and we have devoted particular attention to the definition of a galaxy's stellar mass: we have quantified the effects that stellar mass measurements taken within e.g. different 3D spherical apertures (see Table \ref{tab:defs}) can impart in the shape of the stellar-to-halo mass relation (Figure \ref{fig:sm2hm}, top panels) and chiefly the high mass end of the galaxy stellar mass function (Section \ref{sec:gsmf} and Figure \ref{fig:mf}).\\

\noindent Highlights of our quantitative findings are summarized as follows.
\begin{itemize}

\item We have quantified the 3D total stellar mass profiles of massive galaxies (diffuse mass, excluding satellites), both by stacking the average distribution of stellar mass in bins of halo mass and by inspecting individual profiles. We find that galaxies residing in less massive haloes have significantly more {\it centrally concentrated} stellar mass distributions when compared to more massive galaxies, even when the radial apertures are renormalized to the virial radius of the host haloes. By this we do not mean that they have larger Sersic indexes, but rather that, while more than 90\% of the total stellar mass of a $10^{12}\,\MSUN$ halo is within 10\% of its virial radius, there is at least another 30\% of stellar mass beyond 10\% of the virial radius of a $10^{14}\,\MSUN$ group (Figure \ref{fig:profiles_prototypes}). \\

\item We find that the 3D stellar enclosed-mass radial profiles beyond about 1 kpc from the centers (Figures \ref{fig:profiles_prototypes} and \ref{fig:stellarprofiles}, top left), normalized by both the virial radius and the total mass, exhibit the same functional shape for all halo masses between $10^{12}$ and $10^{15}\MSUN$. We capture this with a two-parameter sigmoid function (or smooth step function, Eq.~(\ref{eq:logistic})), whose best-fit parameters reveal remarkably clear trends with halo mass: more massive galaxies have shallower stellar profiles (smaller steepness) and larger midpoint pivot parameters than galaxies residing in less massive haloes. The latter parameter is de-facto the 3D stellar half mass radius of the galaxy when its stellar mass is measured out to the farthest extent of its underlying DM halo (Figure \ref{fig:stellarprofilesfits_1}): for halo (stellar) masses $\ga 10^{12} \MSUN$ ($\ga 5\times 10^{10}\MSUN$), the 3D stellar half mass radius of TNG galaxies lies on a tight power-law relation with halo mass (or halo virial radius), with $r^{\rm stars}_{\rm 0.5}\propto (\MTC)^{0.41-0.53}$ and with a $\sim 0.16$ dex scatter. \\

\item The stellar mass density profile at large distances (tens of kpc away from the centers) can also be well approximated by a simple functional form across halo masses: a power law function of clustercentric distance. Confirming the results of \cite{Pillepich:2014bb} and expanding them to the most massive clusters in the Universe, we find that the extended stellar envelopes (`stellar haloes', `intra-cluster light' or `intra-halo light') that surround the most massive galaxies can be almost as shallow as the underlying dark-matter, with slopes in the range $-3.5 \la \alpha_{\rm 3D} \la -3$ for $10^{15}\MSUN$ haloes. These can be compared to stellar mass density slopes of $\alpha_{\rm 3D} \sim -5$ for the diffuse mass in the outskirts of $10^{12}\MSUN$ haloes (Section \ref{sec:profiles} and Figure \ref{fig:stellarprofilesfits_2}). \\

\item In the studied mass range $10^{13}\MSUN \leq \MFC \leq 10^{15}\MSUN$ and for $z\la 1$, the stellar mass content in different cluster components (central galaxy, outskirts/ICL, satellites, and total) scales as a power law of the total halo mass, independently of the exact operational definition (Figure \ref{fig:stellarmasses}). However, the exact shape and scatter of the scaling relations with halo mass do depend on the component definitions. We find that the stellar mass of TNG central galaxies measured within 3D 30 kpc (100 kpc) scales as $(\MFC)^{\rm 0.49 ~(0.59)}$ and that the total stellar mass (central + ICL + satellites) can be as steep as  $\propto (\MFC)^{0.84}$, with even steeper scaling relations for the stellar mass in the ICL ($>$ 30 or 100 kpc) as well as the mass in satellite galaxies (Figure \ref{fig:stellarmasses} and Table \ref{tab:massfits}).\\

\item In relative terms, for $10^{15}\MSUN$ haloes, up to 90 (80) per cent of their stellar mass is found beyond clustercentric distances of 30 (100) kpc, both in diffuse intra-halo light and satellite galaxies (Figure \ref{fig:fractions}, top panel). As the number of satellite galaxies is a steep function of halo mass (Figure \ref{fig:stats}), haloes more massive than about $5\times 10^{14}\MSUN$ have more mass in satellite galaxies than in diffuse material (central galaxy or ICL; figure \ref{fig:fractions}, bottom panel). In fact, the amount of ICL relative to the total diffuse mass grows with halo/galaxy mass (Figure \ref{fig:fractions}, third panel from the top): haloes more massive than about $5\times 10^{14}\MSUN$ have more diffuse stellar mass outside 100 kpc than within 100 kpc.\\

\item While it has been long recognized that massive haloes are less efficient than MW-like galaxies at making stars, the exact functional form and magnitude of the stellar mass to halo mass relation strongly depends on the definition of a central galaxy's stellar mass (Figure \ref{fig:sm2hm}). For $10^{13}\MSUN$  haloes, the baryonic conversion efficiency can differ by more than a factor of 2 for different operational definitions of central galaxy stellar mass: this discrepancy increases to a factor of a few at larger masses, modifying the overall slope of the SMHM relation at the high mass end. If we consider the total stellar mass (central+ICL+satellites), the stellar-to-halo mass fraction decreases by only a factor of two between Milky Way mass haloes and the most massive clusters in the Universe. \\

\item The halo-to-halo variation (intrinsic scatter) in all the aforementioned relations is remarkably small for all halo masses larger than $10^{13}\MSUN$. According to the TNG simulations, the scatter in stellar masses at fixed halo mass is $0.15\pm0.03$ dex for haloes larger than $10^{12.5} \MSUN$ and between $z\sim1$ and today (Figure \ref{fig:sm2hm}, bottom left, and Table \ref{tab:massfits}). 
At $z=0$ the different cluster components are ranked in their scatter: the total stellar mass in the whole halo (central + ICL + satellites) has the smallest scatter ($\la$ 0.1 dex), followed by the total diffuse component (central+ICL, 0.11 dex), the central galaxy and the stellar mass in the outskirts ($\sim$ 0.13 dex), and finally the stellar mass locked in satellites, with the largest halo-to-halo variation, slightly larger than 0.2 dex.\\

\item In Section \ref{sec:theory}, we have shown in quantitative terms the extent to which massive galaxies that form at the centers of massive haloes are the culmination of the hierarchical growth of structure: for $\MTC\ga 10^{14}\MSUN$, 80 per cent of their total stellar mass has been accreted via merging and stripping of smaller luminous galaxies (Figure \ref{fig:exsitu}). This ex-situ fraction exceeds 90 per cent when considering the stellar mass in the outskirts, e.g. beyond clustercentric distances of 100 kpc. In fact, the contribution of ex-situ stars to the stellar mass content of the most massive central galaxies is as high as 60 per cent even within their innermost 10 kpc.\\

\item Not surprisingly, for very massive objects the bulk of this accreted material has been brought by rather massive galaxies (Figure \ref{fig:progmass}): we find that 90 per cent of the accreted stellar mass in galaxies larger than about $5\times 10^{11}\MSUN$ in stars ($\MTC \ga 10^{14}\MSUN$) has been acquired from galaxies of at least $3-5 \times 10^{10}\MSUN$, i.e. galaxies as massive as our current Milky Way. For the stars in the ICL, this picture is less top heavy: at clustercentric distances larger than e.g. 100 kpc, 90 per cent of the accreted material has been stripped from galaxies at least as massive as $5-8 \times 10^9\MSUN$, making the outer cluster envelopes the smooth bridge between the most massive and rare galaxies in the Universe and the many more numerous dwarf ellipticals at the lower end of the galaxy mass function. 

\end{itemize}

All in all our analysis confirms that halo mass is a very good predictor of stellar mass, and vice versa, i.e. that by measuring the stellar mass within e.g. 30 kpc in massive galaxies ($\ga$ a few $10^{11}\MSUN$), the total mass of the underlying DM halo can be known to about 0.2 dex precision. In fact, we have shown that halo mass alone is a very good predictor of the whole stellar mass profile of massive galaxies beyond a few kpc from the center, at least on average and at 5-10\% accuracy (Section \ref{sec:profiles}, Table \ref{tab:profilefits} and Figure \ref{fig:stellarprofilesfits_1}). To provide a new comprehensive benchmark for comparison to observations and other models, we have supplied fitting formulae for the 3D stellar enclosed-mass profiles out to large radii of TNG groups and clusters at $z=0$, profiles that can be fully computed from either the total halo mass or e.g. the stellar mass measured within 30 kpc. 

By exploring the distribution of stellar material throughout the entirety of dark matter haloes, we have focused on results from the TNG simulations in a regime where our model is fully predictive. The spatial distribution of the stellar mass in massive haloes is a non-trivial outcome of the model, as it depends on many physical processes including the stripping and tidal destruction of infalling satellites within the framework of hierarchical structure formation. In fact, here we have pushed our analysis to a mass regime ($\MTC \ga 10^{14}\MSUN$) where our galaxy physics model was never directly calibrated, or indeed even ever run at all before the execution of the TNG300 simulation itself. 

In fact, all the considerations above demonstrate that the high mass end of the galaxy and halo mass functions is an exquisitely precise and challenging test bed for galaxy physics models. If the stellar mass in the most massive objects in the Universe is 80 per cent accreted, not only it is important to identify the right quenching mechanisms of its central (in-situ) galaxy, but it becomes even more relevant how the mechanisms that regulate star formation act across the whole spectrum of accreted galaxies. 
Improvements in the adopted AGN feedback mechanisms at low accretion rates \citep{Weinberger:2017} and the refined galactic winds regulating lower mass galaxies \citep{Pillepich:2017} give us confidence in the TNG model outcome, both in terms of mass budget within the central galaxy as well as in the distant, low surface brightness outskirts. This can be appreciated through the comparisons of TNG galaxies to observational and semi-empirical constraints, including the galaxy color bimodality distribution at low redshift (\textcolor{blue}{Nelson et al. 2017}), the shapes of the stellar-to-halo mass relation (Figure \ref{fig:sm2hm}) and the galaxy stellar mass functions across redshift (Figure \ref{fig:mf}). While the latter confirms the good agreement of the abundance (and hence SMHM) of TNG galaxies below $\la 10^{11}\MSUN$ in stellar mass, above this limit our preliminary comparisons to observational data indicates that the TNG relation between centrals stellar mass ($< 30$ kpc) and halo mass at $z=0$ is steeper than the one inferred by e.g. \cite{Kravtsov:2014}. It remains to be determined whether this is a fundamental limitation of the model or more subtle issues lie in the determination of the measured stellar masses. From the modeling view point, effects of numerical resolution also impair such comparisons at face value. Throughout this paper we have rescaled the stellar mass results from the TNG300 box to match the known outcome of our model at the superior resolution of the TNG100 simulation (see Appendix \ref{sec:app_res}). While the applied resolution rescaling factors are relatively small (at most $\la 1.4$), such corrections (or lack of thereof) may easily be sufficient to shift the perceived conclusions derived from the comparison to observational data.

However, in closing we emphasize that, to converge towards a consistent and accurate interpretation of models and observations, any definition of a galaxy's stellar mass must be made with caution and explicitly described in all diagnostics. We have attempted to do so throughout this analysis, with the hope to set up a new common standard. In Figure \ref{fig:mf}, we have finally shown the extent to which the shape of the massive end of the galaxy stellar mass function at recent times changes for different (3D spherical) apertures over which we measure TNG stellar masses. While the canonical 30 kpc aperture captures more than 90 per cent of the stellar mass of a galaxy sitting in a $10^{12}\MSUN$ halo (Figure \ref{fig:profiles_prototypes}), this aperture heavily underestimates the more massive objects: at a stellar mass of $10^{12}\MSUN$  there can be a factor of 10 difference in the estimated galaxy abundance for different choices of the central's mass definition. We hence advocate that the most practical and least ambiguous definition for a galaxy stellar mass should be measured within some fixed aperture in physical kiloparsecs (3D spherical or 2D circularized). Echoing previous works, we also discourage attempts to separate the outer component of the BCGs (the ICL) from the central galaxies, if not by means of simple spatial, arbitrary boundary. From our analyses of the stellar mass profiles (Section \ref{sec:profiles}), we could not identify any optimal, qualitative or generalized {\it physical} transition between the inner bright regions of galaxies and their lower surface brightness envelopes -- at least from a population-wide inspection of stellar mass density profiles.

In fact, a pragmatic separation of the cluster components based on fixed apertures is validated by theoretical arguments: all (stellar) haloes, regardless of their mass, formed by smooth accretion and merging, even though in different relative amounts at different masses. At the high mass end, stellar mass accretion is the dominant mechanism for the build up of {\it both} the inner and the most distant stellar components of a cluster: this makes the distinction between a central galaxy and its ICL rather arbitrary. 

\section*{Acknowledgements}

The authors thank the anonymous referee for the useful and detailed comments and suggestions. A special grateful note goes to Vicente Rodriguez-Gomez for sharing his codes and data files for the merger trees and stellar assembly catalogs of the IllustrisTNG simulations. It is also our pleasure to thank Arjen van der Wel, Alexie Leauthaud, Song Huang, Andrej Kravtsov, Benedetta Vulcani, Joe Mohr, Veronica Strazzullo, Alex Saro, Greg Rudnick, and Glenn van de Ven for useful conversations. AP thanks Ben Moster for sharing the observational data collected for the galaxy stellar mass functions at different redshifts and the organizers and participants of the Galapagos conference ``On the Origin (and Evolution) of Baryonic Galaxy haloes'' and of ``The Galaxy-Halo Connection'' KITP program for inspiring input. 
VS, RW, and RP acknowledge support through the European Research Council under ERC-StG grant EXAGAL-308037 and would like to thank the Klaus Tschira Foundation. PT acknowledges support from NASA through Hubble Fellowship grant HST-HF-51384.001-A awarded by the Space Telescope Science Institute, which is operated by the Association of Universities for Research in Astronomy, Inc., for NASA, under contract NAS5-26555. SG, through the Flatiron Institute is supported by the Simons Foundation and by the Hubble Fellowship grant HST-HF2-51384.001-A through NASA. JPN is supported by the NSF AARF award AST-1402480 and MV acknowledges support through an MIT RSC award, the support of the Alfred P. Sloan Foundation, and support by NASA ATP grant NNX17AG29G. 
The flagship simulations of the IllustrisTNG project used in this work have been run on the HazelHen Cray XC40-system at the High Performance Computing Center Stuttgart as part of project GCS- ILLU of the Gauss centers for Supercomputing (GCS). Ancillary and test runs of the project were also run on the Stampede supercomputer at TACC/XSEDE (allocation AST140063), at the Hydra and Draco supercomputers at the Max Planck Computing and Data Facility, and on the MIT/Harvard computing facilities supported by FAS and MIT MKI.

\bibliographystyle{mnras}
\bibliography{TNGPaper_I}


\appendix
\label{sec:appendix}

\section{Resolution studies of the diagnostics presented in this paper}
\label{sec:app_res}

In Section \ref{sec:res} we discussed the issue of numerical convergence, which has already been explored in some depth for the TNG model in \citet[][Appendix A]{Pillepich:2017} and \citet[][Appendix B]{Weinberger:2017}. Broadly speaking, any given property of a simulated galaxy or galaxy halo may not be necessarily fully converged at the particular mass/spatial resolutions available in TNG100(-1) and TNG300(-1). We generally find that both stellar masses and star formation rates increase with better resolution, for dark matter haloes of a fixed mass. These effects are more severe at low masses, where systems may be `resolved' with only 100 or 1000 resolution elements. And because the extended stellar components of galaxy groups and clusters, the main topic of this paper, are built up from the accreted remnants of less massive galaxy mergers, we add some additional details to better understand the relevant impact.

Here, we demonstrate how we take full advantage of all runs of the IllustrisTNG series presented in this paper (see Table \ref{tab:sims}).

\subsection{Stellar masses}
 
\begin{figure}
\centering
\includegraphics[width=8.3cm]{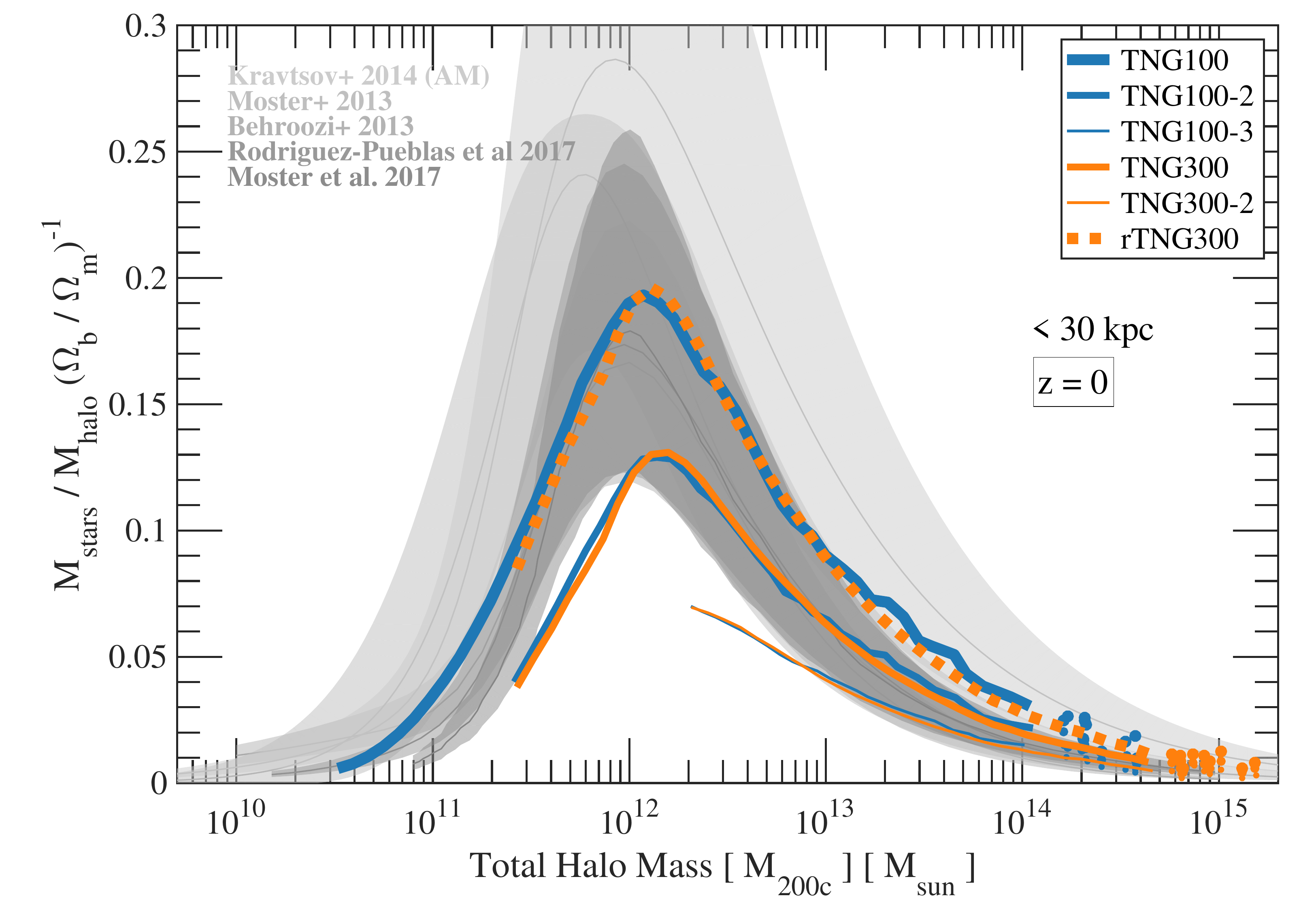}
\includegraphics[width=8.3cm]{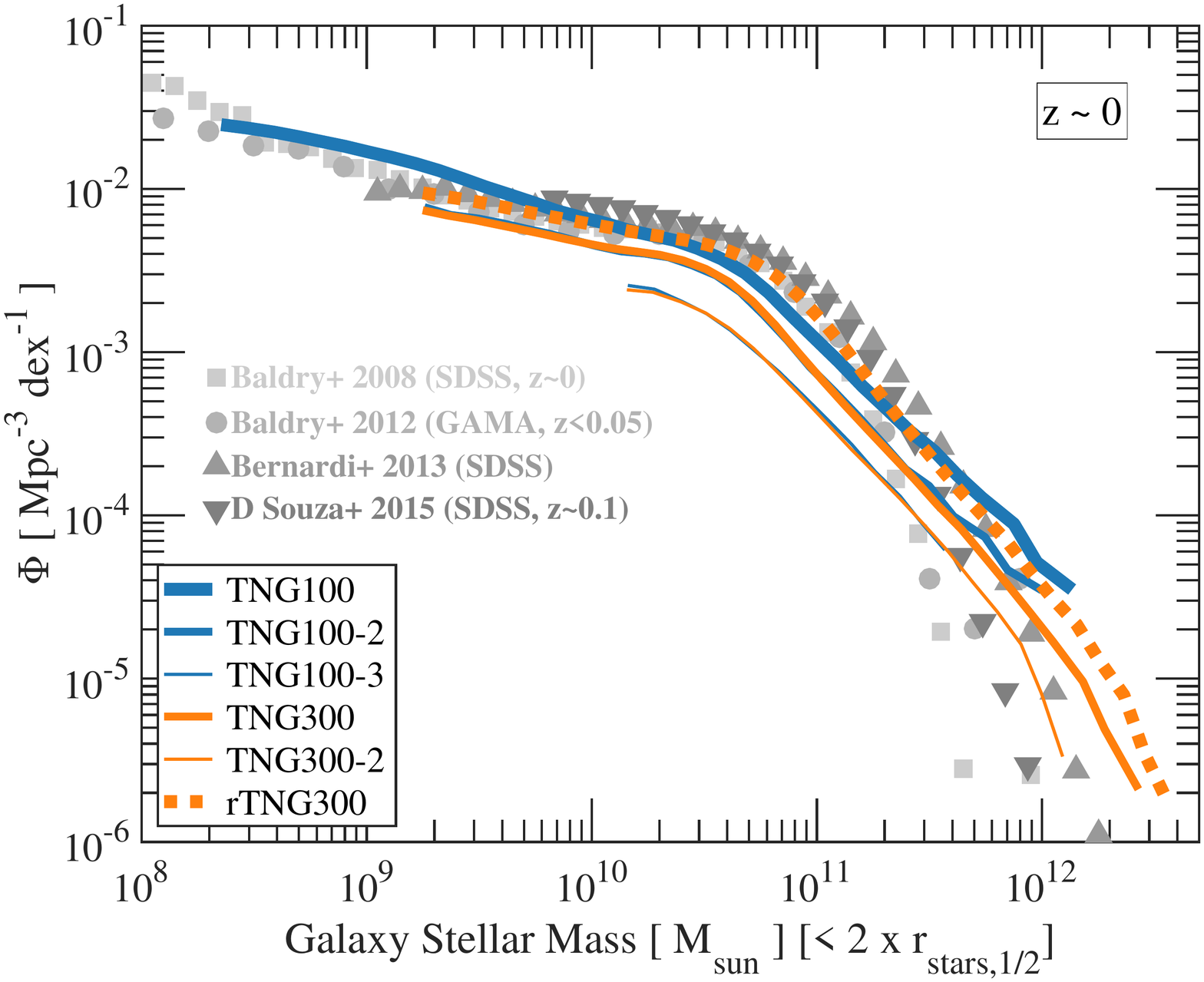}
\caption{\label{fig:resolution} Resolution effects and comparison across cosmological volumes on some of the galaxy statistics presented in this paper.  Top:  SMHM relation at $z=0$; bottom: galaxy stellar mass function at $z=0$. In each panel, we show the results from  all our resolution runs (see Table \ref{tab:sims}). As a reminder, TNG100-2 and the flagship TNG300 run share the same spatial and mass resolutions, as well as TNG100-3 and TNG300-2, respectively 8 and 64 times worse particle mass resolution than the highest resolution run of the series: TNG100. Better resolution implies larger stellar masses at fixed halo mass (top), but different volumes at the same resolution return perfectly consistent results. We therefore use the comparisons on a population basis between TNG100 and TNG100-2 to estimate by how much TNG300' stellar masses are underestimated. Throughout this paper, curves denoted `rTNG300'  (here dotted) represent the TNG300 results {\it multiplied} by a resolution correction function informed by the comparison between TNG100 and TNG100-2 on the same statistics or observable.}
\end{figure}

Fig.~\ref{fig:resolution} shows the convergence of the stellar mass to halo mass relation (top) and galaxy stellar mass function (bottom) at $z=0$, following from Figures \ref{fig:sm2hm} and \ref{fig:mf}. All three resolution levels of TNG100 (blue lines, varying thickness) together with the overlapping resolutions of TNG300 as well as rTNG300 (orange lines) are included. TNG100-2 lies exactly on top of TNG300-1, and likewise for TNG100-3 versus TNG300-2, indicating the same resolution pairs realized in different volumes behave identically. For the integral galaxy stellar mass within a given aperture, the inspection of the ratio between TNG100(-1) and TNG100-2 reveals that, the stellar masses of galaxies in haloes $\geq 10^{12.5}\MSUN$ in our flagship TNG300 run are underestimated in comparison to TNG100 by about 40\%, with no appreciable mass trend or dependence on the aperture over which the stellar mass is measured. A simple, constant multiplicative factor of $\simeq$ 1.4 applied to e.g. TNG300-1 results, shifting them to the TNG100-1 equivalents, would therefore offer an alternative, zeroth order correction for the observed stellar mass trends with resolution. Such ratio drops at larger redshifts, being as small as 1.3 (1.1) at $z\sim1$ (2). For smaller halo masses, however, the discrepancy due to resolution is larger and mass dependent. In this paper, for any observable involving estimating stellar masses ($M_{\rm s}$) as a function of (bins of) halo mass ($M_{\rm h}$) at a given redshift, we hence apply a mass-modulated rescaling procedure to derive rTNG300, as follows. For halo mass (bins) $M_{\rm h}\la 10^{14}\MSUN$
\begin{small}
\begin{equation}
\label{eq:appendix}
M_{\rm s}(M_{\rm h}; \mbox{rTNG300}) = M_{\rm s}(M_{\rm h}; \mbox{TNG300}) \times \frac{M_{\rm s}(M_{\rm h}; \mbox{TNG100-1})}{M_{\rm s}(M_{\rm h};\mbox{TNG100-2})},
\end{equation}
\end{small}
whereas for halo mass (bins) $M_{\rm h}\geq 10^{14}\MSUN$, the last fractional term of the equation is averaged across all haloes in the range $10^{13}\MSUN \leq M_{\rm h} \leq 10^{14}\MSUN$.
De facto, for the groups and clusters of galaxies studied in this paper, the rescaling reduces to a simple correction by 30-50 per cent at $z=0$. 

In the case of the SMHM relation (Figure \ref{fig:resolution}, top), rTNG300 lies exactly on top of TNG100-1, as desired. The only discrepancy is at the high-mass end of TNG100 where the small volume of TNG100 leads to fluctuations from low number statistics. A similar procedure is applied to the observables of e.g. Figure \ref{fig:stellarmasses}.
When presenting $z=0.5$ or $z=1$ results, we have derived and applied the corrections at those times, with decreasing importance towards high redshift.

For the galaxy stellar mass function $\phi$ at any given time (see Figure \ref{fig:resolution}, bottom panel), the resolution-rescaled realization of TNG300 (rTNG300) is obtained by correcting on a halo-by-halo basis the stellar masses of each TNG300 haloes, according to Eq.~(\ref{eq:appendix}).  
This procedure returns a very similar answer to the alternative and more straightforward multiplication of the TNG300 galaxy stellar mass function by a mass-constant factor, namely: $\phi(\mbox{rTNG300}) = \phi(\mbox{TNG300}) \times \langle \phi(\mbox{TNG100-1}) / \phi(\mbox{TNG100-2})\rangle$, where angle brackets denote averages across the entire galaxy stellar mass range.

 \begin{figure}
\centering
\includegraphics[width=8.3cm]{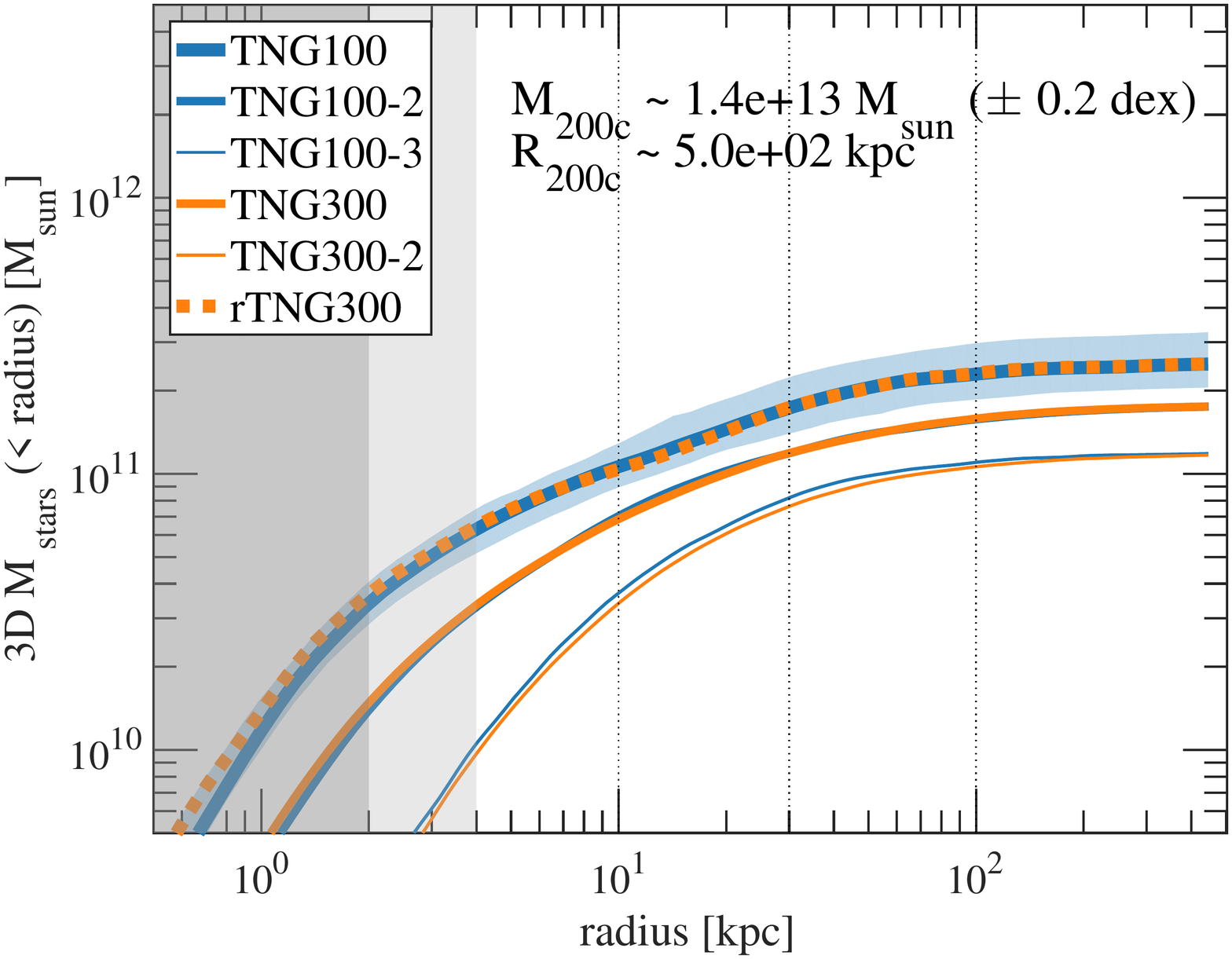}
\includegraphics[width=8.3cm]{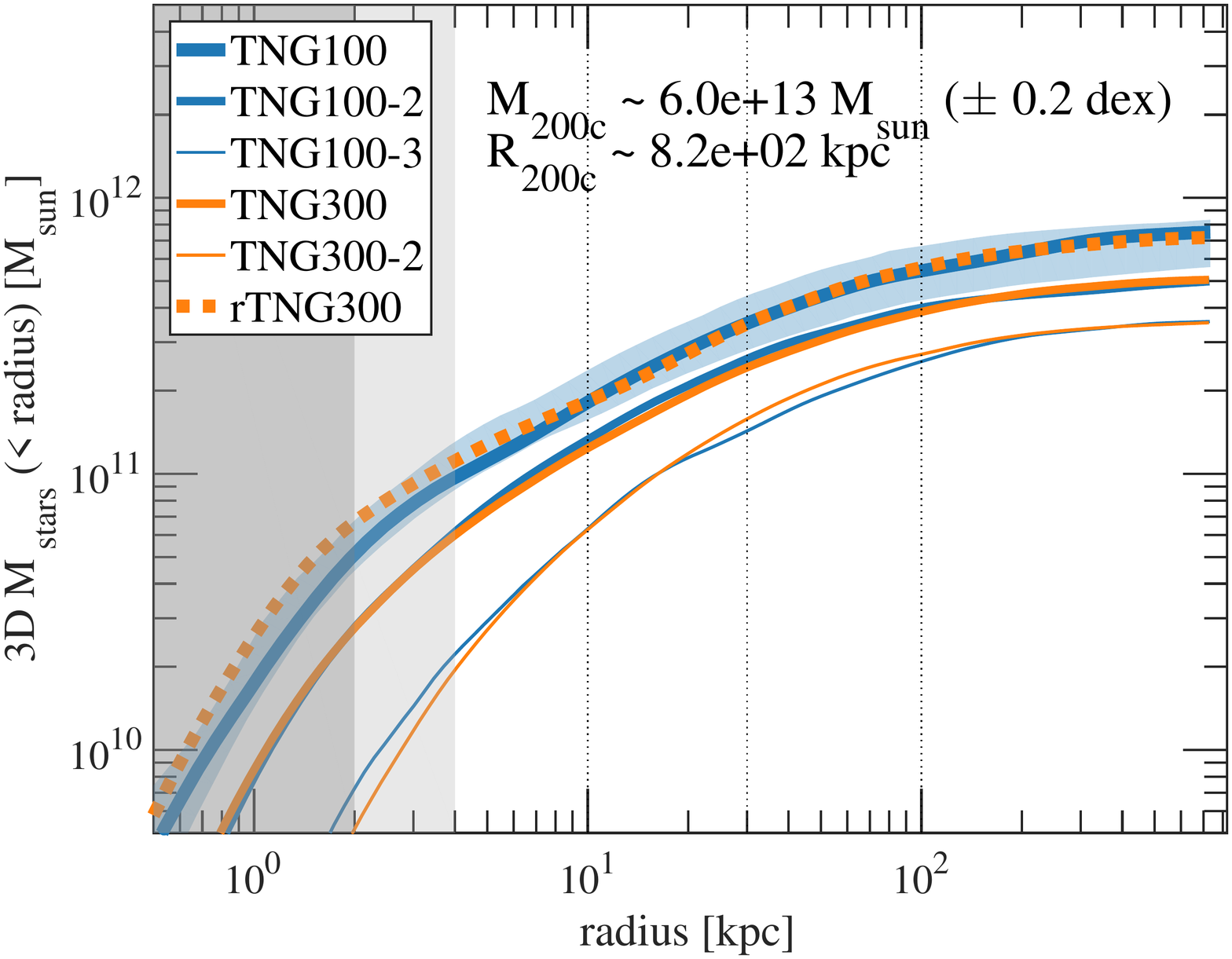}
\caption{\label{fig:resolution_2} Resolution effects and comparison across cosmological volumes on some of the galaxy statistics presented in this paper: here we focus on the stellar mass profiles in two mass bins where TNG100(-1) and TNG100-2 offer good statistics (annotations as in Figure \ref{fig:stellarprofiles}). The shaded areas below $\la 2-4$ kpc indicate the regions where our spatial resolutions do not allow us precise statements.}
\end{figure}

 \subsection{Stellar mass profiles}
 
Finally, Figure \ref{fig:resolution_2} shows two examples of stacked, 3D, enclosed stellar mass profiles in analogy to Figure \ref{fig:stellarprofiles}. We include all three resolution levels of TNG100 (blue lines, varying thickness) and the two highest resolution of TNG300 (thin orange curves) together with the rescaled rTNG300 (thick dotted orange). We reassuringly see that the stacked profiles of TNG300(-1) are identical to those of TNG100-2 in the same halo mass bins, as these two simulations have the same numerical resolution. Likewise, the two stacked profiles at TNG300-2 resolution are in good agreement with those from TNG100-3. For the profiles, we use information also on the radial behaviour of TNG100-2 vs TNG100(-1) to obtain a resolution-rescaled version of TNG300.  
Namely, we measure the ratio of the profiles in TNG100(-1) to TNG100-2 in bins of radial distance, and hence the rescaling factor is allowed to be modulated with cluster centric distance and to differ at different radii. The standard rescaling prescribed in Eq.~\ref{eq:appendix} would simply underestimate in a non-negligible fashion the rescaling at a few kpc from the centres, i.e. rTNG300 profiles would appear too low at radii smaller than a few kpc, and visibly so in the not normalised plots of Figures~\ref{fig:stellarprofiles} and \ref{fig:resolution_2}.
With the appropriate procedure, the two rTNG300 profiles are in excellent agreement with the TNG100-1 results, particularly at masses where enough TNG100-1 haloes exist to robustly estimate the mass distributions in the average. For halo masses above $10^{14}\MSUN$, we apply a radial-dependent correction informed by the comparison of the stacked profiles in TNG100(-1) and TNG100-2 averaged in the well-sampled mass bin $10^{13}\MSUN \leq M_{\rm h} \leq 10^{14}\MSUN$.

In Figure \ref{fig:resolution_2}, the only visible discrepancy is at small radii, less than a few kpc and $<$ 1 kpc in particular. In fact, different resolutions may disagree on the exact shapes and on the first derivatives of the stellar profiles in certain distance ranges. Even with a rescaling procedure we cannot a priori fully trust the results at TNG100-2 resolution in the innermost regions of galaxies. Here we observe that the stellar masses of the inner bulges of rTNG300 profiles may indeed be too large due to a breakdown of the rescaling procedure at the halo centers. With our best stellar softening length fixed at about 0.7 (1.4) kpc at our the resolutions of TNG100 (TNG300), we cannot in any case claim any quantitative or qualitative numerical convergence at radii smaller than about 2.8 times the softening length. Although the rescaled profiles of TNG300 agree with the profiles of TNG100 to better than 10-20 per cent also in the $2-4$ kpc range, we fix at 2 and 4 kpc our spatial resolution limits - indicated as shaded areas. We will restrict our investigation, including the various quantitative fits to stellar mass profiles, to outside the $2-4$ kpc regime, where we expect of order a few per cent accuracy or better.

When both stellar mass and radial apertures are renormalized to the total stellar mass and the virial radius, respectively (as in Fig.~\ref{fig:stellarprofiles}, top left panel), the large radii behaviour (i.e. beyond one per cent of the virial radius) is well captured also at TNG300 resolution, at least in the overlapping mass bins we can test (not shown). Analogously, for the 3D stellar mass density profile (Fig.~\ref{fig:stellarprofiles}, top right panel), rTNG300 seems to return a more faithful estimate of the amplitude (although not necessarily of the shape) of the profiles within about 10 kpc, while providing large scale profiles in very good agreement with its lower version TNG300.
\newline

In closing, it is worthwhile to emphasise that, while the incomplete resolution convergence of TNG300 is undoubtably a limitation of our model, the needed rescaling factors of $\la 1.4$ described thus far are rather small. In practice, they are comparable or smaller than the ratios between different choices of a galaxy's stellar mass definition (e.g. within 30 kpc vs. within 10 kpc for $10^{14}\MSUN$ haloes, see e.g. Fig.~\ref{fig:sm2hm}). Else, the massive end of the $z=0$ GSMF is modified to a lesser degree by the resolution rescaling described in this Appendix than the current discrepancies across different observational measurements above galaxy stellar masses of $10^{11}\MSUN$. We argue that the determinism of our simulated results -- both at equal resolutions in different (large) volumes and at different resolutions in the same volume -- is a powerful feature of our numerical model: it allows us to infer to a good level of accuracy how the model functions at increasingly better resolution by knowing how it performs at lower resolutions.

\end{document}